\documentclass[a4paper,11pt]{article}
\pdfoutput=1 

\usepackage{jheppub} 

\usepackage[T1]{fontenc} 
\usepackage{blindtext}
\usepackage{epstopdf}
\usepackage{graphicx}
\usepackage{epsfig}
\usepackage{dcolumn}  
\usepackage{bm}    
\usepackage{caption}
\usepackage{subcaption}
\usepackage{amssymb} 
\usepackage{tikz}
\usetikzlibrary{arrows}
\usepackage{nccmath}
\usepackage{xcolor}
\usepackage{epstopdf}
\usepackage{graphicx,wrapfig,lipsum}
\usepackage{epsfig}
\usepackage{amsmath,bm}
\usepackage{amsfonts}  
\usepackage{amsmath}  
\usepackage{slashed}  
\usepackage{enumitem}
\usepackage[mathscr]{euscript}
\usepackage{tabu}
\usepackage{epsfig}
\makeatletter
\g@addto@macro\bfseries{\boldmath}
\makeatother

\def\l1{{{1-loop}}}

\def\n1{\Bigg|_{n=1}}

\def\n{{(n)}}

\def\Im{{\text{Im}}}

\usepackage[T1]{fontenc} 
\usepackage{tikz}
\usepackage{amsmath,amssymb}
\usepackage{relsize}
\usepackage{latexsym}
\usepackage{leftidx}
\usepackage{xcolor}
\usepackage{diagbox}
\usepackage[T1]{fontenc}
\usepackage{array}
\usepackage{makecell}
\usepackage{csquotes}
\usepackage{tikz}
\usepackage{enumitem}
\usepackage{setspace}
\usepackage{multirow}
\usepackage{amsmath,amssymb}
\usepackage{relsize}
\usepackage{latexsym}
\usepackage{leftidx}
\usepackage{xcolor}
\usepackage{csquotes}
\usepackage{tikz}
\usetikzlibrary{decorations.pathmorphing}
\usetikzlibrary{arrows.meta}
\usepackage{enumitem}
\usetikzlibrary{decorations.markings}
\usetikzlibrary{decorations.pathmorphing}
\usetikzlibrary{decorations.markings}
\usetikzlibrary{decorations.pathmorphing}
\usepackage{pifont}
\usepackage{hyperref}
\usepackage{url}

\usepackage{bookmark}

\title{\textbf{\textsf{One point functions in large  $N$  vector models at finite chemical potential
}}}
\author{Justin R. David, Srijan Kumar}
\affiliation{\vspace{.1cm} Centre for High Energy Physics, \\ Indian Institute of Science,\\
	C. V. Raman Avenue, Bangalore 560012, India.}
\emailAdd{justin@iisc.ac.in, srijankumar@iisc.ac.in}
\abstract{
	We evaluate the thermal one point function of higher spin currents 
	in the critical 
	model of  $U(N)$ complex scalars interacting with a quartic potential 
	and the $U(N)$  Gross-Neveu model of Dirac fermions at large $N$  and strong coupling using the Euclidean 
	inversion formula. 
	These models are considered 
	in odd space time dimensions $d$ and held at  finite temperature and finite  real chemical potential $\mu$ measured 
	in units of the temperature. 
	We show that these one point functions simplify both at large spin and large $d$.
	At large spin, the one point functions behave as though the theory is free,  the chemical potential 
	appears through a simple pre-factor which is either 
	$\cosh\mu$  or $\sinh\mu $ depending on whether the spin is even  or odd. 
	At large $d$, but at finite spin and chemical potential, the 1-point functions  are suppressed  exponentially in
	$d$  compared to the free theory.   We study a fixed point of the critical  Gross-Neveu model in $d=3$ with 1-point 
	functions  exhibiting  a branch cut  in the chemical potential plane. The  critical exponent 
	for the free energy or the pressure 
	at the branch point  is $3/2$
	which coincides with the mean field exponent of the  Lee-Yang edge singularity for repulsive core
	interactions.}
\begin{document}
	\maketitle
	\flushbottom
	
	\section{Introduction}

	Studying large $N$ 
	conformal field theories at finite temperature  is relevant to understand black hole physics. This connection arises 
	from the AdS/CFT correspondence which relates certain large $N$ theories at finite temperature to black holes in 
	$AdS$. Similarly when such  CFT's are also held at finite chemical potential, they are dual to charged black holes in 
	$AdS$. Among large $N$ CFT's,  vector models 
	are solvable and  are dual to Vasiliev's higher spin theories in $AdS$ \cite{Giombi:2012ms}. 
	The thermal properties of these theories are therefore 
	of interest to  shed light on possible higher-spin black holes solutions in higher spin gravity.

	In this paper we use the Euclidean inversion formula obtained in \cite{Iliesiu:2018fao,Petkou:2018ynm,David:2023uya} 
	to study the behaviour of one point functions
	of higher spin currents
	in the  model of $N$   complex bosons in the fundamental of $U(N)$  interacting with a quartic potential
	 and the $U(N)$ Gross-Neveu model of Dirac fermions. 
	The model of  scalars and the $U(N)$ Gross-Neveu model of Dirac fermions are important examples of 
	quantum field theories. 
	Apart from being holographic duals to Vasiliev theories, 
	these models are relevant to understand critical phenomenon,   quantum 
	chromodynamics and large $N$ behaviour. 
	Though these models were introduced  and primarily studied in $d=3$ and $d=2$ space time dimensions, they 
	have been explored in higher dimensions to study  renormalization group flows and non-trivial fixed points.
	\cite{Vasiliev:1981dg,Lang:1992zw, Petkou:1994ad,Petkou:1995vu,Fei:2014yja,Fei:2014xta,Stergiou:2015roa, Guerrieri:2016whh,Gliozzi:2016ysv,Gliozzi:2017hni,Filothodoros:2018pdj,Giombi:2019upv}. 
	 These models with their quartic interactions are non-renormalizable in higher 
	dimensions, the existence of these fixed points are inferred through an analysis which involves the Hubbard-Stratonovich
	transformation and large $N$.

	We  study the  thermal one point functions of higher spin currents 
	analytically for both   the critical  scalar theory  and the critical Gross-Neveu models in $ d=2k+1$ 
	dimensions at large $N$ and at strong coupling.   
	We  generalise the analysis to the case when these systems are held at 
	non-zero chemical potential. These one point functions determine the higher spin charges of possible charged 
	black hole solutions dual to the CFT thermal states in higher spin theory.

	The  Euclidean inversion formula  in \cite{Iliesiu:2018fao} and that 
	developed   for CFT's with fermions in \cite{Petkou:2018ynm,David:2023uya} 
	 is suitable to  obtain the expressions for thermal one point 
	functions of higher spin currents and can be generalised for non-zero real  chemical potentials. 
	In \cite{David:2023uya}, there were observations made  based on studying the one points functions numerically about their
	behaviour at large spin and large $d$. 
	In this paper we derive these observations analytically by studying the asymptotic behaviour of the one point functions and 
	generalise the observations to non-zero real chemical potentials. 
	The one point functions that we wish to study in detail in these theories are 
	\begin{eqnarray}
		a_{{\cal O} }[0, l] &=& 
		\langle {\cal O}[0, l ] \rangle_{\beta, \mu } = \langle \tilde \phi^\dagger D_{\nu_1}  D_{\nu_2} \cdots D_{\nu_l }\tilde \phi \rangle_{\beta , \mu } , \\ 
		\nonumber
		a_{{\cal O}_+ } [0, l]&=& 
		\langle {\cal O}_+ [0, l ] \rangle_{\beta, \mu } = \langle \tilde \psi^\dagger  \gamma_{\nu_1} D_{\nu_2} D_{\nu_3} \cdots 
		D_{\nu_l}\tilde \psi \rangle_{\beta, \mu} .
	\end{eqnarray}
	where these currents are symmetric traceless tensors of rank $l$. 
	$\mu $ is  the chemical potential measured in units of the temperature, $\tilde\phi$ are 
	complex  bosons, and $\tilde\psi$ are Dirac fermions transforming under $U(N)$. 
	The covariant derivative in the Euclidean 
	time direction is given by  $D_\tau = \partial_\tau   -\mu $ and  in the spatial directions 
	we have  $D_i = \partial_i$.

	The resulting one point functions can be cast in terms of sums of polylogarithms just as in the case of the stress 
	tensor of the $O(N)$ model which was first observed by \cite{Sachdev:1993pr}. 
	However this does not reveal much insight. 
	Early in the 
	 conformal bootstrap program it was noticed
	  that  anomalous dimensions as well as correlators
	   simplify for operators at large spin \cite{Fitzpatrick:2012yx,Komargodski:2012ek}. 
	  With this intuition, we examine the one point functions at large spin. 
	We  show that the one point functions in both the models 
	simplify and they  are independent of the thermal mass of the theories at large spin $l$. 
	We obtain the following universal behaviour for large spins in the critical $O(N)$ model at large $N$ and 
	at strong coupling. 
	\begin{eqnarray}
		\lim_{l \rightarrow \infty} a_{ {\cal O}   }[0, l]=   a^{\rm free}_{ {\cal O} \    }[0, \infty]\times  \begin{cases}
			\cosh \mu , \qquad l \;\;{\rm even}, \\
			\sinh \mu, \qquad l \;\; {\rm odd}.
		\end{cases} 
	\end{eqnarray}
	where $ a^{\rm free}_{ {\cal O}   }[0, \infty]$ is the one point functions of the corresponding free  massless theory
	at large spin. It is important to note that this result holds for the well studied  model in 
	$d=3$ space-time dimensions. As far as we are aware, this has not been observed earlier in the literature. 
	Similarly for the large $N$  critical Gross-Neveu model at strong coupling we obtain 
	\begin{eqnarray}
		\lim_{l \rightarrow \infty} a_{{\cal O}_+  }[0, l] = a^ {\rm free}_{ {\cal O}_+   }[0, \infty]\times  \begin{cases}
			
			\cosh \mu , \qquad l \;\;{\rm even}, \\
			\sinh \mu, \qquad l \;\; {\rm odd}.
		\end{cases} 
	\end{eqnarray}
	where $a^{\rm free}_{ {\cal O}_+   }[0, \infty]$ is the one point functions of the  theory
	of massless fermions at large $l$.

	Next we study the behaviour of the thermal one point functions  at large $d$.   
	This question is motivated by the 
	observations  that  conformal blocks  at large $d$  simplify \cite{Fitzpatrick:2013sya}. 
and the  recent conjectures in \cite{Gadde:2020nwg,Gadde:2023daq} 
	that conformal field theories at large $d$ are trivial, or non-unitary.  The large $d$ limit also is useful 
	 in holography, see \cite{Emparan:2020inr} for a review of large $d$ methods in 
	gravity and \cite{Giataganas:2021jbj,David:2022nfn} for applications in holography. 
	There has been   no study using explicit examples of CFT's.
	To obtain the thermal one point functions,  it is necessary to solve 
	the gap equation and obtain the thermal mass.  
	The gap equation at large $N$ and strong coupling coincides for both the  
	case of complex scalars 
	 and the Gross-Neveu model. At large $d= 2k+1$, it   is given by 
	\begin{equation}
		\frac{(m_{\rm th}) ^{k-\frac{1}{2}}  {\pi}}{\Gamma( \frac{1}{2} + k)}-{2^{k+\frac{3}{2}} K_{k-\frac{1}{2}}(m_{\rm th})}\cosh(\mu )=0.
	\end{equation}
	Here $k$ is odd for the bosonic model and even for the fermionic one and $m_{\rm th}$ is the thermal mass in units of the 
	temperature. 
	This  transcendental equation admits a solution for the thermal mass which scales with the dimension $k$ and is given by
	\begin{equation}
		\lim_{k\rightarrow \infty} m_{\rm th} = 0.69486 k + 0.313322\log(\cosh \mu) + 0.0560686 .
	\end{equation}
	Here again  $k$ is odd for the bosonic model and $k$ even for the fermionic model. 
	The thermal one point functions also admit a uniform expression for both the models.
	For the $O(N)$ model we have 
	\begin{eqnarray} \nonumber
		\lim_{k\rightarrow \infty} \frac{ a_{{\cal O}} [0, l]  }{ a_{ {\cal O}   }^{\rm free}[0, l] }
		&=&  \Big[ 1.54 \exp\big(- 0.11k + 0.10 l - 0.098 \log \cosh\mu  - 0.60 \big)\Big]  \\ 
		&& \qquad \times 
		\begin{cases}
			\cosh \mu , \;\; l \;\;{\rm even}. \\
			\sinh \mu, \;\;  l \;\; {\rm odd}.
		\end{cases} 
	\end{eqnarray}
	The expression for the Gross-Neveu model is identical to the above expression, except that now $k$ needs to be 
	odd.  Here both the spin $l$ and the chemical potential $\mu$ are held fixed. 
	Note that the one point functions are exponentially suppressed at large dimensions and therefore the one 
	point functions vanish when compared to the theory at the Gaussian-fixed point. 
	This conclusion is consistent with the conjectures  of \cite{Gadde:2020nwg,Gadde:2023daq} 
	which  states that the conformal field theories are trivial 
	or non-unitary at large $d$.

	In arriving at these conclusions we test the expressions for the one point functions  of the stress tensor obtained directly 
	using the partition function at large $N$ and strong coupling of these models  held at finite chemical potential 
	against that obtained  from  the inversion formula. We also check that the gap equation  obtained from the saddle point of the 
	partition function agrees with that obtained by demanding that the expectation value of the spin zero fields
	$\langle  |\phi|^2 \rangle_{\beta, \mu} $ in the  model of complex scalars or $\langle \bar\psi \psi \rangle_{\beta, \mu }$ 
	in the Gross-Neveu model vanishes. 
	
	Since the case of $d=3$ is special both for the bosonic theory and the Gross-Neveu model, we study this case 
	in detail. 
	For the model of $U(N)$ scalars, we see that since there is no Bose-Einstein condensation in $d=3$, we can 
	take chemical potentials to be arbitrary large. 
	We obtain the leading behaviour of the one point functions  of the higher spin-currents at large values of the chemical 
	potential. For the case of the Gross-Neveu model, we study a fixed point at which the 
	thermal mass is imaginary. 
	This fixed point was first noticed in \cite{Petkou:2000xx}, 
	where it was argued that it can be related to a Yang-Lee edge singularity.  
	We study this fixed point with the chemical potential turned on and show that  all the one point functions 
	exhibit a branch cut in the chemical potential plane. The 
	critical exponent of the pressure or the stress tensor at the branch point is $3/2$. 
	This coincides with the mean field theory exponent at the  Yang-Lee edge singularity of systems whose
	interactions have a repulsive core singularity \cite{10.1063/1.470178}. 
	Though the thermal mass is imaginary at this fixed point, 
	 the one point functions of the operators ${\cal O}_+[0,l]$, which 
	includes the stress tensor are all real. The property relies on the symmetry $m_{\rm th} \rightarrow -m_{\rm th}$ 
	satisfied by  the  one point functions in the complex $m_{\rm th}$ plane, 
	which in turn is due to non-trivial  mathematical identities  satisfied by Bernoulli polynomials. 
	This is proved  in the appendix \ref{symmetry}.

	This paper is organised as follows. In section \ref{unmodel}
	 we study the  model of $N$ complex scalars in the fundamental of $U(N)$ 
	 which is followed by the section \ref{secgnmodel} on 
	the Gross-Neveu model. Section  \ref{conclude} contains our conclusions.
	The appendix \ref{gap part} 
	 contains the derivation of the gap equation and the stress tensor from the partition function of these 
	models at large $N$ and at strong coupling.  Appendix \ref{symmetry} discusses the symmetry of the one point functions 
	of the Gross-Neveu model under $m_{\rm th} \rightarrow -m_{\rm th}$.

	\section{Scalars in the fundamental of $U(N)$} \label{unmodel}

	The Euclidean inversion formula  was used in \cite{Iliesiu:2018fao} to obtain one point functions of  arbitrary 
	spin bi-linears  in  the $O(N)$ model in $d=3$ and  generalised to higher dimensions in \cite{Petkou:2018ynm}. 
	Recently in \cite{Karydas:2023ufs}, the inversion formula was  used on   the 
	massive thermal 2-point function of free  complex scalars with chemical 
	potential to read out  the  would be higher spin one-point functions had the theory been  conformal. 
	The authors then examine  the spin-0 and spin-1,  one point functions and show 
	that they agree  with the explicit calculation in 
	\cite{Petkou:2021zhg}, for the  theory at trivial fixed point with zero thermal mass. 
	They demonstrate that 
the traceless spin-2 current from the inversion formula agrees with that constructed from the 
	linear combination of the Hamiltonian, spin-1 and spin-0 current  at the trivial fixed point.  
	
	In this section we derive the inversion formula for one point functions of higher spin bi-linears of  $N$ 
	scalars in the fundamental of $U(N)$  with chemical potential, interacting through  a quartic potential. 
	Our crucial new input  is that demanding the vanishing of the 
	 expectation value of the spin-0 current   is equivalent to imposing
	the gap equation derived from the large $N$, strong coupling saddle point analysis of the 
	$U(N)$  model of complex scalars  interacting with a quartic potential. 
	This condition, then ensures the theory is a thermal conformal field theory even at non-zero thermal mass. 
	This fact was noticed in the absence of 
	 the chemical potential 
	 in \cite{Iliesiu:2018fao,Petkou:2018ynm,David:2023uya}.  Here we observe that it continues to hold in the presence of 
	 chemical potential. 
	 In fact once the gap equation is 
	satisfied,  the thermal mass in units of temperature or $m_{\rm th} \beta$ and the dimensionless 
	chemical potential $\mu \beta$ are related. 
	This relation ensures that the stress tensor and the free energy  $F$ are related as 
	\begin{equation} \label{partstresrel}
	T_{00} = (d-1) F.
	\end{equation}
	which is a signature of a thermal CFT.  Furthermore, 
the stress tensor obtained from the partition function 
	coincides with the spin-2 conserved current obtained from the inversion formula.
	The gap equation also 
	 ensures that all  higher spin point function scale with temperature according to the conformal dimensions
	as they should in a  thermal  CFT.  Therefore the gap equation is crucial to  ensure that the results for the spin-$s$ one point functions are that of a  thermal CFT. 
	
	We use the gap equation and the one point function to define thermal CFT's with real 
	 chemical potentials   and study their behaviour in odd space time dimensions.  In odd dimensions on can set the 
	 physical scale in the theory to a critical value, so that the theory behaves as a thermal CFT. 
	 In even space-time dimensions the large $N$ vector model  at finite temperature
	generates a scale and the gap equation depends logarithmically on the scale. 
	and is in general not a  thermal CFT, as can be seen for the $O(N)$ model in $d=4, d=2$ in \cite{Moshe:2003xn,Romatschke:2019gck}.    We have reviewed these features in appendices \ref{appendixa2} and \ref{A.2.1}. 
	Therefore we restrict our study to odd space time dimensions. 
	
	We begin with a detailed study of the model in $d=3$.   This  model has been extensively discussed in the literature. 
The value of the one point function of the stress tensor was obtained in 
\cite{Chubukov:1993aau,Sachdev:1993pr}.
However there has been 
few studies of  model at real values of chemical potential that we are aware of.
Earlier studies of 
 \cite{Filothodoros:2016txa,Filothodoros:2018pdj,Alvarez-Gaume:2019biu}  examined the model at imaginary values of the chemical potential. 
	The gap equation admits a  real solution for the thermal mass
	 at arbitrary  real values of chemical potential.   Since there is no Bose-Einstein condensation in $d=3$, we are able to tune the chemical potential to arbitrarily  large values. We  then study the behaviour of all higher spin one point functions as function of both the  chemical potential and spin. 
	 
	The only other  odd dimension which is within the unitary window is $d=5$
	\footnote{The composite field $|\phi|^2$ 
	has dimensions $2$ in all $d$ at the non-trivial fixed point. For unitarity we have the bound 
	$2\geq \frac{d}{2} -1$}. Here the gap equation admits complex conjugate pairs as solution for the thermal mass
	$m_{\rm th}$  at zero chemical potential \cite{Petkou:2018ynm,Giombi:2019upv}. 
	We have also verified that at each of these solutions, the free energy is complex. 
	Therefore, 
	 we should  first develop methods to understand these solutions so that they  are physical. One suggestion in 
	 \cite{Giombi:2019upv} is to sum over each of the solutions. 
	 We do not address this  in this paper and leave it for future work.

	The theory in  odd dimensions $d>5$ is non-unitary.  However, there is one real solution of the gap equation for 
	$d=2k +1$ for all $k$ odd. We obtain the expression for the thermal mass or the solution to the gap equation
	 at large $k$. The stress tensor  is real 
	has all the features of a thermal CFT  at 
	this saddle point. We also 
 demonstrate that the 
	ratio of thermal one point function of arbitrary spin bi-linears to the free Gaussian theory is exponentially suppressed in 
	$k$, once $k$ is sufficiently large even in the presence of chemical potential.

	The subsection is organised as follows, in section \ref{chemtwist}, 
	we recall how chemical potentials can be introduced by considering twisted boundary conditions with the 
	$U(N)$ model as an example. 
	In section \ref{3dsection}, we consider the model  $U(N)$ scalars in $d=3$ and develop the Euclidean inversion formula 
	for the higher spin currents from the thermal 2-point function at finite chemical potential. 
	We show that the gap equation for the large $N$ critical model coincides with the requirement that the 
	one point function of the spin-0 bi-linear vanishes.  The gap equation is derived from the partition function of the 
	model with quartic interaction at large $N$ and at large coupling in appendix \ref{unbfd}. 
	We then study the behaviour of the one-point functions of higher spin bi-linears.
	In $d=3$, there is no  Bose-Einstein condensation, therefore the chemical potential can be taken to infinity.
	In sub-sections \ref{largchar}, \ref{largel}
	we study the behaviour of the one point function of higher spin currents  for 2 limiting situations:
	large chemical potential, large spin.  In section \ref{greatert3}
	we  study the  model in $d=2k+1$ dimensions  for $k$ odd  and solve the 
	 gap equation at large $k$. We  obtain the  higher spin one-point functions and study their behaviour at large $k$ and spin. 
	The bosonic model for $k\geq 3$  is not unitary, nevertheless the theory exhibits an interesting fixed point at which 
	all thermal  one-point function of higher spin bi-linears  behave as  expected of a thermal CFT. 
	
		\subsection{Chemical potential and twisted boundary conditions} \label{chemtwist}
		
	Considering physical systems at finite chemical potential is an important deformation to  study 
	in the phase diagram of a given system. 
	In QCD,  turning on chemical potential has been studied extensively so as to understand the complete 
	phase diagram of QCD.  
	For conformal field theories which admit $AdS$ duals,  turning on chemical potential is dual to considering 
	a charged black hole in $AdS$
	and is  therefore  an important deformation to investigate. 
	Furthermore in this specific context of the use of Euclidean inversion formula other than the work in 
	\cite{Karydas:2023ufs}, there has been no other study of the Euclidean inversion formula with finite chemical potential. 
	A standard  approach to study field theories at finite chemical potential is to consider imaginary chemical potential 
	and then analytically continue results to real values of the chemical potential \cite{Alford:1998sd,deForcrand:2002hgr,DElia:2002tig,Karbstein:2006er}.  
	As long as one restricts the values of $\mu$ to be within phase boundaries, such analytical continuations can be used to 
	retrieve information of thermodynamic observables. 	 
	
	To be specific, let us consider the bosonic model of $N$ complex scalars whose action is given by, 
	\begin{eqnarray}\label{actbos1}
		S = \int d^{2k+1} x \left[  |( \partial_\tau  -i 
		\hat \mu  ) \tilde \phi |^2  +\partial_i \tilde\phi \partial_i \tilde\phi^* 
		+ \frac{\lambda}{N} |\tilde\phi|^4 \right].
	\end{eqnarray}
	Here $ \tilde\phi$ is a  $N$ dimensional vector of complex scalars and $\tilde\phi^* \tilde\phi = \sum_{a=1}^N \tilde\phi^{a\, *} \tilde\phi^a$, 
	the label  $i= 1, 2,   \cdots 2k$ refers to the spatial directions.  As we will see, the theory in odd space-time dimensions admits 
	a thermal CFT in the large $N$ strong coupling limit. In even space-time dimensions the theory at finite temperature
	generates a scale due to the cut off dependence  and is in general not a  thermal CFT, as can be seen for the $O(N)$ model in $d=4, 2$ in \cite{Moshe:2003xn}. 
	The Euclidean direction $\tau$ is identified using periodic boundary conditions
	\begin{equation}\label{pbc}
		\tilde\phi( \tau +\beta, x) = \tilde \phi( \tau, x) ,
	\end{equation}
	$\hat \mu $  in (\ref{actbos1})
	 is the chemical potential,  when $\hat\mu $ is real, the system is held at purely imaginary potential. 
	$m_{\rm th}$ is the thermal mass which will be  determined by the gap equation. 
	This equation is essentially the saddle point equation  
	obtained at large $N$ and large $\lambda$.   
	The gap equation relates $m_{\rm th}\beta$ to the chemical potential $\hat\mu\beta$.	
	We will  evaluate  the thermal expectation values of the traceless symmetric spin-$l$ bilinears
	\begin{equation} \label{untwist1pt}
		\langle {\cal O}[n, l ] \rangle_{\beta, \hat\mu}
		= \langle \tilde\phi^* D_{\mu_1} D_{\mu_2} \cdots  D_{\mu_l}  D^{2n}  \tilde\phi\rangle_{(\beta, \hat\mu)} ,
	\end{equation}
	where $D_\tau = \partial_\tau - i \hat\mu $ and $D_i = \partial_i$. 
	We study the behaviour of these  expectation values at large spin $l$ and large dimensions $k$ at the non-trivial critical 
	point where $m_{\rm th}$ satisfies the gap equation.  Our strategy is to use purely imaginary chemical potential during the 
	analysis of the Euclidean inversion formula and then analytically continue to real chemical potential while studying the gap equations and the one point functions.

	It is convenient to perform the re-definition for the bosonic field so as to absorb the chemical potential present in the covariant 
	derivative in terms of twisted boundary conditions. 
	Consider the field re-definition
	\begin{equation}\label{redef}
		\tilde  \phi (\tau, x) = e^{i \hat\mu  \tau} \phi(\tau, x) ,
	\end{equation}
	then note that the field  $ \phi$ obeys twisted boundary conditions which is inherited from the periodic boundary conditions 
	(\ref{pbc}) and the field re-definition (\ref{redef}). 
	\begin{equation}\label{twistbc}
		\phi(\tau + \beta, x) = e^{-i \hat\mu \beta } \phi( \tau, x) .
	\end{equation}
	Substituting the field redefinition, we see that the action of the covariant derivative becomes an ordinary derivative on 
	$\phi$
	\begin{equation}
		D_\tau \tilde \phi (\tau, x) = e^{i \hat\mu\tau} \partial_\tau  \phi( \tau, x) .
	\end{equation}
	Therefore in terms of the twisted field $\phi$, the spin-$l$ currents are given by 
	\begin{equation}
		\langle {\cal O}[n, l ] \rangle_{\beta, \hat\mu} = \langle \phi^* \partial_{\mu_1} \partial_{\mu_2} \cdots \partial_{\mu_l}  \partial^{2n} \phi\rangle_{(\beta, \hat\mu) } .
	\end{equation}
	Observe that the derivatives have become ordinary derivatives on the RHS. 
	To obtain the one point functions, we apply the  Euclidean inversion formula,  on the thermal 
	two point function of the twisted bosons.

	\subsection{The model in $d=3$ } \label{3dsection}
	
	The action  in $d=3$ is given by 
	\begin{align} \label{3dlag}
		S=\int d^3 x \bigg[|(\partial_\tau-i\hat \mu){\tilde \phi}|^2+\partial_i \tilde\phi \partial_i \tilde\phi^*+\frac{\lambda}{N}|{\tilde\phi}|^4\bigg].
	\end{align}
	where, ${\tilde \phi}$ is a complex scalar transforming in the fundamental of $U(N)$ and $i=1,2$. 
	We consider the theory on $S^1\times R^2$ along with the chemical potential $\hat \mu$. 
	As presented in the Lagrangian the chemical potential is purely imaginary, but once we obtain results for physical observables like energy density or thermal one point functions we will analytically continue  to real chemical potential by the replacement 
	$\hat \mu \rightarrow i \mu $. 
	In appendix \ref{unbfd}, we have used the Hubbard-Stratonovich trick to obtain the free energy at large $N$. 
	The saddle point analysis at large $N$ leads to the following gap equation  which determines the thermal mass 
	in terms of the chemical potential
	\begin{eqnarray} \label{gapeq3db}
	m_{\rm th} \beta + \log\Big( 1- e^{-\beta( m_{\rm th} +\mu) } \Big) + \log\Big( 1-e^{-\beta ( m_{\rm th} - \mu)} \Big)  = 0.
	\end{eqnarray}
	Here we have used equation (\ref{appengapeq}) with $k=1$ and analytically continued the chemical potential to real values. 
	This equation admits a real solution for the thermal mass for all values of $\mu$, which is given by 
	\begin{eqnarray} \label{solgap3d}
	\beta m_{\rm th} = \log\frac{ ( 2\cosh\beta \mu +1) + \sqrt{ ( 2\cosh\beta \mu +1)^2 -4}}{ 2} .
	\end{eqnarray}
	Note that the thermal mass is an even function of the chemical potential and it 
	 reduces to the well known value \cite{Sachdev_1992}  when the chemical potential vanishes
	\begin{eqnarray}\label{solgapmu3d}
	\lim_{\mu\beta\rightarrow 0} \beta m_{\rm th}  = \log\Big( \frac{ 3+ \sqrt{5}}{2} \Big)  + \frac{(\beta\mu)^2}{\sqrt{5} }+ 
	O((  \beta\mu)^4 ).
	\end{eqnarray}
	When the chemical potential is large we obtain 
	\begin{eqnarray}\label{large mu expan of mth b}
	\lim_{\mu\beta \rightarrow \pm\infty} \beta m_{\rm th} = \beta |\mu|  + e^{-\beta |\mu| } -\frac{1}{2} e^{-2\beta |\mu| } + \cdots .
	\end{eqnarray}
	It is important to note that the chemical potential can be taken to be arbitrarily large as there is no Bose-Einstein condensation 
	in $d=3$.  The free energy  density and the energy density  at the saddle point  are related  by  \ref{stressfreerel}, 
	\begin{eqnarray}
	F &=& -\frac{1}{\beta N V_{2} } \log Z = \frac{T_{00}}{2}.
	\end{eqnarray}
	This is the canonical relation between  the Free energy density and the energy density for a thermal CFT in $3$ dimensions. 
	The energy density $T_{00}$ is given by  (\ref{T00})
	\begin{eqnarray}\label{3dstressb}
	T_{00} &=& -\frac{T^3}{3\pi}  \Big[   3 {\rm Li}_3( e^{-\beta( m_{\rm th} +\mu ) } ) 
	+ 3 {\rm Li}_3( e^{-\beta( m_{\rm th} -\mu ) } )  \\ \nonumber
	 &&\qquad\qquad +   3 m_{\rm th} \beta {\rm Li}_2( e^{-\beta( m_{\rm th} +\mu ) } ) + 3 m_{\rm th}\beta {\rm Li}_2( e^{-\beta( m_{\rm th} -\mu ) } ) 
	 \nonumber \\  \nonumber
	 &&\qquad 
	  - ( m_{\rm th} \beta)^2 \log( 1 -  e^{-\beta( m_{\rm th} +\mu )})
	   - ( m_{\rm th} \beta)^2 \log( 1 -  e^{-\beta( m_{\rm th} -\mu )} \Big].
	\end{eqnarray}
	Here we need to substitute the thermal mass in terms of the chemical potential from (\ref{solgap3d}). 
	We have also divided the stress tensor by $N$. 
	This expression is the generalization of the one point function of the stress tensor obtain in \cite{Sachdev:1993pr} 
	for non-zero values 
	of real chemical potential. 
	It is instructive to study the stress tensor in various limits, note that it is an even function of the chemical potential. 
	At small values of the chemical potential, it admits the expansion
	\begin{eqnarray} \label{musmall}
		\lim_{\mu\beta \rightarrow 0} T_{00} &=& T^3 \Big[ - \frac{8}{5\pi} \zeta(3)   + \frac{ \log( 930249 -416020\sqrt{5})}{6\sqrt{5} } (\mu \beta)^2 + \cdots \Big].
	\end{eqnarray}
	As expected at zero chemical potential one obtains the familiar result for the stress tensor of the non-trivial fixed point 
	of the $O(2N)$ model \footnote{ This is because we have  $N$ complex bosons. }. 
	The first correction is also negative, the energy density continues to monotonically decrease  as the chemical potential is 
	increased \footnote{A positive energy density in Minkowski signature translates to a negative energy density in Euclidean signature $T_{00} |_{\rm Euclidean} = (i^2) T_{00}|_{\rm Minkowski} $. }
	At large values of the chemical potential the stress tensor behaves as 
	\begin{eqnarray}
	\lim_{|\mu|\beta \rightarrow \infty} 
	T_{00} &=& T^3 \Big[ -\frac{ (\beta|\mu|)^3}{3\pi} -\frac{\pi}{6} (\beta|\mu|)  -\frac{\zeta(3)}{\pi}  + O( e^{-\beta |\mu |} ) \Big], \\ \nonumber
	&=& |\mu|^3 \Big[ - \frac{1}{3\pi}  -\frac{\pi}{6} \frac{T^2}{|\mu|^2} - \frac{ \zeta(3)}{\pi} \frac{ T^3}{|\mu|^3}  
	+  O( e^{-\beta |\mu |} ) \Big].
	\end{eqnarray}
	The leading dependence is a cubic power in chemical potential which is expected since 
	the chemical potential now sets the scale for the one point function. 
	Finally for reference, we write down the energy density for the Gaussian fixed point, that is $m_{\rm th} =0, \mu =0$, which is the trivial solution of the gap equation in (\ref{gapeq3db}). 
	\begin{eqnarray}\label{gauss}
	T_{00}|_{m_{\rm th} =0, \mu =0} = -T^3 \frac{2\zeta(3)}{\pi}.
	\end{eqnarray}
	As a simple consistency check, we can read out 
the ratio of the stress tensors at the non-trivial fixed point with vanishing 
 chemical potential to the  Gaussian fixed point can be read out from (\ref{musmall}) and (\ref{gauss}). 
 The ratio turns out to be the $4/5$ which 
 agrees with the known result in literature \cite{Sachdev:1993pr}. 

Let us also examine the expression for the charge density  given in (\ref{apenchargeb}) for $d=3$
\begin{eqnarray} \label{defq3db}
J &=&  \frac{1}{\beta}\partial_\mu \log(Z), \\ \nonumber
&=& \frac{T^2 }{2\pi } \Big[  m_{\rm th} \beta (  \log( 1- e^{-\beta( m_{\rm th} +\mu) } ) 
-  \log( 1- e^{-\beta( m_{\rm th} -\mu) } ))  \\ \nonumber
&& \qquad \qquad + {\rm Li}_2(  e^{-\beta( m_{\rm th} -\mu) }) - {\rm Li}_2 (  e^{-\beta( m_{\rm th} +\mu) }) \Big].
\end{eqnarray}
Here we have analytically continued to real values of the chemical potential by replacing $\hat \mu \rightarrow  i\mu$. 
At this point we should mention that we are studying the system in the grand canonical ensemble where we specify the 
chemical potential instead of the charge. 
If we were to hold the charge fixed, we need to introduce  the term $\int d^3 x J \mu  $  
in the action in (\ref{3dlag}). 
This will then result in a second gap equation in which we minimise with respect to the chemical potential, 
which leads to the equation  given as in (\ref{defq3db}). 
We would then have to use this equation and write the one point functions in terms of the charge. 
This approach is followed in \cite{Filothodoros:2016txa,Filothodoros:2018pdj}
\footnote{See equations (3.19)-(3.23) for the bosons and (3.1) and (3.5) for the 
Gross-Neveu model. Observe that the second gap equation in this approach
 results in relating the charge to expectation value of the $U(1)$ current just as in (\ref{defq3db}).}.
Expanding the charge density at small values of chemical potential we obtain 
\begin{eqnarray}
\lim_{\mu\beta \rightarrow 0} J &=& 
\frac{T^2}{\pi } \Big[ \sqrt{5} \log\big( \frac{ 1+\sqrt{5}}{2} \big) \mu\beta   +\frac{1}{30} \Big( 
5 - 2\sqrt{5} \log\big( \frac{ 1+\sqrt{5}}{2} \big) \Big) (\mu\beta)^3 + \cdots \Big].
\end{eqnarray}
where we have used (\ref{apenchargeb}).  Note that the charge is an odd function of $\mu$. 
At large values of the chemical potential we obtain the following expansion
\begin{eqnarray} \label{largemucur}
\lim_{|\mu|\beta \rightarrow \infty} J &=& \frac{\mu^2{\rm sgn}(\mu) }{2\pi  } \Big[1 + \frac{\pi^2}{6} \frac{T^2}{\mu^2} + O (e^{-\beta|\mu|} )  \Big].
\end{eqnarray}
As expected,  at large values of chemical potential the scale is set by $\mu$ and therefore the charge density 
grows quadratically in $\mu$.

	\subsubsection{The Euclidean inversion formula}
	
	The starting point to apply the OPE inversion formula to obtain thermal one point functions of arbitrary 
	spin bi-linears is the two point function of the fundamental field $\tilde \phi$.
	As we discussed in section (\ref{chemtwist}), to obtain one point functions of currents as  given in (\ref{untwist1pt}) containing covariant derivatives, it is sufficient to look at twisted bosons $\phi$  which satisfy the boundary conditions in (\ref{untwist1pt}). 
	The two-point function for twisted bosons at finite temperature with thermal mass $m_{\rm th}$ has the form,
	\begin{align}\label{3d2ptfn}
		 G(\tau,\vec{x})=\langle \phi^{*} (\tau,\vec{x})\phi(0,0)\rangle=\sum_{n=-\infty}^\infty \int \frac{d^2k}{(2\pi)^2}
		  \frac{e^{i(\tilde \omega_n \tau+\vec{k}\cdot\vec{x})}}{\tilde\omega_n^2+\vec{k}^2+m_{th}^2}.
	\end{align}
	Here we have set $\beta=1$ and 
	\begin{equation} \label{twistfreq}
	\tilde \omega_n=2\pi n- \hat\mu.
	\end{equation}
	where $\hat \mu$ is the purely imaginary chemical potential. 
	Observe that this two point function obeys the twisted boundary condition
	\begin{eqnarray}
	 G(\tau +1,\vec{x}) = e^{-i\hat \mu}   G(\tau,\vec{x}) .
	\end{eqnarray}
	We use Poisson resummation to convert the Matsubara sum  over frequencies 
	 into sum over images in $\tau$. After which we carry out the integration
	\begin{align} 
		 G(\tau,\vec{x})=\sum_{n=-\infty}^\infty e^{-i \hat \mu\tau} \int\frac{d\omega d^2k}{(2\pi)^3} \frac{e^{i\vec{k}\cdot\vec{x}}e^{i\omega(\tau-n)}}{(\omega-\hat \mu)^2+\vec{k}^2+m_{\rm th}^2}.
	\end{align}
	Now by shifting the integration variable $\omega$ by $\omega'=\omega-\hat \mu$, 
	the two-point function is recasted into the form,
	\begin{align} \nonumber
		 G(\tau,\vec{x})&=\sum_{n=-\infty}^\infty e^{-in \hat \mu}\int\frac{d\omega d^2k}{(2\pi)^3} \frac{e^{i\vec{k}\cdot\vec{x}}e^{i\omega(\tau-n)}}{\omega^2+\vec{k}^2+m_{\rm th}^2},\\
		&= \frac{1}{4\pi}\sum_{m=-\infty}^\infty e^{-im\hat \mu}\,\frac{e^{-\big(m_{th}\sqrt{(m-z)(m-\bar z)}\,\big)}}{\sqrt{(m-z)(m-\bar z)}}, \label{prop 3d}
	\end{align}
	where,
	\begin{align} \label{defwplane}
		z=\tau+ix=r w,\qquad \bar z=\tau-ix=rw^{-1}.
	\end{align}
	We have used the rotational invariance in the spatial directions to choose $\vec x$ to be along a particular axis and 
	$x \equiv|\vec x|$.  
	Using conformal symmetry and rotational invariance in the spatial directions 
	we can write the two point function in (\ref{3d2ptfn}) in terms of its OPE expansion  as 
	\begin{align}\label{3dopeexp}
		G(\tau,\vec{x})=\sum_\mathcal{O}\frac{a_\mathcal{O}}{\beta^\Delta} C_l^{(\nu)}(\eta) |x|^{\Delta-2\Delta_\phi}.
	\end{align}
	where the coefficients $a_{\cal O}$ are proportional to the one point functions  of all operators ${\cal O}$ which appear in the 
	$\phi\times \phi$ OPE, 
	\begin{equation} \label{abrelation}
	a_{\cal O} = \frac{ f_{\phi\phi {\cal O} } b_{\cal O} }{ c_{\cal O}} \frac{ l!}{ 2 ^l ( \nu)_l }, 
	\qquad \nu = \frac{d-2}{2} =\frac{1}{2}, 
	\end{equation}
	$f_{\phi \phi {\cal O}}$ are the structure constants and $c_{\cal O}$ are the normalization of the two point functions of the 
	operator ${\cal O}$ of spin $l$ and   the Pochhammer symbol  is defined as
	\begin{equation}
	(a)_n = \frac{\Gamma( a+n) }{\Gamma(a) }.
	\end{equation}
	The one-point functions are given by 
	\begin{eqnarray}
	\langle {\cal O }^{\mu_1 \mu_2 \cdots \mu_{l}} (x)  \rangle =  b_{\cal O} T^{\Delta} 
	( e^{\mu_1} e^{\mu_2} \cdots e^{\mu_l}  - {\rm Traces} ) .
	\end{eqnarray}
	and $e^\mu$ is the unit vector along the $\tau$ direction. 
	In the large $N$ limit for the model in (\ref{3dlag})
	 the only operators that occur in the the OPE expansion  of a pair of $\phi$'s are bi-linears of the 
	form 
	\begin{eqnarray}
	{\cal O}[n, l]  = \phi^* \partial_{\mu_1} \partial_{\mu_2 } \cdots \partial_{\mu_l} \partial^{2n} \phi,
	\end{eqnarray}
	which is a $l$-rank traceless symmetric tensor of dimensions $\Delta_{\cal O} = 1 + l + 2n $ in the large $N$ limit. 
	Finally in (\ref{3dopeexp}), the $C_l^{(\nu)}(\eta)$ are Gegenbauer polynomials and 
	\begin{equation}
	\eta = \frac{\tau}{|x|}.
	\end{equation}

	The Euclidean 
	 inversion formula developed in \cite{Iliesiu:2018fao}   is an expression for  the thermal 1-point functions 
	 given the thermal 2-point function. 
	 It can be applied once 
	 certain analytical properties of the 2-point function  and fall off conditions in the  complex $z$ plane  are satisfied.
	The analytic behaviour of the 
	  propagator   \eqref{prop 3d}
	  in the $z$ plane is same as that when $\mu=0$ for each $m$ and therefore we can 
	  apply the  Euclidean inversion formula as explained in detail in 
	 \cite{Iliesiu:2018fao,David:2023uya} to obtain the one point functions. 

First the  spectral function  $\hat a(\Delta, l )$ is introduced in the $\Delta$-plane 
 from which the one point functions are read out as the negative residue at $\Delta =\Delta_{\cal O}$ as follows
	\begin{align} \label{3dadef}
		a_\mathcal{O}[n,l]=\hat a( \Delta_{} , l)|_{{\rm Res\ at\ }\Delta=\Delta_\mathcal{O}} &= - [\hat a_{\rm disc} ( \Delta_{}, l) + \theta( l_0 -l) \hat a_{\rm arcs} ( \Delta_{} ,l)]|_{{\rm Res\ at\ }\Delta=\Delta_\mathcal{O}} ,
	\end{align}
	Here 
	 $ a_{\rm disc}(\Delta,l) $ comes from the discontinuity of the propagator in the $ z $-plane and is given by,
	\begin{align}\label{a disc}
		&\hat a_{\rm disc}(\Delta,l)=K_l \int_0^1 \frac{d\bar z}{\bar z}\int_1^{1/\bar z}\frac{dz}{z} (z \bar z)^{\Delta_\phi-\frac{\Delta}{2}-\nu}(z-\bar z)^{2\nu} F_l\bigg(\sqrt{\frac{z}{\bar z}}\bigg) {\rm Disc} [{G}(z,\bar z)], \\
		&\text{with,}\quad \nonumber
		K_l=\frac{\Gamma(l+1)\Gamma(\nu)}{4\pi \Gamma(l+\nu)},
		\qquad F_l(w) = w^{l+ 2\nu } {}_2 F_1( l + 2\nu , \nu , l + \nu +1, w^2) .
	\end{align}
	and,
	\begin{align} \label{Disc}
		{\rm Disc}[ G(z, \bar z) ] = \frac{1}{i} \big(  G( z +i \epsilon , \bar z) - G( z-i\epsilon, \bar z) \big) .
	\end{align}
	While rest of the contribution $ a_{\rm arc}(\Delta,l) $ involves the contribution at infinity in the $z$-plane. 
	This contribution is non-zero only for angular momenta less than a given angular momentum $l_0$. 
	The 
	 integration along a circle of infinite radius in the complex $ w $ plane is given by 
	\begin{eqnarray}\label{db2arc}
		\hat a_{\rm arcs} (\Delta_{},  l)  &=&  2 K_l   \int_0^1 \frac{dr}{r^{\Delta_{}  +1 - 2\Delta_\phi} }  \times  \\ \nonumber
		&& \oint \frac{dw}{i w} \lim_{|w| \rightarrow \infty} 
		\left[ \Big( \frac{ w - w^{-1} }{i} \Big)^{2\nu} 
		F_l(w^{-1}) e^{i\pi\nu}  {G}( r, w) 
		\right].
	\end{eqnarray}
	The set of  formulae in (\ref{3dadef}), (\ref{a disc}), (\ref{Disc}), (\ref{db2arc}) together form 
	 the Euclidean inversion formula and we need to apply these on the twisted Green's function in (\ref{prop 3d}).

	Let us proceed to evaluate $\hat a_{\rm disc}(\Delta,l)$. For this we need  the discontinuity of the propagator \eqref{prop 3d} across its branch cuts using equation \eqref{Disc}.
	\begin{align}
		{\rm Disc} [G(z,\bar z)]=\frac{1}{2 \pi}\sum_{\substack{m=-\infty\\m\ne 0}}^\infty e^{-i m\hat \mu}\,\frac{\cos \left(m_{th} \sqrt{(z-m) (m-\bar z)}\right)}{\sqrt{(z-m) (m-\bar z)}}.
	\end{align}
	We substitute this expression into the  equation \eqref{a disc} to obtain, 
	\begin{align}\label{a++a-}
		\hat a_{\rm disc}(\Delta,l)=\hat a_{\rm disc}^{+}(\Delta,l)+\hat a_{\rm disc}^{-}(\Delta,l).
	\end{align}
	Here we have separated  the contribution from  terms with 
	positive $ m $ terms and negative $ m $  using the notation $a^{\pm}_{\rm disc}(\Delta,l)$ defined as,
	\begin{align}
		\hat a_{\rm disc}^{\pm}(\Delta,l)=\frac{K_l}{2\pi}\sum_{m=1}^{ \infty}\int_0^1 \frac{d\bar z}{\bar z}\int_m^{{\rm max}(m,\frac{1}{\bar z})}\frac{d z}{ z}\frac{ (z-\bar z) F_l\left(\sqrt{\frac{\bar z}{z}}\right) (z \bar z)^{-\frac{\Delta }{2}} e^{\mp i \hat \mu  m} }{ \sqrt{(z\mp m) (\pm m-\bar z)}}\nonumber\\
		\times\cos \left(m_{th} \sqrt{(z\mp m) (\pm m-\bar z)}\right).
	\end{align}
	It is easy to show that,
	\begin{align}\label{a-}
		a_{\rm disc}^{-}(\Delta,l)=(-1)^la^+_{\rm disc}(\Delta,l)|_{\hat \mu\to -\hat \mu}.
	\end{align}
	Now we use the transformation, 
	\begin{align}
		\bar z=z'\bar z'\qquad {\rm and} \qquad z=mz',
	\end{align}
	to obtain, 
	\begin{align}
		\hat a^+_{\rm disc}(\Delta,l)=-\frac{K_l}{2\pi}\sum_{m=1}^\infty\int_0^1 d\bar z\int_1^{{\rm max}(1,\frac{1}{m\sqrt{\bar z}})} dz\frac{(\bar z-1) e^{-i \mu  m} F_l(\sqrt{\bar z})  \cos \left(m m_{\rm th} \sqrt{(1-z) (z \bar z-1)}\right)}{\bar z \left(m^2 z^2 \bar z\right)^{\frac{\Delta }{2}} \sqrt{(1-z) (z \bar z-1)}}.
	\end{align}
	As mentioned earlier the thermal 1-point functions are computed as the negative residue of the spectral function $ \hat a_{\rm disc}(\Delta,l) $ at the operator's dimension $\Delta=\Delta_\mathcal{O} $.  We perform the integral by first 
	expanding the integrand in $\bar z$.  This decouples the $z$ and $\bar z$ integration as upper limit of the  $z$ integral is 
	taken to $\infty$. Performing the integral on the leading order term in the $\bar z$ integration we obtain 
	\begin{align}
		\hat	a_{\rm disc}^{+(0)}=\frac{2K_l}{\pi}\sum_{m=1}^\infty\frac{e^{-im\hat  \mu}}{m^\Delta(-\Delta +l+1)}\int_0^\infty dy \frac{   \cos (m m_{\rm th} y)}{(1+y^2)^\Delta}.
	\end{align}
	where $ y=\sqrt{z-1} $.  Therefore the residue at  $ \Delta=l+1 $ is,
	\begin{align}\label{Bessel form}
		-\hat a_{\rm disc}^{+(0)}|_{{\rm Res\ at\ }\Delta=l+1}=\frac{K_l m_{\rm th}^{l+\frac{1}{2}}}{\pi^{\frac{1}{2}}2^{l-\frac{1}{2}}l!}\sum_{m=1}^\infty\frac{  e^{-i \hat \mu  m} }{ \sqrt{m }}K_{l+\frac{1}{2}}(m m_{\rm th}).
	\end{align}
	We can perform the sum over $m$ by first expressing 
	 the Bessel function with half-integer order as a finite sum \footnote{\begin{equation} \label{besli}
			K_{l +\frac{1}{2}}(x) =  e^{-x} \sum_{n=0}^l \frac{ \sqrt{\pi} ( l + 1- n)_{2n}}{ (2x)^{n + \frac{1}{2}} n!}, \qquad l \in \mathbb{Z}.
	\end{equation}}. We obtain
	\begin{align}\label{a_O}
		a_\mathcal{O}[0,l]&=-(a^{+{(0)}}_{\rm disc}+a^{-(0)}_{\rm disc})|_{{\rm Res\ at\ }\Delta=l+1}\nonumber\\
		&=\frac{K_l}{ l!}\sum_{n=0}^l\frac{ m_{\rm th}^{l-n} (l-n+1)_{2 n} }{2^{l+n} n! }\times\Big(\text{Li}_{n+1}(e^{-m_{\rm th}-i \hat \mu })+(-1)^l \text{Li}_{n+1}(e^{-m_{\rm th}+i \hat \mu })\Big).
	\end{align}
	
	\subsection*{Arc Contribution}
	
	To complete the analysis of inversion formula we need to evaluate the  contribution from the arc at infinity
	$a_{\rm  arc}(\Delta,l)$, which  given by the formula \eqref{db2arc}. This 
involves integration along the circle of infinite radius in the complex $w$-plane.  At $|w|\to \infty$, the 2-point function \eqref{prop 3d} has only non-vanishing contribution from the $m=0$ term.
	\begin{align}
		\lim_{|w|\to \infty}	G(z,\bar z)=\frac{e^{-m_{th}r}}{4\pi r}.
	\end{align}
	Substituting  this expression into \eqref{db2arc} and the $w$ integral turns out to be non-zero only for $l=0$.
	\begin{align}
		\hat a_{\rm arc}(\Delta,l)=K_0\int_0^1 dr \frac{e^{-m_{\rm th}r}}{r^{\Delta+1}}.
	\end{align}
	Thus,
	\begin{align}
		-\hat a_{\rm arc}|_{{\rm Res \ at\ }\Delta=1}=-K_0 m_{\rm th}.
	\end{align}
	The contributions due to disc and arc part to the residue at the pole $\Delta=1$, sum up to give the expectation value of 
	$|\phi|^2$.  At the non-trivial fixed point, this operator should not be present in the spectrum or the $U(N)$ symmetry should be unbroken. 
	This leads to the following equation 
	\begin{align}\label{gap}
		&	-(\hat a_{\rm disc}+\hat a_{\rm arc})|_{{\rm Res\ at \ }\Delta=1}=0,\nonumber\\
		&	m_{\rm th }+\log(1-e^{-m_{\rm th}-i\hat \mu})+\log(1-e^{-m_{\rm th}+i\hat \mu})=0.
	\end{align}
	We can now analytically continue this to real values of the chemical potential by the replacement 
	$\hat \mu \rightarrow i \mu$. 
	This leads us to the equation 
	\begin{equation}
	m_{\rm th }+\log(1-e^{-m_{\rm th}-\mu})+\log(1-e^{-m_{\rm th}+ \mu})=0.
	\end{equation}
	Comparing this  equation to the gap equation obtained from the partition function 
	(\ref{gapeq3db}), we see that they precisely coincide. 
	Thus the condition that the expectation value $\langle |\phi|^2 \rangle$ vanishes, 
	 ensures that the gap equation is obtained even when
	chemical potential is turned on. 
	
	For $l>0$  there  are no contributions form the arcs at infinity, therefore from (\ref{a_O}), the one point functions for real values of
	chemical potential is given by 
		\begin{align}  \nonumber
		a_\mathcal{O}[0,l]
		&=\frac{K_l}{ l!}\sum_{n=0}^l\frac{ m_{\rm th}^{l-n} (l-n+1)_{2 n} }{2^{l+n} n! }\times\bigg(\text{Li}_{n+1}
		\left(e^{-m_{\rm th}+  \mu }\right)+(-1)^l \text{Li}_{n+1}\left(e^{-m_{\rm th}-  \mu }\right)\bigg), \nonumber \\ 
		  &\qquad\qquad\qquad\qquad\qquad l>0 . \label{ao3drealm}
	\end{align}

		Before we proceed to study these one point functions, let us compare the  one point functions  of the 
		spin-1 and spin-2 currents 
		against that obtained from the partition function. 
		Comparing (\ref{ao3drealm}) at $l=2$ with that of the stress tensor in (\ref{3dstressb}), we find 
		\begin{eqnarray}
		T_{00} = -4 a_{{\cal O}}[0, 2].
		\end{eqnarray}
		The proportionality constant is expected since the one point functions $a_{\cal O}$ are only proportional to the 
		one point functions $b_{\cal O}$ by (\ref{abrelation}) which involves the structure constants, which in this case is 
		determined by the stress tensor Ward identity. 
		 Indeed the fact that there is 
		negative sign relating these expectation values was noted in \cite{Iliesiu:2018fao}. 
		Similarly comparing the $l=1$ expression from (\ref{ao3drealm}) to the charge density in (\ref{defq3db})  we see that 
		\begin{eqnarray}
		J =  2a_{{\cal O}}[0, 1].
		\end{eqnarray}
		At this point it is important to emphasize that the expressions for the $T_{00}$ and $J$ in (\ref{3dstressb}) and (\ref{defq3db}) were obtained
		from the partition function which was written  untwisted field $\tilde \phi$. 
		However ${\cal O}[0, l ]$ in (\ref{ao3drealm}) was obtained using the Euclidean 
		 inversion formula on twisted bosons.  It is expected that they agree and therefore it serves as a simple 
		 consistency check for the approach of using the Euclidean inversion formula on twisted bosons. 
		
	\subsubsection{1-point functions at large chemical potential} \label{largchar}

	In this subsection we study the behaviour of the higher spin one point functions in (\ref{ao3drealm}) at large values of 
	chemical potential.  For this we need the behaviour of the thermal mass at large chemical potential given in 
	\eqref{large mu expan of mth b}. 
	Let us first examine the limits  for large positive values of the chemical potential,
	\begin{eqnarray}
	\lim_{\mu\rightarrow \infty} {\rm Li}_{\;1}( e^{-m_{\rm th} + \mu } )  &=&
	 - \lim_{\mu\rightarrow \infty} ( 1- e^{ - m_{\rm th}  + \mu } )  , 
	\\ \nonumber
	&=& \mu +  O(e^{-\mu}),   \\ \nonumber
	\lim_{\mu\rightarrow \infty} {\rm Li}_{\; n+1}( e^{-m_{\rm th} + \mu } )  &=& \zeta(n+1) + O (e^{-\mu}) , 
	\qquad \qquad {\rm for}\; n\geq 1 .
	\end{eqnarray}
	We have used the limiting behaviour of the thermal mass in \eqref{large mu expan of mth b} to arrive at  this behaviour. 
	Using such limits in the one point functions (\ref{ao3drealm}),  we obtain the following.
	\begin{eqnarray} \label{largemucur2}
	\lim_{|\mu \beta | \rightarrow \infty} a_{\cal O}[0, l ] &=&
	\frac{K_l |\mu|^{l+1}  \big({\rm sgn}(\mu) \big)^l  }{ 2^l l!} \left[ 1  + \sum_{n =1}^l \frac{   (l - n +1)_{2n} \zeta(n+1)  }{ 2^n n!} \frac{T^{n+1} }{ |\mu|^{n+1} } + O(e^{-\beta |\mu|} )\right] .\nonumber \\ 
	\end{eqnarray}
	Here we have reinstated the dependence on the temperature. 
	Observe that at large chemical potentials,  first correction always occurs at the order 
	$\frac{T^2}{|\mu|^2}$, the linear term is missing.  The dependence $|\mu|^{l+1}$ in (\ref{largemucur}) and (\ref{largemucur2}) 
	arises because the theory is conformal and at large $\mu$, the chemical potential is the relevant scale. 
	Then by conformal invariance, the dimension of the operator determines the power $l+1$ in the expectation value.

	\subsubsection{The large spin limit}
	
	The  behaviour of thermal one point functions of bi-linears with large spin has been studied for the 
	$O(N)$ model in $d= 2k+1$ dimensions with $k=1,  3, 7, \cdots$ in  \cite{David:2023uya} numerically in the 
	absence of chemical potential. 
	This involved solving the gap equation for the $O(N)$ model numerically and then using the numerical 
	value of the thermal mass at the non-trivial fixed point   to study the 
	one point functions are large values of $l$. 
	The key observation in  \cite{David:2023uya}  is that in any given dimension $d$, 
	the one point function of bi-linear operators    at the non-trivial fixed point approach  that of the Gaussian or the 
	trivial fixed point of the theory at sufficiently large values of the spin, $l$. 
	
	In this section, we study the behaviour of the expectation values of large spin bi-linears in  $d=3$ model of 
	$N$ complex scalars transforming as fundamental of $U(N)$ at finite values of the chemical potential. 
	Our analysis will be analytical and we will show that the result here reproduces the numerical observations of
	 \cite{David:2023uya}  on setting the chemical potential to zero.

We begin from the expression for the 1-point functions in terms of Bessel functions given in 
	 \eqref{Bessel form},
	\begin{align}\label{Bessel form 1}
		-\hat a_{\rm disc}^{+(0)}|_{{\rm Res\ at\ }\Delta=l+1}=\frac{K_l m_{\rm th}^{l+\frac{1}{2}}}{\pi^{\frac{1}{2}}2^{l-\frac{1}{2}}l!}\sum_{m=1}^\infty\frac{  e^{-i \hat \mu  m} }{ \sqrt{m }}K_{l+\frac{1}{2}}(m m_{\rm th}).
	\end{align}
	At large orders and fixed argument the asymptotic form of the Bessel function is given by the following formula
	\begin{align}\label{largeobessel}
		\lim_{l\rightarrow \infty} K_{l+\frac{1}{2}}(m m_{\rm th})= \sqrt{\frac{\pi}{2(l+\frac{1}{2})}}\bigg(\frac{e m m_{\rm th}}{2(l+\frac{1}{2})}\bigg)^{-l-\frac{1}{2}}.
	\end{align}
	Note that the  asymptotic form of the Bessel function ensures the the expression in (\ref{Bessel form 1}) is independent of 
	the thermal mass. 
	Using the expression of $K_l$ from (\ref{a disc}), we have  the limit 
	\begin{equation}
	\lim_{l \rightarrow \infty} \frac{2 K_l ( 2l +1)^l }{ ( 2 e)^{l +\frac{1}{2} } l!} = \frac{1}{4\pi} .
	\end{equation}
	Now substituting the asymptotic form for the Bessel function in (\ref{largeobessel}) and using the above limit we obtain 
	\begin{eqnarray} \label{limitl}
	\lim_{l\rightarrow\infty} -\hat a_{\rm disc}^{+(0)}|_{{\rm Res\ at\ }\Delta=l+1} =\frac{1}{4\pi} e^{-i\hat \mu}.
	\end{eqnarray}
	Here we have taken the large $l$ limit term by term in the sum and kept only the $m=1$ term since the rest of the terms 
	are exponentially suppressed in $l$. 
	The fact that we can take the large $l$ limit inside the sum over $m$ will be justified when we verify the final results
	using numerics. 
	We substitute this limit  (\ref{limitl}) in 
	  \eqref{a++a-} and  use the relation \eqref{a-}. This results in the following expression for the leading 
	  contribution to the 1-point function of bi-linears of large spin, 
	\begin{align}
		\lim_{l\rightarrow \infty} a_\mathcal{O}[0,l] =
	\frac{1}{4\pi}(e^{-i\hat \mu}+(-1)^le^{i\hat \mu}).
	\end{align}
Therefore,  at large $l$
	\begin{align}
		a_{\mathcal{O}}[0,l]&=\frac{1}{2\pi}\cos \hat \mu\quad {\rm for }\ l\ {\rm even},\\ \nonumber
		&=\frac{1}{2\pi}\sin \hat \mu\quad {\rm for }\ l\ {\rm odd}.
	\end{align}
	Note that at large spin
	the 1-point function which is independent of the thermal mass and just depends on the odd or even property of the spin. 
	Its dependence on the chemical potential is through 
	a trignometric function. The prefactor $ 1/2\pi $ can be identified with the 1-point function of the bi-linear 
	at the Gaussian  or the Stefan-Boltzman fixed point of the theory.  This can be seen as follows. 
	The Stefan-Boltzmann limit of the 1-point functions 
	can be obtained  taking both $\hat \mu$ and $m_{\rm th} $ to be zero in the formula \eqref{a_O}, this results in 
	\begin{align}
		a_\mathcal{O}^{\rm free}[0,l]
		&=(1+(-1)^l)  \frac{1}{4\pi } \zeta(l+1).
	\end{align}
As expected it is only the even spins which have non-trivial expectation values at the Gaussian fixed point. 
Now taking the large spin limit, we obtain 
	\begin{eqnarray}
		\lim_{l \rightarrow \infty} a_\mathcal{O}^{\rm free}[0,l] &=&\frac{1}{2\pi}, \qquad  l \in 2 \mathbb{Z} , \\ \nonumber
		&\equiv& a_\mathcal{O}^{\rm free}[0, \infty].
	\end{eqnarray}
	Therefore 
	 at large $l$, the 1-point functions at the non-trivial fixed point  which we denote by  $ a_\mathcal{O}^{\rm crit }[0,l] $ 
	 can be written as 
	\begin{align}
		\lim_{l \rightarrow \infty} 
		a_{\mathcal{O}} [0,l]&=   a_\mathcal{O}^{\rm free}[0, \infty]  \cos \hat\mu\quad {\rm for }\ l\ {\rm even},\\ \nonumber
		&= a_\mathcal{O}^{\rm free}[0, \infty] \sin \hat\mu \quad {\rm for }\ l\ {\rm odd}.
	\end{align}
	For real chemical potentials we analytically continue the the asymptotic formula for the 1 point function at large $l$, to have,
	\begin{align}\label{re ch}
		\lim_{l \rightarrow\infty}
		a_{\mathcal{O}}^{\rm crit}[0,l]&=  a_\mathcal{O}^{\rm free}[0, \infty] \cosh \mu\quad {\rm for }\ l\ {\rm even},\\ \nonumber
		&= a_\mathcal{O}^{\rm free}[0, \infty] \ \sinh \mu \quad {\rm for }\ l\ {\rm odd}.
	\end{align}
	Note that from this expression,  it is easy to see that the numerical observations in \cite{David:2023uya} is evident. 
	 The 1-point functions  at large spin
	reduce to the 
	Gaussian values on setting the chemical potential to zero. 
	We have obtained these properties for the one point functions $a_{\mathcal{O}}$, which include the 
	structure constants involved in the OPE expansion.  However we have made similar 
	statements for the one point functions $b_{\mathcal{O}}$ in the introduction. 
	This is because these  one point functions  are related by relation (\ref{abrelation}). 
	For vector models  the structure constants, conformal dimensions of the fundamental field, as well the normalization of  its 2-point functions do not change from the free field values  at large $N$, see for instance in \cite{Fei:2014yja,Moshe:2003xn,Vasiliev:1981dg,Lang:1992zw,Petkou:1994ad,Petkou:1995vu}. 
	This ensures the equations in (\ref{re ch}) can be written for the one point functions $b_{\mathcal{O}}$ as well. 
	
\begin{figure}[t]
		
		\begin{subfigure}{.475\linewidth}
			\includegraphics[width=.9\linewidth]{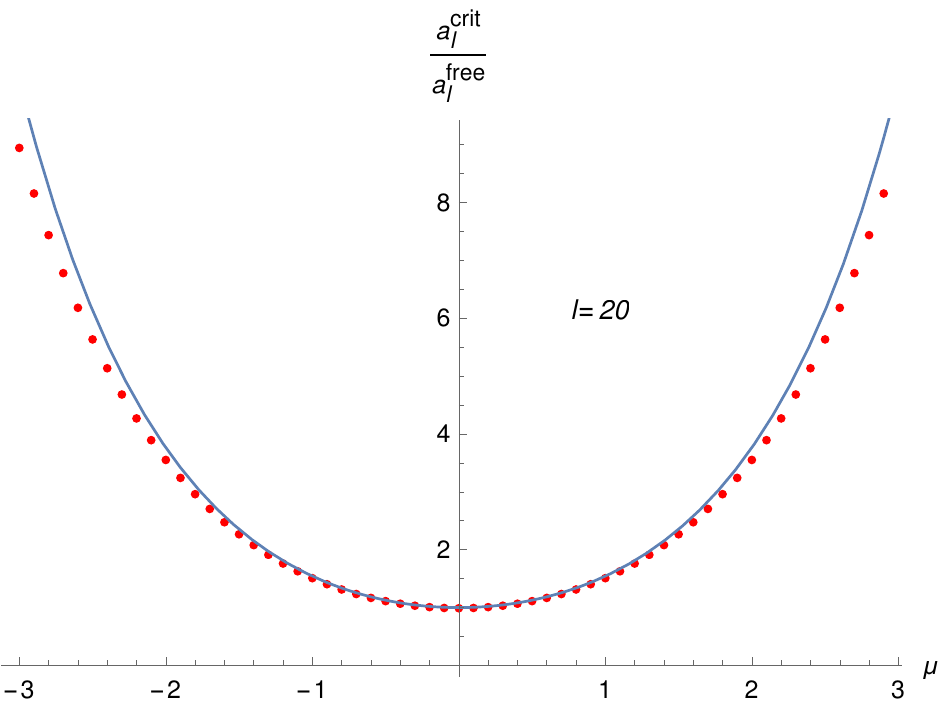}
			\caption{}
		\end{subfigure}\hfill 
		\begin{subfigure}{.475\linewidth}
			\includegraphics[width=.9\linewidth]{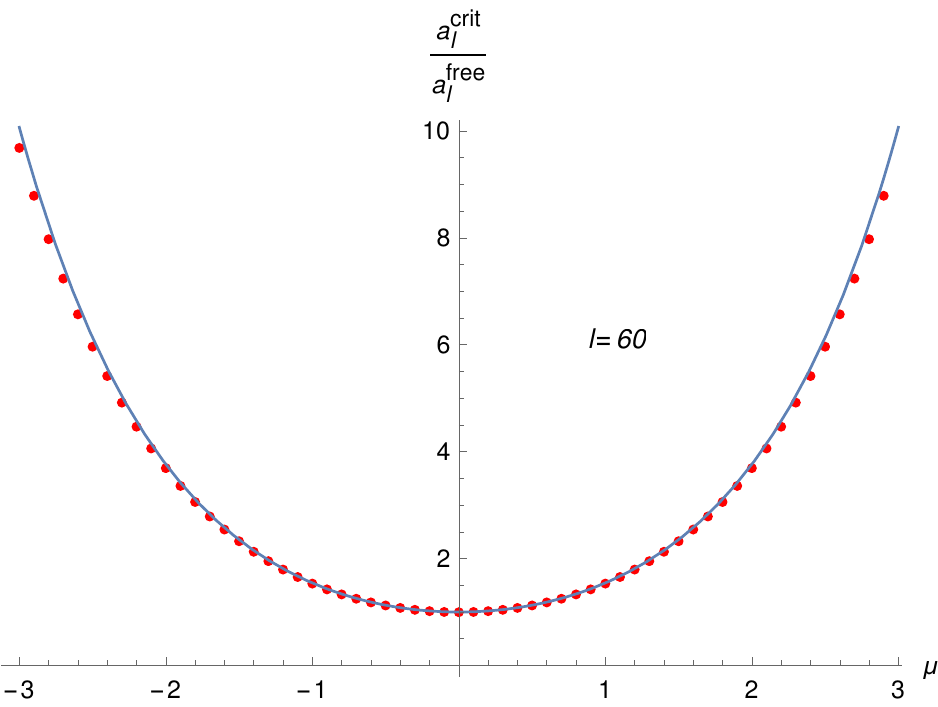}
			\caption{}
		\end{subfigure}
		\par\bigskip
		\par\bigskip
		\par\bigskip
		\begin{subfigure}{.475\linewidth}
			\includegraphics[width=.9\linewidth]{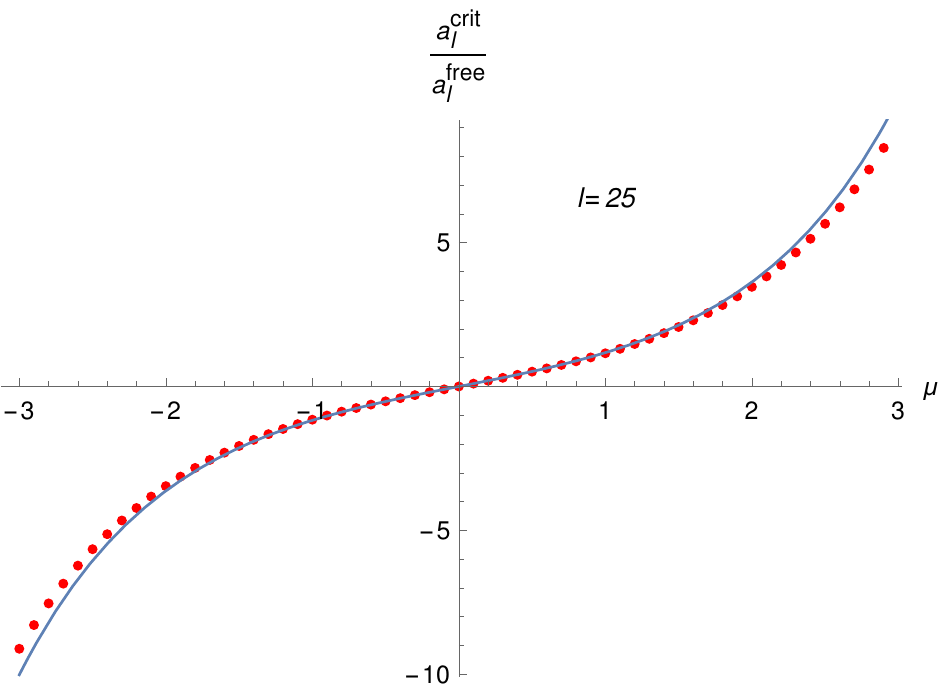}
			\caption{}
		\end{subfigure}\hfill 
		\begin{subfigure}{.475\linewidth}
			\includegraphics[width=.9\linewidth]{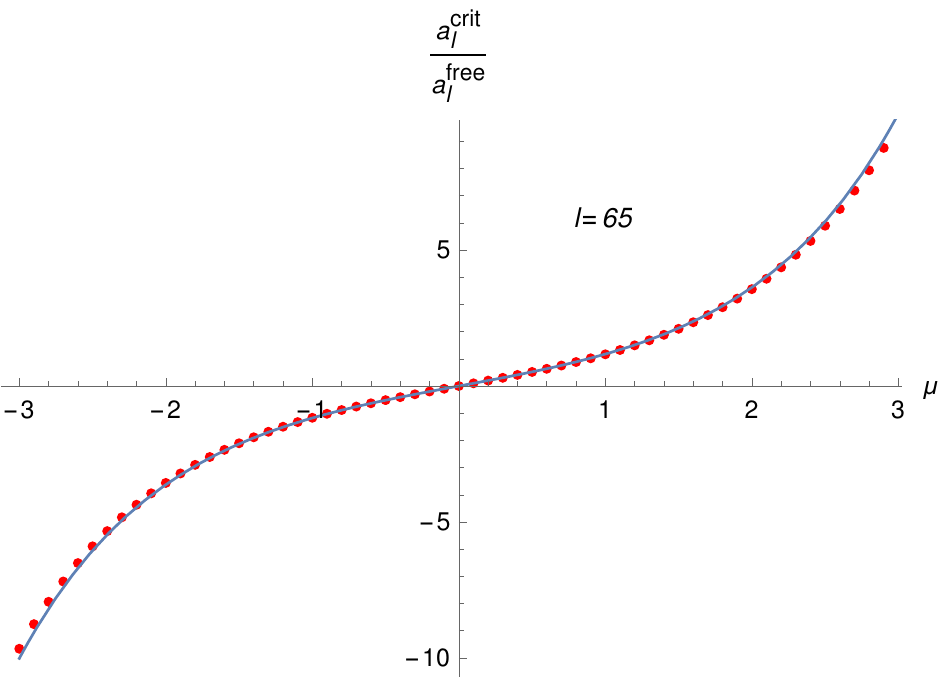}
			\caption{}
		\end{subfigure}
		
		\caption{In this plot of the ratio of 1-point functions at critical to free theory versus 
		real chemical potential, we compare the asymptotic formula of the one point function at large $l$ against numerical values of $a_{\mathcal{O}}[0,l]$ from the formula \eqref{ao3drealm}. The red points are the numerically computed values of 
		$2\pi  a_\mathcal{O}[0,l]$ from the equation \eqref{ao3drealm} while the blue solid curves refer to the graph of $\cosh \mu$ in figures  (a) and (b)  and $\sinh \mu$ in figures  (c) and (d). 
		The 1-point function is well approximated   by the formula \eqref{re ch} when $l\gg\mu$. $a_l^{\rm crit}$ and $a_l^{\rm free}$ denote the 1-point function for critical and free theory respectively in the figure.}
		\label{fig:sinh}
	\end{figure}

	As we have seen, the derivation leading up to the above expression for the large spin limit of the 1-point function involved 
	taking the limit inside the sum in (\ref{Bessel form 1}) and analytical continuation in (\ref{re ch}). 
	Therefore it is important to check the validity of the expression using numerics. 
	This is done in the graphs given in 
	 given in figure \ref{fig:sinh}. The red dots in the figures are numerical values of the one point functions evaluated using 
	 (\ref{ao3drealm})  
	 by substituting the numerical value of $m_{\rm th}$  in (\ref{solgap3d}) for various value of the chemical potential. 
	 This is compared to the expressions in 
	 (\ref{re ch}) for spins $l=20, 60, 25, 65$.  The graphs demonstrate that the large spin result in (\ref{re ch}) agrees with the numerics 
	 to a high degree of accuracy. 
	 The large spin  behaviour of these one point functions 
	 may be useful to study the thermal behaviour of Vasiliev's higher spin theory on $AdS_4$ which is the holographic dual to the 
	 $O(N)$ model at the non-trivial fixed point. 
	
	\subsection{The model in  $d= 2k +1$ with $k$ odd} \label{greatert3}
	
	As mentioned  in the introduction to  section \ref{unmodel},  
	the $O(N)$ model for $d=5$ admits only complex conjugate solutions 
	for the thermal mass, this results in complex stress tensors. However it was noticed in \cite{Petkou:2018ynm} 
	and again confirmed in \cite{David:2023uya} , that
	for $d=2k+1$ with $k$ odd, the gap equation admits a real solution. 
	Though for $k\geq 3$, these models  at the interacting fixed point 
	are non-unitary, they allow the study of  thermal  conformal field theories in higher 
	dimensions. They serve as concrete  examples to study the conjectures and observations in \cite{Fitzpatrick:2013sya,Gadde:2020nwg,Gadde:2023daq}. 
	
	We begin first by generalising the Euclidean inversion formula to extract thermal 1-point functions of bi-linears for the theory
	described by the Lagrangian in (\ref{actbos1})  at finite temperature 
	 and held at finite chemical potential. 
	 As discussed in section \ref{chemtwist}, we will develop the Euclidean inversion formula for 
	 twisted bosons. 
	 The calculation is identical to that done for $d=3$, but we just need to keep track of the dependence of the dimensions. 
	 Again we will demonstrate that the  condition for the vanishing of the  expectation value of $|\phi|^2$ is identical to the gap 
	 equation \ref{appengapeq} obtained from the partition function. 
	 After studying the one point function for large spins where we observe the same phenomenon as seen in $d=3$ we 
	 proceed to solve the gap equation for large dimensions.
	 We then use this solution to show that  the 1-point function of bi-linear operators vanish exponentially in $k$. 
	 This phenomenon was observed numerically in \cite{David:2023uya} and here we prove this analytically using the asymptotic
	 solution of the gap equation at large dimensions. 

	\subsubsection{Thermal 1-pt functions}

	At finite temperature the twisted 2-point function for complex $U(N)$ scalars at the critical point with chemical potential for any odd dimensions $d =2k +1$ is given by,
	\begin{align}
		\tilde G(\tau,\vec{x})=\langle \phi^*(\tau,\vec{x})\phi(0,0)\rangle=\sum_{n=-\infty}^\infty \int \frac{d^{d-1}k}{(2\pi)^{d-1}} \frac{e^{i(\tilde \omega_n \tau+\vec{k}\cdot\vec{x})}}{\tilde\omega_n^2+\vec{k}^2+m_{th}^2},
	\end{align}
	Here $\tilde \omega_n$ is defined in (\ref{twistfreq}) and 
	 $m_{\rm th}$ is thermal mass. 
	 The thermal mass will be determined by demanding that the thermal 1-point function of $|\phi|^2$ vanishes. 
	 This defines the  conformal fixed points of the theory.  We 
	  sum over Matsubara frequencies using Poisson re-summation, which transforms the two point function to a
	    sum over images in the imaginary time direction. This results in 
	\begin{align}
		 G(\tau,\vec{x})&=\sum_{n=-\infty}^\infty e^{-i \mu\tau} \int\frac{d\omega d^{d-1}k}{(2\pi)^d} \frac{e^{i\vec{k}\cdot\vec{x}}e^{i\omega(\tau-n)}}{(\omega-\mu)^2+\vec{k}^2+m_{\rm th}^2}, \\ \nonumber
		&=\sum_{n=-\infty}^\infty e^{-in\mu} \int \frac{d\omega d^{d-1}k}{(2\pi)^d} \frac{e^{i\vec{k}\cdot\vec{x}}e^{i\tilde \omega(\tau-n)}}{\tilde \omega^2+\vec{k}^2+m_{\rm th}^2}.
	\end{align}
	The integral can be written in a closed form and we obtain the two point function of twisted bosons in position representation
	\begin{align}\label{prop d}
		{G}(\tau,\vec{x})=\sum_{m=-\infty}^\infty \frac{e^{-im \mu}}{(2\pi)^{k+\frac{1}{2}}} m_{\rm th}^{k-\frac{1}{2}} \frac{K_{k-\frac{1}{2}}(m_{\rm th}\sqrt{(m-z)(m-\bar z)}\,)}{[(m-z)(m-\bar z)]^{\frac{k}{2}-\frac{1}{4}}}.
	\end{align}
	The discontinuity of the twisted Green's function in  the $z$-plane can be obtained using the identity
	\begin{align}
		{\rm Disc} \bigg[\frac{K_{k-\frac{1}{2}}(m_{\rm th}\sqrt{(m-z)(m-\bar z)}\,)}{[(m-z)(m-\bar z)]^{\frac{k}{2}-\frac{1}{4}}}\bigg]=\pi \frac{J_{\frac{1}{2}-k}(m_{\rm th}\sqrt{(z-m)(m-\bar z)})}{[(z-m)(m-\bar z)]^{\frac{k}{2}-\frac{1}{4}}}.
	\end{align}
	We can now substitute the above expression for the discontinuity into the Euclidean inversion formula  \eqref{a disc},
	\begin{align} \label{disccontri}
		\hat a_{\rm disc}(\Delta,J)=\pi K_l \sum_{m=-\infty}^\infty \frac{e^{-im \mu}m_{\rm th}^{k-\frac{1}{2}}}{(2\pi)^{k+\frac{1}{2}}} \int_0^1 \frac{d\bar z}{\bar z}\int_1^{1/\bar z}\frac{dz}{z} (z \bar z)^{\Delta_\phi-\frac{\Delta}{2}-\nu}(z-\bar z)^{2\nu} F_l\bigg(\sqrt{\frac{z}{\bar z}}\bigg) \times\nonumber\\
		\frac{ J_{\frac{1}{2}-k}(m_{\rm th}\sqrt{(z-m)(m-\bar z)})}{[(z-m)(m-\bar z)]^{\frac{k}{2}-\frac{1}{4}}}.
	\end{align}
	Following the same steps as in the case of $d=3$, we change variables of integration to 
	\begin{align}
		\bar z=z'\bar z'\qquad {\rm and} \qquad z=mz',
	\end{align}
	and obtain 
	\begin{align}\nonumber
		\hat a_{\rm disc}^+(\Delta,l)=\pi K_l \sum_{m=1}^\infty \frac{e^{-im \mu}m_{\rm th}^{k-\frac{1}{2}}}{(2\pi)^{k+\frac{1}{2}}m^{\Delta+\frac{1}{2}-k}} \int_0^1 {d\bar z}\int_1^{1/\bar z} dz \frac{z  (z-z \bar z)^{2 k-1} F_l(\sqrt{\bar z}) }{[(z-1) (1-z \bar z)]^{\frac{1}{4} (2k-1)}\left(z^2 \bar z\right)^{\frac{\Delta }{2}+1}}\\
		\times J_{\frac{1}{2}-k}\left( mm_{\rm th} \sqrt{ (1-z) (z \bar z-1)}\right).
	\end{align}
	Finally we 
	expand the integrand in small $\bar z$, this  decouples the integrals and then the $\bar z$ integral is performed which picks out the  poles in the $\Delta$-plane.  The leading  term  resulting from   the small $\bar z$ expansion is given by,
	\begin{align}
		\hat a_{\rm disc}^{+(0)}(\Delta,J)=2\Big(\frac{m_{\rm th}}{2\pi}\Big)^{k-\frac{1}{2}}K_l\sum_{m=1}^\infty\frac{e^{-i\mu m}}{m^{\Delta+\frac{1}{2}-k}}\int_{0}^\infty dy\frac{y^{\frac{3}{2}-k} J_{\frac{1}{2}-k}(m m_{\rm th} y)  }{(-\Delta +2 k+l-1)(1+y^2)^{\Delta+2-2k}}.
	\end{align}
	Performing the integral we obtain 
	\begin{align}\label{Bessel form d}
		-\hat a^{+(0)}_{\rm disc}|_{{\rm Res\ at\ } \Delta=2k+l-1}
		&=K_l\frac{m_{\rm th}^{k+l-\frac{1}{2}}}{2^{k+l-\frac{3}{2}}\pi^{k-\frac{1}{2}}l!}\sum_{m=1}^\infty e^{-im\mu} m^{-k+\frac{1}{2}} K_{k+l-\frac{1}{2}}(m m_{\rm th}),\\
		&=K_l\sum_{n=0}^{k+l-1}\frac{ m_{\rm th}^{k+l-n-1} (k+l-n)_{2 n} \text{Li}_{k+n}\left(e^{-m_{\rm th}-i \mu }\right)}{2^{k+l+n-1}\pi^{k-1}n! \Gamma (l+1)}.
	\end{align}
	Combining the terms from both the  $m>0$ and $m<0$ in the sum of (\ref{disccontri}), we obtain the contribution across the 
	discontinuities to be 
	\begin{align}\label{a l d}
		a_\mathcal{O}[0,l]=K_l\sum_{n=0}^{k+l-1}\frac{ m_{\rm th}^{k+l-n-1} (k+l-n)_{2 n} }{2^{k+l+n-1}\pi^{k-1}n! \Gamma (l+1)}\Big[\text{Li}_{k+n}\left(e^{-m_{\rm th}-i \mu }\right)+(-1)^l\text{Li}_{k+n}\left(e^{-m_{\rm th}+i \mu }\right)\Big].
	\end{align}
	
	\subsection*{Arc Contribution}
	Just as in the case 
	of $ d=3 $  we must  compute the arc contribution to the inversion formula  by 
	integrating along the circle of infinite radius in complex $ w $-plane defined in (\ref{defwplane}). 
	Here again only the $ m=0 $ term from the 2-point function \eqref{prop d} survives along this circle at infinity
	\begin{align}
		\lim_{|w|\to \infty} g(r,w)=\frac{m_{\rm th}^{k-\frac{1}{2}}}{(2\pi)^{k+\frac{1}{2}}}\frac{K_{k-\frac{1}{2}}(m_{\rm th} r)}{r^{k-\frac{1}{2}}}.
	\end{align}
	Further using the  formula \eqref{db2arc}, it can be seen in the integral along the arc, it is only the mode 
	 $ l=0 $  contributes.  Finally, we can 
	extend the limit of $ r $-integral till infinity as it doesn't alter the pole structure of $ \hat a(\Delta,0) $ or the residue,
	\begin{align}
		\hat a_{\rm arc}(\Delta,0)&=2 K_l\bigg(\frac{m_{\rm th}}{2\pi}\bigg)^{k-\frac{1}{2}} \int_0^\infty dr\frac{ K_{k-\frac{1}{2}}(m_{\rm th}r)}{r^{\Delta-k+\frac{3}{2}}}.
	\end{align}
	The integral is performed easily and results in 
	\begin{align}
		-\hat a_{\rm arc}|_{{\rm Res\ at\ }\Delta=d-2}=\frac{m_{\rm th}^{2 k-1} \Gamma \big(\frac{1}{2}-k\big)}{(2\sqrt{\pi})^{2k-1}}.
	\end{align}
	
	The arc contribution occurs for the 1-point function of $|\phi|^2$. Combining the contribution from the discontinuity for the one point function of $|\phi|^2$ along with the arc and demanding that the thermal one point function vanishes, we obtain 
	\begin{align}  \label{gap eq d}
		&-(\hat a_{\rm disc}(\Delta,0)+\hat a_{\rm arc}(\Delta,0))|_{{\rm Res\ at\ }\Delta=d-2}=0,\\ \nonumber
		&
		\sum_{n=0}^{k-1}\frac{m_{\rm th}^{k-n-1} (k-n)_{2 n} \left[\text{Li}_{k+n}\left(e^{i \mu -m_{\rm th}}\right)+\text{Li}_{k+n}\left(e^{-m_{\rm th}-i \mu }\right)\right]}{\pi^{k-1}2^{k+n-1}n!}+\frac{m_{\rm th}^{2 k-1} \Gamma \big(\frac{1}{2}-k\big)}{(2\sqrt{\pi})^{2k-1}}=0.
	\end{align}
	Comparing this equation to the gap equation (\ref{appengapeq}) which was  obtained from the partition function at large $N$ and strong 
	coupling we see that they precisely agree.  
	
	\subsubsection{The large spin limit} \label{largel}
	
In this sub-section we study the behaviour of the one point functions of the model at large spin. 
We will again observe that the one point functions simplify just as we have seen for the model in $d=3$. 
Let us begin with  the expression of the one point function in terms of the Bessel functions given in \eqref{Bessel form d},
	\begin{align}\label{Bessel form d 1}
		-\hat a^{+(0)}_{\rm disc}|_{{\rm Res\ at\ } \Delta=2k+l-1}
		&=K_l\frac{m_{\rm th}^{k+l-\frac{1}{2}}}{2^{k+l-\frac{3}{2}}\pi^{k-\frac{1}{2}}l!}\sum_{m=1}^\infty e^{-im\hat \mu} m^{-k+\frac{1}{2}} K_{k+l-\frac{1}{2}}(m m_{\rm th}).
	\end{align}
	The asymptotic behaviour of the BesselK at large orders, but fixed argument  is given by the formula,
	\begin{align}
		\lim_{l\rightarrow \infty} 
		K_{k+l-\frac{1}{2}}(m m_{\rm th}) =
		\sqrt{\frac{\pi}{2(k+l-\frac{1}{2})}}\bigg(\frac{e m m_{\rm th}}{2(k+l-\frac{1}{2})}\bigg)^{-(k+l-\frac{1}{2})}.
	\end{align}
	Following the same steps as in the case of $d=3$ by substituting the 
	asymptotic formula of Bessel function in \eqref{Bessel form d 1}, taking the large $l$ limit inside the sum over $m$ and retaining the term leading order in $l$ we obtain the following expression for the 1-point function in the 
	large $l$ limit
	\begin{align} \label{larglk}
		\lim_{l \rightarrow\infty} 
		a_\mathcal{O}[0,l]=\frac{\Gamma \left(k-\frac{1}{2}\right)}{4\pi^{k+\frac{1}{2}}} (e^{-i\hat\mu}+(-1)^le^{i\hat\mu}) .
	\end{align}
	Here we have retained only the $m=1$ term as the rest of the terms in the sum  (\ref{Bessel form d 1}) are exponentially suppressed in $l$. 
	We can compare the pre-factor that occurs in (\ref{larglk}) with the 1-point function of bi-linears of large spin in 
	 the Gaussian theory. For this we examine (\ref{a l d})  at $m_{\rm th}=0, \mu=0$
	 \begin{eqnarray} \label{freek1pt}
	 a_{\cal O}^{\rm free} [0, l] =  \frac{  \Gamma( k -\frac{1}{2} ) }{ 4\pi^{k-\frac{1}{2} } }  \zeta( 2k +l -1)  \times\Big( 1+ (-1)^l  \Big).
	 \end{eqnarray}
	 Taking the large spin limit of the one point functions in the free theory we obtain 
	 \begin{eqnarray}
	 \lim_{l\rightarrow \infty}  a_{\cal O}^{\rm free} [0, l]  &=&  \frac{  \Gamma[(k -\frac{1}{2} ) }{ 2\pi^{k-\frac{1}{2} } } , \qquad l \in 2 \mathbb{Z}, \\ \nonumber
	 &\equiv& a_{\cal O}^{\rm free} [0, \infty]  .
	 \end{eqnarray}
	 Now comparing the expression for the  large spin limit of the 
	 one point function at the non-trivial fixed points in (\ref{freek1pt}) and using the above definition 
	 we can write
	 \begin{eqnarray} \label{arbkll}
	 \lim_{l \rightarrow\infty} 
		a_\mathcal{O}[0,l]  &=&  a_{\cal O}^{\rm free} [0, \infty]  \cosh\mu , \qquad  {\rm for}\quad  l \in 2 \mathbb{Z}, \\ \nonumber
		 \lim_{l \rightarrow\infty} 
		a_\mathcal{O}[0,l] &=&  a_{\cal O}^{\rm free} [0, \infty]  \sinh\mu , \qquad  {\rm for}\quad  l \in 2 \mathbb{Z} +1,
		\end{eqnarray}
		Here we have analytically continued to real chemical potential. 
	Therefore we see that at large $l$, the one point functions at the non-trivial fixed point 
	factorize into a pre-factor which can be identified with the one point function of the free theory. The dependence 
	of the chemical potential is through the elementary hyperbolic functions $\cosh\mu$ or $\sinh\mu$.
	As we have mentioned earlier the result in (\ref{arbkll})  is also true for the one point functions $b_{\cal O}[0,l]$ since these
	are related by structure constants and normalizations which do not change from that of the free theory at large $N$. 

	\subsubsection{The large $d$ limit} \label{largedun}
	
	One point functions of bi-linear currents were studied numerically 
	in $d =2k +1$ dimensions with $k$ odd  for the $O(N)$ model and $k$ even for the Gross-Neveu at large $k$ and fixed spin
	 \cite{David:2023uya}.
	This involved numerically solving the gap equation for each case and evaluating the one point function. 
	It was seen that the ratio of one point function at the critical point to that of the 
	Gaussian fixed point  tends to zero on increasing the dimensions, this conclusion was arrived at by 
	observing the trends to the maximum dimensions of  around $k=39, 40 $ for the $O(N)$ and Gross-Neveu model respectively. 
	In this section we will demonstrate this observation analytically. 
	To do this we solve the gap equation analytically in the large $k$ limit. 
	
	We begin first with the gap equation for the model of $U(N)$ scalars in the absence of chemical potential, then the model 
	is essentially the $O(2N)$ model. 
	The gap equation  at zero chemical potential can be read out from (\ref{gap eq d}) and  is given by 
	\begin{align}\label{gap eq b}
		\sum_{n=0}^{k-1}\frac{m_{\rm th}^{k-n-1} (k-n)_{2 n} \text{Li}_{k+n}\left(e^{-m_{\rm th}}\right)}{\pi^{k-1}2^{k+n-2}n!}
		+\frac{m_{\rm th}^{2 k-1} \Gamma \big(\frac{1}{2}-k\big)}{(2\sqrt{\pi})^{2k-1}}=0.
	\end{align}
	At large $k$, the polylogarithm functions can be approximated as,
	\begin{align}
		\lim_{k\rightarrow \infty} {\rm Li}_{k+n}(e^{-x}) = e^{-x} + O(e^{(-k \log 2) } ) . \label{Li asymp}
	\end{align}
Substituting this  approximation in the gap equation we arrive at the  equation
	\begin{align} \label{firstlkb}
		{2^{k+\frac{3}{2}} K_{k-\frac{1}{2}}(m_{\rm th})}+	\frac{\pi(m_{\rm th}) ^{k-\frac{1}{2}}  }{\Gamma( \frac{1}{2} + k) \cos\pi k}=0.
	\end{align}
	Here we have used the series representation of the BesselK to  write the sum over $n$ in (\ref{gap eq b}).
	Since $k$ is odd, we can set $\cos\pi k =-1$. 
	To simplify the equation further, we can appeal to the asymptotic expansion for the Bessel function at large $k$. 
	However from the numerical analysis of 
	\cite{David:2023uya}  we observe that $m_{th}$ also increases linearly with $k$ for large $k$. 
	Therefore we look for the asymptotic expansion in which its  argument along with the order are large. 
	This is given by \cite{NIST:DLMF}
	\begin{align}\label{bessel asymp}
		&\lim_{\nu\to \infty}	K_\nu(\nu z)\sim_{}\Big(\frac{\pi}{2 \nu}\Big)^{1/2} \frac{e^{-\nu \eta}}{(1+z^2)^{1/4}}, \\
		&{\rm where,}\quad
		\eta=\sqrt{1+z^2}+\log\bigg[\frac{z}{1+\sqrt{1+z^2}}\bigg]\nonumber.
	\end{align}
	For the Bessel function in (\ref{firstlkb})  $\nu$ and the $z$ are given by,
	\begin{align}
		\nu\equiv k-\frac{1}{2} \qquad {\rm and}\qquad z\equiv\frac{m_{\rm th}}{k-\frac{1}{2}},
	\end{align}
	Therefore using 
	\eqref{bessel asymp} we obtain 
	\begin{align}
		\lim_{k\rightarrow \infty} {2^{k+\frac{3}{2}} K_{k-\frac{1}{2}}(m_{\rm th})}\sim\frac{4 \sqrt{\pi } e^{-\frac{1}{2} \sqrt{(1-2 k)^2+4 m_{\rm th}^2}} \left({2 k-1+\sqrt{(1-2 k)^2+4 m_{{\rm th}}^2}}\right)^{k-\frac{1}{2}}}{m_{\rm th}^{k-\frac{1}{2}}[{(1-2 k)^2+4 m_{\rm th}^2}]^{1/4}}.
	\end{align}
Let us now proceed with the following  ansatz for the thermal mass
	\begin{align} \label{thermmassan}
		m_{\rm th}=\alpha k+\beta \log k+c,
	\end{align}
	where $ \alpha,\beta  $ and $ c $ are constants.
	Consider the 
	 the ratio of the 1st term to the 2nd term from the gap equation \eqref{gap eq d}, we obtain 
	  at large $k$,
	\begin{eqnarray}\label{gap eq rat}\nonumber
		&& \lim_{k\rightarrow \infty} \frac{{2^{k+\frac{3}{2}} K_{k-\frac{1}{2}}(m_{\rm th})}}{\frac{\pi(m_{\rm th}) ^{k-\frac{1}{2}}  }{\Gamma( \frac{1}{2} + k) \cos\pi k}} \\ \nonumber
		&& \qquad\qquad\qquad\qquad \sim -
		\exp\Bigg\{\Big[-\sqrt{\alpha ^2+1}+\log \left(2 \left(\sqrt{\alpha ^2+1}+1\right)\right)-2 \log (\alpha )-1\Big]k  \\
		&&-\Big[\frac{(1+\sqrt{\alpha ^2+1}) \beta}{ \alpha }\Big]\log k+\log \left(\frac{4}{({\alpha ^2+1})^{1/4}}\right)+\frac{\alpha  \log \left(\frac{\sqrt{\alpha ^2+1}-1}{2}\right)-2c (\sqrt{\alpha ^2+1} -1)}{2 \alpha }\nonumber\\
	&& \qquad\qquad\qquad	+O\Big(\frac{1}{k}\Big)\Bigg\} .
	\end{eqnarray}
	The ratio in the above equation has to be $-1$ to satisfy the gap equation at large $k$.  Note we have used $\cos \pi k =-1$ as 
	$k$ is odd. 
	Thus at large $ k $ the coefficients associated with $ k, \log k $, and the constant term  in the exponent 
	should vanish. 
	These 
	  conditions  result in the following equations which can be used  to determine the constants $ \alpha,\beta $ and $ c $
	  in the ansatz (\ref{thermmassan})
	\begin{eqnarray}
		\log \left[2 \left(\sqrt{\alpha ^2+1}+1\right)\right]-\sqrt{\alpha ^2+1}-2 \log (\alpha )-1&=&0,
		\qquad	\\ \nonumber
		 \beta&=&0, \\ \nonumber
		\qquad \log \left(\frac{4}{({\alpha ^2+1})^{1/4}}\right)+\frac{1}{2} \Bigg(\log \Big(\frac{\sqrt{\alpha ^2+1}-1}{2}\Big)-\frac{2 (\sqrt{\alpha ^2+1}+1) c}{\alpha }\Bigg)&=&0.
	\end{eqnarray}
	Since the equation for $\alpha$ is decoupled, we can solve the 1st equation and then substitute the value of 
	$\alpha$ in the 3rd to determine $c$. 
	This yields the following asymptotic solution for the thermal mass for large values of $k$. 
	\begin{align}\label{.69 k}
		m_{\rm th}=0.69486\, k+0.0560686\quad{\rm for}\ k\in2\mathbb{ Z}+1.
	\end{align}
	Note that due to the presence of the factor $\cos(\pi k)$ in the expression \eqref{gap eq rat}, the gap equation has real root for $m_{\rm th} $ when $k$ is odd integer.
	To verify the asymptotic solution in (\ref{.69 k}), let us compare it with the numerical solution of the gap equation 
	obtained using Mathematica. This comparison is done in table \ref{tablemth}. Observe that  when $k$ reaches around 
	$25$, this is accurate up to $.02\%$ error. 
	
	\begin{table}
		\begin{center}
			\begin{tabular}{|c |c |c |c|}\hline
				$ k $& $ m_{\rm th} $  & Asymptotic value & $ \% $ error\\\hline 
				1 & 0.962424 & 0.750928 & 21.9753 \\
				3 & 2.17756 & 2.14065 & 1.69491 \\
				5 & 3.55044 & 3.53037 & 0.565411 \\
				7 & 4.93425 & 4.92009 & 0.286982 \\
				9 & 6.32077 & 6.30981 & 0.173384 \\
				11 & 7.70846 & 7.69953 & 0.115946 \\
				13 & 9.09679 & 9.08925 & 0.0829414 \\
				15 & 10.4855 & 10.479 & 0.0622524 \\
				17 & 11.8744 & 11.8687 & 0.0484366 \\
				19 & 13.2635 & 13.2584 & 0.0387561 \\
				21 & 14.6528 & 14.6481 & 0.0317117 \\
				23 & 16.0421 & 16.0378 & 0.0264264 \\
				25 & 17.4315 & 17.4276 & 0.0223601 \\
				\hline
			\end{tabular}
			\caption{This table demonstrates  the accuracy of the asymptotic formula of thermal mass \eqref{.69 k} at large $ k $ by comparing its values with the numerical solution of the gap equation at different large values of  dimensions $d = 2k +1 $.} \label{tablemth}
		\end{center}
	\end{table}

\begin{figure}\begin{center}
			\includegraphics[scale=.58]{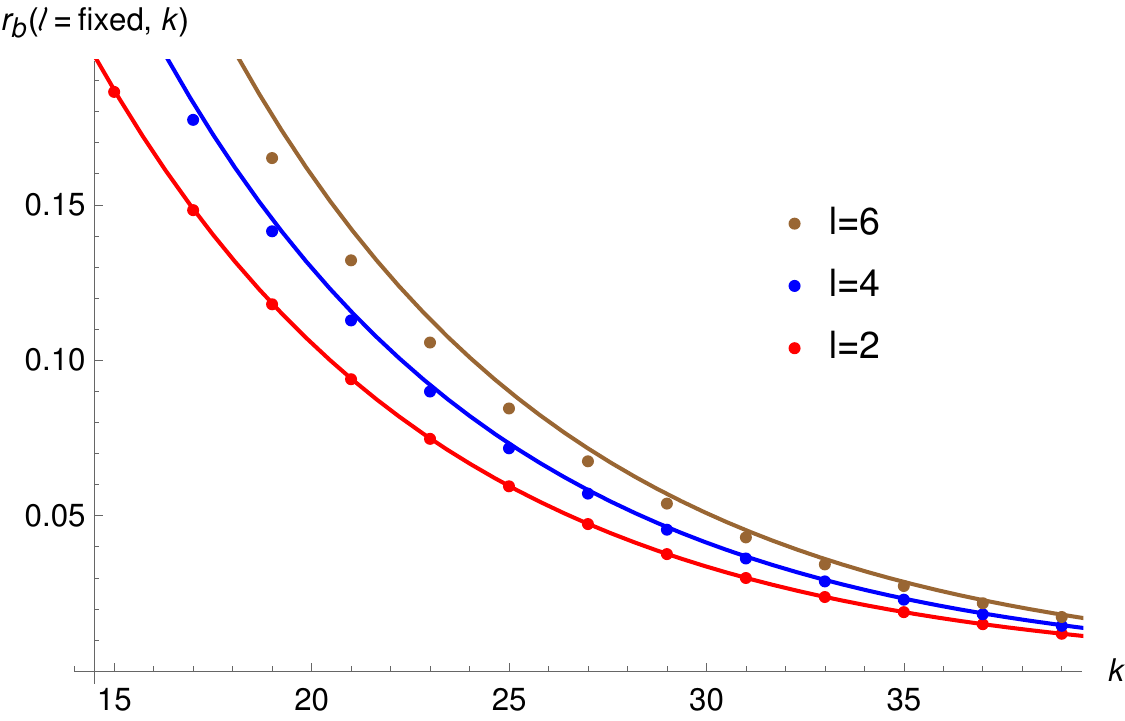}
			\caption{ At $\mu=0$, $r_b(l,k)$ is evaluated by solving the gap equation \eqref{gap eq b} numerically and plotted against $k$ for various fixed values of $l$; the red, blue and brown dots represent the numerical values of $r_b$ for $ l=2,4,6 $ respectively.  The red, blue and brown solid curves in the figure refer to the plot of the large $k$ asymptotic formula of $r_b$ \eqref{exp} for $ l=2,4 $ and $6$ respectively. It is shown that the asymptotic formula works well for $l\ll k$.	 }\label{large k b}
		\end{center}
	\end{figure}

We can now proceed to substitute the asymptotic formula \eqref{.69 k}  in the 1-point function. 
At zero chemical potential, the one point function of the bi-linears are obtained by setting 
 $\hat\mu=0$ in the equation. Only currents with even spin have non-zero expectation values, this is given by 
  \eqref{a l d},
	\begin{align}\label{1 pt fn free}
		a_\mathcal{O}[0,l] &=\frac{1}{\pi^k (k-\frac{1}{2})_l}\sum_{n=0}^{k+l-1}\frac{ m_{\rm th}^{k+l-n-1} (k+l-n)_{2 n} }{2^{k+l+n}n! }\text{Li}_{k+n}\left(e^{-m_{\rm th}}\right), 
		\qquad l \in 2 \mathbb{Z}.
	\end{align}
	We would like to evaluate the ratio of one point functions at the non-trivial fixed point to the Gaussian fixed point. 
	The one point function at the Gaussian fixed point  is given by setting $m_{\rm th} =0$ in the above equation
\begin{align} \label{gaussval}
		a_\mathcal{O}^{\rm free}[0,l]=\frac{\Gamma \left(k-\frac{1}{2}\right)}{2 \pi ^{k+\frac{1}{2}} } \zeta(2k+l-1), 
		\qquad  l \in 2 \mathbb{Z}.
	\end{align}
	Let us define the ratio of the one point functions in (\ref{1 pt fn free}) to the corresponding one at the 
	Gaussian fixed point 
\begin{align}\label{r_b}
		r_b(l,k)=\frac{a_\mathcal{O}^{\rm crit}[0,l]}{a_\mathcal{O}^{\rm free}[0,l]},
	\end{align}
	Here  $a_\mathcal{O}^{\rm crit}[0,l]$ refers to the 1-point function at the 
	critical value of the thermal mass which satisfies the gap equation and $a_{\mathcal{O}}^{\rm free }[0,l] $ is the 
	1-point function at the Stefan-Boltzmann limit obtained by taking $m_{\rm th}\to 0 $ in \eqref{1 pt fn free} given in (\ref{gaussval}). 
	We now proceed to use asymptotic value
	 of $m_{\rm th }$ at large $k$  from (\ref{.69 k}) to obtain the large  $k$ limit  for the ratio in (\ref{r_b}) keeping $l$ fixed. 
	Here we again use the fact that  the large order 
 polylogarithm functions appearing in the one point function \eqref{1 pt fn free} can be approximated by (\ref{Li asymp}), 
	Then the  1-point function at large $k$ is given by,
	\begin{align}
		\lim_{k\rightarrow \infty} 
		a_\mathcal{O}[0,l]=\Big(\frac{m_{\rm th}}{2}\Big)^{k+l-\frac{1}{2}}\frac{ K_{k+l-\frac{1}{2}}(m_{\rm th})}{\pi^{k+\frac{1}{2}}(k-\frac{1}{2})_l}, \qquad l \in 2 \mathbb{Z}.
	\end{align}
	Further using the asymptotic formula \eqref{bessel asymp} for bessel functions 
	and the asymptotic solution of the gap equation \eqref{.69 k}  we obtain the following asymptotic value of the 
	 ratio $r_b(l,k)$ defined in \eqref{r_b}  
	\begin{eqnarray}\label{exp}
		\lim_{k\rightarrow\infty} 
		r_b(l,k) &=&1.53742\times\exp\bigg[-0.114385 k+0.10333 l-0.597829+O(k^{-1})\bigg] \nonumber \\ 
		&& \qquad \qquad \times\Big(1+O(k^{-1}) \Big), \\ \nonumber
		&&  \qquad \qquad \ \qquad \qquad  l \in 2\mathbb{Z}.
	\end{eqnarray}
	From this explicit expression, it is easy to see that the ratio vanishes exponentially in $k$.  It can also been seen for 
	higher values of the spin  $l$,   the same suppression is achieved at larger values of $k$  due to positive coefficient 
	of $l$ in the exponential. 
	These characteristics of the ratio was noted in the numerical study of \cite{David:2023uya}. 
	In the figure \ref{large k b}, we have plotted  the ratio   $r_b(l,k) $  given by the the asymptotic formula in (\ref{exp}) 
	against the same ratio calculated by substituting the 
	 values of the thermal mass by solving  \eqref{gap eq b} numerically. 
	 We can see that the asymptotic expression (\ref{exp}) is in good agreement with the numerics. 

Let us now turn on the chemical potential and study the ratio \eqref{r_b} and large dimensions. We repeat the same steps 
as in the case when $\mu=0$.  The gap equation in presence of chemical potential given by \eqref{gap eq d} takes following form at large $ k $
	\begin{align} \label{gapeqchempot}
		{2^{k+\frac{3}{2}}   K_{k-\frac{1}{2}}(m_{\rm th})}\cosh (\mu )+	\frac{m_{\rm th} ^{k-\frac{1}{2}}  \pi}{\Gamma( \frac{1}{2} + k) \cos\pi k}=0.
	\end{align}
	Here we have used the asymptotic formula  \eqref{Li asymp}  and taken the 
chemical potential $\mu$  to be real 
\footnote{Note that when $m_{\rm th}=0$, it is easy to see that the gap equation in large dimensions 
 (\ref{gapeqchempot}), admits  the solution for purely imaginary chemical potential 
with $\hat\mu = i \mu = \frac{\pi}{2}$. This was observed in \cite{Filothodoros:2018pdj}.}.
 We proceed by approximating the Bessel functions at large order by the formula \eqref{bessel asymp} and  then consider  the ansatz,
	\begin{align}
		m_{\rm th}=\alpha  k+\beta  \log (k)+c,
	\end{align}
	where $ \alpha,\beta, c  $ are constants.  The ratio of the 1st term to the 2nd term of the gap equation
	(\ref{gapeqchempot})  at  large $ k $ is given by 
	\begin{align}\label{ratio mu}
		&\lim_{k\rightarrow\infty} \frac{{2^{k+\frac{3}{2}}   K_{k-\frac{1}{2}}(m_{\rm th})}\cos (\mu )}{\Big(\frac{m_{\rm th} ^{k-\frac{1}{2}}  \pi}{\Gamma( \frac{1}{2} + k) \cos\pi k}\Big)}\sim\\
		&\exp\Bigg\{-\Big[1+\sqrt{\alpha ^2+1}-\log \Big({\frac{2}{\alpha^2} (1+\sqrt{\alpha ^2+1})}\Big)\Big]k 
		-(1+\sqrt{\alpha ^2+1}) \frac{\beta}{\alpha}\log k   + \nonumber\\
		&\log \left(\frac{4 \cosh \mu }{({\alpha ^2+1})^{1/4}}\right)+\frac{1}{2} \log \Big(\frac{\sqrt{\alpha ^2+1}-1}{2}\Big)-{\frac{c}{\alpha} (\sqrt{\alpha ^2+1}+1) } +O\Big(\frac{1}{k}\Big)\Bigg\}\nonumber\\ \nonumber
		&\qquad\qquad\qquad\qquad\qquad\qquad  \times\cos(\pi k) \Big(1+O(k^{-1})\Big).
	\end{align}
	The gap equation is satisfied when the right hand side  of the equation \eqref{ratio mu} is $ -1 $.
	This implies that to the leading orders in $k$, we need the coefficient of $k, \log k$ and the constant in the exponential 
	of equation (\ref{ratio mu}) to vanish. 
	Setting the coefficient of $k$ and $\log k$ to vanish,  we obtain 
	\begin{align}\label{.69 k 1}
		\alpha=0.69486 \qquad{\rm and }\qquad\beta=0.
	\end{align}
	The dependence of the  chemical potential
	 $\mu$ appears in constant term,  setting this term to zero we obtain the equation 
	\begin{align}
		\log \left(\frac{4 \cosh \mu }{({\alpha ^2+1})^{1/4}}\right)+\frac{1}{2} \bigg(\log \Big(\frac{\sqrt{\alpha ^2+1}-1}{2}\Big)-\frac{2 c (\sqrt{\alpha ^2+1}+1) }{\alpha }\bigg)=0.
	\end{align}
	Substituting the value of $\alpha$ from \eqref{.69 k 1} in the above equation we can determine the constant $c$. 
	This leads to the following asymptotic value for the thermal mass at large dimensions. 
	\begin{align} \label{mthlkchem}
		m_{\rm th}=0.69486 k+0.313322 \log (\cosh \mu )+0.0560686\quad {\rm for}\ k\in2 \mathbb{Z}+1.
	\end{align}
	Note that we have kept the chemical potential to be fixed and finite in this analysis. 
	Now we are ready to investigate the large $k$ behaviour of the 1-point functions, again we follow the similar steps as before. First we approximate the polylogarithm functions appearing in the expression for the 1-point function using (\ref{Li asymp}).
Then taking the large $k$ limit in  \eqref{a l d} and also analytically continuing to real chemical potential, we obtain 
	\begin{align}
		\lim_{k\rightarrow\infty}
		a_\mathcal{O}[0,l]=\Big(\frac{m_{\rm th}}{2}\Big)^{k+l-\frac{1}{2}}\frac{ K_{k+l-\frac{1}{2}}(m_{\rm th})}{2\pi^{k+\frac{1}{2}}(k-\frac{1}{2})_l}(e^{\mu}+(-1)^le^{-\mu}).
	\end{align}
	Substituting the large $ k $ asymptotic solution for the thermal mass  (\ref{mthlkchem})
	in the above expression, we can evaluate the ratio $ r_b $ which yields 
	\begin{align} \label{asymkb}
		\lim_{k\rightarrow \infty} r_b(l,k) &=1.53742 \exp \Big[-0.114385 k+0.10333 l -0.0981709 \log (\cosh \mu )
		\\ \nonumber  & -0.597829 
		+O(k^{-1})\Big]
		 \times \frac{e^{\mu}+(-1)^le^{-\mu}}{2}\Big(1+O(k^{-1})\Big).
	\end{align}
	From this expression it is clear that  at large $ k $ even in presence of chemical potential, the one point function decays
	exponentially in $k$. 
The leading dependence on the chemical potential is at the sub-leading order in $k$. 
Again we emphasize the equation (\ref{asymkb}) also holds for the ratios of the one point functions $b_{\cal O}[0, l]$, since 
they are proportional to $a_{\cal O}[0, l]$. The constant of proportionality involves only normalizations and structure constants 
which are identical to that of the free theory at large $N$.

	\section{The Gross-Neveu model} \label{secgnmodel}
	
	The Gross-Neveu model is an important model both in particle and condensed matter 
	physics.  The  phenomenon of dynamical symmetry breaking and dynamical Higgs mechanism 
	can be exactly demonstrated in this model for in $d=2$ space-time dimensions \cite{Gross:1974jv}.  
	In $d=3$, the model is perturbatively non-renormalizable, however it is known to have a 
	UV fixed point \cite{Rosenstein:1988pt}, 
	which renders the theory to be finite and is a good toy model for asymptotic 
	safety of gravity \cite{Braun:2010tt}.  The theory is non-unitary for $d>4$, 
	In this section we study the behaviour of the one point function of the 
	 higher spin currents of  the Gross-Neveu model at finite chemical potential  in $d=3$
	 as well as in 
	 $d =2k+1$ 
	 dimensions with $k$ even.  We focus on the behaviour of these one point functions 
	 at the non-trivial fixed point. 
	
	The model is defined by the action 
	\begin{eqnarray}
	S = \int d^{2k+1} x \left[ { \tilde \psi}^\dagger  \gamma^\mu \partial_\mu \tilde \psi +\mu{\tilde \psi}^\dagger  \gamma^0  
	{ \tilde \psi}
	  -\frac{\lambda}{N} ({\tilde \psi}^\dagger \tilde \psi)^2  \right].
	\end{eqnarray}
	where $\tilde \psi$ is a Dirac fermion  in $2k+1$ dimensions 
	transforming in the fundamental of $U(N)$  and the gamma matrices satisfy the 
	relation 
	\begin{eqnarray}
	\{\gamma^\mu, \gamma^\nu \} = 2\delta^{\mu\nu}, \qquad \mu,\nu = 1,\cdots ,2k+1.
	\end{eqnarray}
	In \cite{David:2023uya}, thermal  one point functions of the  
	following bi-linears were obtained using the Euclidean inversion formula for the model at large $N$ and strong coupling
	$\lambda$
	\begin{eqnarray} \label{fermcurrent}
	{\cal O}_{+}[n, l ] &=& {\tilde \psi }^\dagger \gamma_{\mu_1} D_{\mu_2} \cdots D_{\mu_l}( D^2)^n  \tilde\psi,  \\ \nonumber
	{\cal O}_{-}[n, l]  &=& {\tilde \psi }^\dagger  \gamma^\mu D_\mu D_{\mu_1} \cdots D_{\mu_l } (D^2)^{n}  \tilde \psi, \\ \nonumber
	{\cal O}_{0}[n, l ] &=& {\tilde \psi }^\dagger  D_{\mu_1} D_{\mu_2} \cdots D_{\mu_l} (D^2)^{n} \tilde  \psi .
	\end{eqnarray}
	Here these  operators are symmetric traceless rank $l$ tensors, 
	the thermal expectation values were obtained in the absence of chemical potential. 
	In the section \ref{gn1ptsec} we generalise this calculation to the case of non-vanishing chemical potential. 
	We show that the condition for vanishing of the expectation value of the operator 
	$ {\tilde \psi}^\dagger \tilde{\psi} $ continues to agree with the 
	the gap equation obtained from the partition function in the presence of the chemical potential. 
	This agreement ensures that the stress tensor obtained from the partition function agrees  with the expectation value 
	of the operator ${\cal O}_+[0, 2]$ and also the canonical  relationship between the partition function  and the 
	stress tensor for a thermal field theory  given in (\ref{partstresrel})
	continues to hold.  The gap equation also ensures all the thermal one point functions  scale with the temperature in accordance with the dimensions of the operator. 
	
	  After the derivation of the thermal one point functions  in the presence of chemical potential for the Gross-Neveu model 
	  in $2k+1$ dimensions, 
	   we focus on the interesting case of $d=3$.  Here apart from the free theory, 
	  there is a  non-trivial solution of the gap equation for purely imaginary thermal mass
	  \footnote{The theory with vanishing  thermal mass and purely imaginary chemical potential was studied in 
	  \cite{Christiansen:1999uv,Filothodoros:2016txa,Filothodoros:2018pdj}. Real chemical potentials were considered in 
	  \cite{Hands:1992ck,Inagaki:1994ec}, but the focus was on  the model at finite coupling not the conformal invariant point. }. 
	  This 
	  was first observed  in \cite{Petkou:2000xx} and the theory was argued to have a 
	  Yang-Lee edge singularity. In  \cite{Diatlyk:2023msc} 
	  the $1/N$ corrections to the one point functions at this  fixed point was obtained. 
	  In section \ref{gn3d} we examine this fixed point in the chemical potential plane. 
	  We show that the thermal one point functions have a branch cut in the $\mu$-plane. 
	  The critical exponent of the pressure or the free energy  at the branch point  is $\frac{3}{2}$. 
	  This coincides with the mean field 
	  theory exponent of the Yang-Lee edge singularity \cite{10.1063/1.470178} for repulsive core interactions.

	  In section \ref{gnlargel}
	  we proceed to study the one point function of the operator ${\cal O}_{+}[0, l ]$ first at large spin $l$ and then 
	  in section \ref{gnlarged}
	  for large dimensions $k$. 
	  Our conclusions are identical to that observed for the bosonic model. 
	  At large spin $l$, the 1-point functions factorise to that of the  free theory times $\cosh\mu$ or $\sinh\mu$ depending 
	  on the spin being  even or odd respectively.
	  For large $d= 2k+1$ with even  $k$ we solve the gap equation analytically and demonstrate that the ratio
	  of the  thermal one point 
	  functions at the non-trivial fixed point to the Gaussian fixed point are suppressed exponentially in $k$.

	\subsection{Thermal one point functions} \label{gn1ptsec}
	
	\subsubsection*{One point functions from the partition function}
	
	The partition function at large $N$ and strong coupling of the Gross-Neveu model has been evaluated in  
	appendix \ref{A.2}
	and is given in (\ref{partitionatm}). 
	The thermal mass is determined from 
	the  saddle point equation or gap equation  (\ref{appengapeqf}) as derived from the partition function
	\begin{eqnarray}
		 &&(2m_{\rm th}\beta )^k \sum _{n=0}^{k-1} \frac{(k-n)_{2n} }{ (2 m_{\rm th}  \beta )^n n!}  \Big[ 
		  \text{Li}_{k+n}(-e^{-m_{\rm th}  \beta - \mu\beta}) + \text{Li}_{k+n}(-e^{-m_{\rm th}  \beta +\mu\beta})
		  \Big]   \nonumber \\
		  && \qquad\qquad\qquad\qquad\qquad\qquad
		   +\frac{(m_{\rm th} \beta)^{2k}}{\sqrt{\pi}} \Gamma\big(\frac{1}{2}-k\big) =0. 
	\end{eqnarray}
	Here we have written the equation for real chemical potential $\mu$ and kept track of the dependence on the 
	temperature.
	This equation relates the thermal mass to the chemical potential and ensures that the theory is conformal when the 
	gap equation is satisfied.  Using the gap equation, 
	the stress tensor evaluated from the partition function is given by 
	\begin{eqnarray} \label{gnstressp}
		T_{00} &=&\frac{1}{\beta^{2k+1}\pi^k}\Big(\frac{k}{2k+1}\Big) \\ \nonumber
		&&\times  \sum_{n=0}^{k+1} \frac{  (k-n+2)_{2n} (\beta  m_{\rm th})^{k-n+1}}{ 2^{n}n! } 
		 \Big[\text{Li}_{k+n}\big(-e^{-m_{\rm th}  \beta -\beta\mu}\big)
		+ \text{Li}_{k+n}\big(-e^{-m_{\rm th}  \beta +\beta\mu}\big)
		 \Big].
	\end{eqnarray}
	This is the stress tensor per unit Dirac fermion,  the  free energy at the critical point satisfies the relation of thermal CFT. 
	\begin{eqnarray}
	F = -\frac{1}{\beta N V_{2k} } \log Z  =\frac{ T_{00}}{2k}.
	\end{eqnarray}
	The expectation value of the  $U(1)$ current is non-vanishing once the chemical potential is turned on. 
	This can also be read out from the partition function by 
	\begin{eqnarray} \label{gnspin1}
	J  &=&  - \frac{1}{\beta NV_{2k} } \frac{\partial}{ \partial\mu} \log Z , \\ \nonumber
	&=& \sum_{n=0}^k \frac{ (k-n+1)_{2 n} (m_{\rm th}  \beta )^{k-n} \Big[\text{Li}_{k+n}\left(-e^{-\beta  (m_{\rm th} - \mu )}\right)-\text{Li}_{k+n}\left(-e^{-\beta  (m_{\rm th}  +\mu )}\right)\Big]}{\beta^{2k-1}\pi^k2^{n+1}n!}.
	\end{eqnarray}
	In the appendix \ref{A.2}  we have also 
	 obtained the expectation value of the current $ \tilde\psi^\dagger D_ \mu  \tilde\psi$ directly from the 
	 partition function,  this is given by 
	 \begin{eqnarray} \label{gnspin11}
	 \tilde J &\equiv&  \langle  \tilde\psi^\dagger D_\mu  \tilde\psi \rangle, \\ \nonumber
	 &=& m_{\rm th}^{2 k}
	 \sum _{n=0}^k \frac{  (k-n+1)_{2 n} \left(\text{Li}_{k+n}\left(-e^{ -m_{\rm th} \beta- \beta  \mu }\right)-\text{Li}_{k+n}\left(-e^{-\beta  m_{\rm th}+ \mu\beta }\right)\right)}{(\beta  m_{\rm th})^{k+n-1}\pi^k2^{n+1}n!}.
	 \end{eqnarray}
	 In the first line, we have used the fact that only the time component  of the current 
	 gets expectation value and defined that as $\tilde J$. 
	 Comparing (\ref{gnspin1}) and (\ref{gnspin11}), we obtain the relation 
	 \begin{eqnarray} \label{relspin1}
	 \tilde J = -m_{\rm th} J.
	 \end{eqnarray}
	 We will see that such relations extend to arbitrary spins as first observed in the absence of chemical potential 
	 in \cite{David:2023uya}. 
	
	\subsubsection*{One point functions from the inversion formula}
	
	To evaluate one point function for higher spin bi-linears we can follow the approach taken for the bosonic model. 
	We introduce chemical potential, by considering twisted boundary conditions. 
	First we consider imaginary chemical potential $\hat \mu = i \mu$, then the action becomes  
	\begin{eqnarray}
	S = \int d^{2k+1} x \left[ { \tilde\psi}^\dagger  \gamma^\nu (\partial_\nu - i \hat \mu \delta_{\nu,\tau} )  \tilde\psi +
	  -\frac{\lambda}{N} ({\tilde \psi}^\dagger \tilde\psi)^2  \right].
	\end{eqnarray}
	Let us define  the fermion
	\begin{equation}
	\tilde\psi =  e^{i \hat \mu \tau} \psi.
	\end{equation}
	It is easy to see that  in terms of the re-defined fermion, the co-variant derivatives in the action and 
	the currents in (\ref{fermcurrent}) reduce to ordinary derivatives. 
	The fermion bi-linears therefore become
	\begin{eqnarray}
	{\cal O}_{+}[n, l ] &=& { \psi }^\dagger \gamma_{\mu_1} \partial_{\mu_2} \cdots \partial_{\mu_l} (\partial^2)^n\psi,  
	\\ \nonumber
	{\cal O}_{-}[n, l]  &=& { \psi }^\dagger  \gamma^\mu \partial_\mu D_{\mu_1} \cdots \partial_{\mu_l } (\partial^2) ^n \psi, \\ \nonumber
	{\cal O}_{0}[n, l ] &=& { \psi }^\dagger  \partial_{\mu_1} \partial_{\mu_2} \cdots \partial_{\mu_l} (\partial^2)^n \psi .
	\end{eqnarray}
	However the anti-periodic boundary conditions of the fermion  $\tilde\psi$  imply twisted boundary conditions on $\psi$
	\begin{eqnarray}
	\psi(\tau + \beta, \vec x) = - e^{-i \hat \mu \beta} \psi( \tau, \vec x) .
	\end{eqnarray}
	Now the problem of obtaining the one point functions reduces to that studied in \cite{David:2023uya}, 
	but with twisted boundary conditions. 
	From the OPE expansion of a pair of fermions, it was shown that the OPE expansion of the following two point functions
	\begin{align}
		g_2(x) &= \langle \psi^\dagger (x) \frac{ \gamma^\mu x_\mu }{|x|} \psi(0) \rangle_{S^1_\beta \times R^{d-1} } , 
		\\ \nonumber
		g_3(x) &= \langle D_\mu \psi^\dagger (x) \gamma^\mu \psi(0) \rangle_{S^1_\beta \times R^{d-1} }.
	\end{align}
contain information of the one point functions ${\cal O}_{+}[n, l ]$ and ${\cal O}_{-}[n, l]$. These OPE expansion in general 
 contain a  linear 
combination of these  1-point functions. 
The OPE expansion of the correlator 
\begin{eqnarray}
	g_1(x)=\langle \psi^\dagger(x)\psi(0)\rangle_{S^1_\beta \times R^{d-1} }.
\end{eqnarray}
is closed and contains information of only the 1-point functions ${\cal O}_{0}[n, l ]$. 
For the class ${\cal O}_{+}[0, l ]$,  the one point functions are decoupled  from  that of ${\cal O}_{-}[n, l ]$
and can be obtained from directly from 
$g_2(x)$ for $l \geq 1$. 
The expectation value of $\bar \psi \psi$ can be obtained from examining  $g_1(x)$ and extracting out 
${\cal O}_{0}[0, 0 ]$.

To proceed, the  thermal 2-point function of the twisted fermion is given by 
	\begin{align} \label{fermicorr}
		\langle \psi_\alpha \psi_\beta^\dagger\rangle =\frac{i}{2^{\frac{d-1}{2}}}\sum_{k_0\in (2n+1) \pi }\int \frac{d^{d-1}k}{(2\pi)^{d-1 }}\frac{\gamma^\mu_{\alpha\beta} \tilde k_\mu+m_{th}\delta_{\alpha\beta}}{\tilde k^2+m_{th}^2} e^{i\tilde k x}.
	\end{align}
	where $\tilde k_\nu=k_\nu-\hat \mu\delta_{\nu 0} $ and $n\in \mathbb{Z}$. $ \alpha,\beta $ denote the spinor indices.  This correlator also satisfies twisted anti-periodic boundary condition along the thermal circle of length $\beta $, which we have 
	chosen to be unity. 
	From this 2-point function, it is easy to construct the 2-point function $g_2( x)$, 
	\begin{align}\nonumber 
		g_2(\tau,\vec{x})&=\frac{i}{|x|}\sum_{k_0\in\pi(2n+1)}\int \frac{d^{d-1}k}{(2\pi)^{d-1 }}\frac{\tilde k_\mu x^\mu}{\tilde k^2+m_{th}^2} e^{i\tilde k x}, \\
		&=\frac{1}{|x|}x^\mu\partial_\mu\sum_{k_0\in\pi(2n+1)}\int \frac{d^{d-1}k}{(2\pi)^{d-1 }}\frac{e^{i\tilde k x}}{\tilde k^2+m_{th}^2} .
		\label{g2express}
	\end{align}
	We use the Poisson re-summation formula to re-cast  the above integral  as sum over images as follows
	\begin{align} \nonumber
		\sum_{k_0\in (2n+1) \pi }\int \frac{d^{d-1}k}{(2\pi)^{d-1 }}\frac{e^{i\tilde k x}}{\tilde k^2+m_{th}^2}&=\sum_{k_0\in\pi(2n+1)}\int \frac{d^{d-1}k}{(2\pi)^{d-1 }}\frac{e^{i((2n+1)\pi-\mu)\tau}e^{i \vec{k}\cdot \vec{x}}}{((2n+1)\pi-\hat \mu)^2+ \vec{k}^2+m_{th}^2}, \\
		&=\sum_{n\in \mathbb{Z}}(-1)^n\int_{-\infty}^\infty d\omega\int\frac{d^{d-1}k}{(2\pi)^{d}}\frac{e^{in\omega}e^{i(\omega-\mu)\tau}e^{i \vec{k}\cdot \vec{x}}}{(\omega- \hat \mu)^2+ \vec{k}^2+m_{th}^2}.
	\end{align}
	The integral can now be done by  shifting the variable  $\omega$ to  $\omega'=\omega-\hat \mu$
	\begin{align} \nonumber
		\sum_{k_0\in\pi(2n+1)}\int \frac{d^{d-1}k}{(2\pi)^{d-1 }}\frac{e^{i\tilde k x}}{\tilde k^2+m_{th}^2}&=\sum_{n\in \mathbb{Z}}(-1)^ne^{in\hat \mu}\int_{-\infty}^\infty d\omega\int\frac{d^{d-1}k}{(2\pi)^{d}}\frac{e^{i\omega(\tau+n)}e^{i \vec{k}\cdot \vec{x}}}{\omega^2+ \vec{k}^2+m_{th}^2}, \\
		&=\sum_{n\in \mathbb{Z}}\frac{(-1)^ne^{in \hat \mu}}{(2\pi)^{d/2}} \bigg(\frac{m_{\rm th}}{|x^{(n)}|}\bigg)^{\frac{d}{2}-1} K_{\frac{d}{2} -1}(m_{\rm th} |x^{(n)}|), 
	\end{align}
	where $x^{(n)} = ( \tau + n , \vec x) $. Performing the derivative in (\ref{g2express}) 
	and  choosing the coordinates as in (\ref{defwplane}), we obtain 
	\begin{align} \label{gng2}
		g_2(z, \bar z) 
		=\sum_{m\in \mathbb{Z}}(-1)^{m+1}e^{im \hat \mu}\big(\frac{m_{\rm th}}{2\pi} \big)^{\frac{d}{2}} 
		\bigg[-\frac{m}{2} \sqrt{\frac{z}{\bar z}} - \frac{m}{2} \sqrt{\frac{\bar z}{ z}}
		+\sqrt{z \bar z}\bigg] 
		\frac{K_{\frac{d}{2}}\left(m_{\rm th} \sqrt{(m-z)(m-\bar z)}\right)}{\big(\sqrt{(m-z)(m-\bar z)}\,\big)^{d/2}}. 
	\end{align}
	The evaluation of  one point functions by applying inversion formula on the above correlator 
	follows the same steps as the case of  zero chemical potential done in \cite{David:2023uya} except for the fact that one has to account for 
	the phase factor $ e^{im \hat \mu} $. The sum over $m$ can be re-organized by expanding the Bessel function using  (\ref{besli}). 
	This shifts the dependence of the chemical potential  in the one point functions into the 
	 the poly-logarithm functions that occur in the absence of chemical potential just as we have seen for the 
	 bosons in section \ref{unmodel}. 
	 In the end we obtain 
	\begin{align} \label{aO+ fer}
		a_{{\cal O}_+}[n=0, l] = &\frac{l  }{ 4 \pi ^k (  k - \frac{1}{2} )_l} \left( \frac{m_{\rm th}}{2} \right)^{l +k -1}\times\nonumber\\ 
		&\sum_{n=0}^{l +k -1} \frac{ (l +k -n)_{2n}}{ (2 m_{\rm th}) ^n n!} [{\rm Li}_{k+n} ( - e^{-m_{\rm th}+\mu} ) 
		+(-1)^l{\rm Li}_{k+n} ( - e^{-m_{\rm th}-\mu} ) ], \nonumber \\
		& l = 1,  2, \cdots.
	\end{align}
	For these values of $l$, there is no contribution from the contours at infinity as shown in \cite{David:2023uya}. 
	We have also replaced purely imaginary potential  $\hat \mu \rightarrow i \mu$. 
	The stress-tensor derived from the partition function is proportional to  the expectation value $a_{{\cal O}_+} [0, 2]$. 
	By comparing with the stress tensor  in (\ref{gnstressp}), this relation  is given by 
	\begin{equation}
	a_{{\cal O}_+ } [0,2]=\frac{1}{2^k k ( 2k -1)} T_{00}.
	\end{equation}
	The  factor $2^k$ occurs since we have  removed it in the construction of the two point function $g_2(x)$. 
	Similarly by comparing the expectation value of the spin one current  $a_{{\cal O}_+} [0, 2]$ with 
	(\ref{gnspin1}), we obtain the relation
	\begin{equation} \label{relu1}
	a_{{\cal O}_+}[0,1]=\frac{1}{2^k(2k-1)} J.
	\end{equation}
	
	The expectation value $a_{{\cal O}_0} [0, l]$     is contained in the OPE expansion 
	of the correlation $g_1(x)$. 
	Using the thermal two point function for the fermions in  (\ref{fermicorr}), we obtain 
	\begin{eqnarray}
	g_1(x) = \sum_{n \in \mathbb{Z}} \frac{(-1)^n e^{ i n \hat \mu} }{ ( 2\pi)^{\frac{d}{2}}} 
	m_{\rm th}^{\frac{d}{2}} | x^{(n) }|^{\frac{ 2 - d}{2}} K_{\frac{ d-2}{2}} \Big( m_{\rm th } |x|^{(n)} \Big) .
	\end{eqnarray}
	Applying the Euclidean inversion formula on this correlator  we obtain the following 1-point functions
	\begin{align} \label{gnao}
		a_{\mathcal{O}_0}[0,l]&= \frac{ m_{\rm th}^{l+k}}{2^{l+k} \pi^k (k+\frac{1}{2})_l } 
		\sum_{n=0}^{l+k-1}\frac{  (l+k-n)_{2 n} (\text{Li}_{k+n}\left(-e^{-m_{th}+\mu}\right)+(-1)^l \text{Li}_{k+n}\left(-e^{-m_{th}-\mu}\right))}{ 2 (2m_{\rm th})^{n} n! }, \nonumber\\ 
		& l = 1,2, \cdots.
	\end{align}
	Comparing this expectation value for $l=1$ with that of $\tilde J$ in (\ref{gnspin11}), we obtain the relation
\begin{eqnarray}\label{relspin11}
a_{{\cal O}_0}[0, 1] = - \frac{\tilde J}{ 2^k ( 2k +1) } .
\end{eqnarray}
	For these expectation values, the contour at infinity in the $w$-plane does not contribute. 
	For the expectation value $\langle \bar\psi \psi \rangle $ which  is  given by $ a_{\mathcal{O}_0}[0,0]$,  the 
	contour at infinity in the $w$-plane contributes. The total contribution from the discontinuity and the arcs at infinity is given by 
	\begin{eqnarray}
	a_{\mathcal{O}_0}[0,0] &=& - [ \hat a_{\rm disc} ( \Delta, 0)  + \hat a_{\rm arc} ( \Delta, 0 ) ]\Big|_{{ \rm Res\;at}\;\Delta=2k}.
	\end{eqnarray}
	The arc contribution can be evaluated using the same steps as in \cite{David:2023uya}. 
	Demanding that the expectation value of this operator vanishes results in 
	\begin{eqnarray} \label{gapinv}
	&& \frac{1}{\pi^k} \Big( \frac{m_{\rm th} }{2} \Big)^k 
	\sum_{n=0}^{k-1}
	\frac{ ( k -n)_{2n} }{ 2  (2m_{ th} )^n} 
	\Big[{\rm Li}_{k+n}( - e^{- m_{\rm th} -\mu })
	+ {\rm Li}_{k+n}( - e^{- m_{\rm th} +\mu }) \Big]  \\ \nonumber
	&& \qquad\qquad\qquad\qquad\qquad+ 
	\frac{m^{2k}}{2^{2k+1}  \pi^{k + \frac{1}{2} }} \Gamma( \frac{1}{2} - k \Big) =0.
	\end{eqnarray} 
	We see that this equation precisely agrees with the gap equation obtained from the partition function. 
	The agreement extends the observation seen earlier by \cite{David:2023uya} in the absence of chemical potential. 
	
	On comparing (\ref{aO+ fer}) and (\ref{gnao}), we see that the two classes of one point functions are related by the equation
	\begin{equation}\label{relcurrent}
	a_{{\cal O}_0}[0, l ] = \frac{ 2k- 1}{ l(2k  + 2 l - 1) }  m_{\rm th}  \, a_{{\cal O}_+}[0, l ] , 
	\qquad\qquad l = 1, 2, \cdots.
	\end{equation}
	Such relations  among the currents were noted in  \cite{David:2023uya} 
	\footnote{In \cite{David:2023uya}, the prefactor on the RHS of (\ref{relcurrent}) was incorrect. It did not contain the 
	ratio $(2k-1)/(l(2k+2l -1))$.  }
	and here this observation extends these 
	relations in the presence of chemical potential. Due to such relations 
	gap equation can also be obtained by 
	demanding the expectation value of the operator $\bar \psi \gamma^\mu D_\mu \psi$ which we call $a_{{\cal O}_{-}} [0,0] $ 
	since this operator is related to $\bar\psi\psi$ by the equation of motion. 
	
	At this point it is good to perform a simple consistency check of the relation in (\ref{relcurrent}) for $l=1$ found using the 
	Euclidean inversion formula  with the 
	relation found in (\ref{relspin1}) obtained  directly from the partition function. 
	Consider (\ref{relcurrent}) at $l=1$, we obtain 
	\begin{equation}
	a_{{\cal O}_0}[0, 1 ] = \frac{ 2 k- 1}{ 2 k + 1 }  m_{\rm th}  \, a_{{\cal O}_+}[0, 1 ].
	\end{equation}
	Now substituting the expression for both the left hand side and the right hand side of the above equation
	 in terms of the respective 
	currents derived from the partition function
	in  (\ref{relu1}) and (\ref{relspin11}) respectively we obtain 
	\begin{equation}
	\tilde J = -m_{\rm th} J, 
	\end{equation}
	which is the same as the one obtained from the partition function analysis in (\ref{relspin1}).
	It is also important to note that the spin one currents $J, \tilde J$ and the spin two current $T_{00}$ and the gap equation
	(\ref{appengapeqf}) 
	 were obtained
	from the partition function at the large $N$ saddle point. The partition function was evaluated using the untwisted field 
	$\tilde\psi$.   However the one point functions $a_{ {\cal O }_+}[0,l], a_{{\cal O}_0} [0, l]$ 
	and the gap equation (\ref{gapinv}) were obtained using 
	twisted fermions. The fact they agree and are consistent is an important check for the approach of using 
	the Euclidean inversion formula using twisted fermions.

	\subsection{The model in $d=3$ and  the Lee-Yang edge singularity}\label{gn3d}
	
	As mentioned earlier, the Gross-Neveu model in 3d is a toy example for asymptotic safety. 
	It has also been useful for modelling systems in condensed matter. 
	In this section we study  this model at finite real chemical potential in detail. 
	Let us begin with the partition function of the model at large $N$  can be obtained using the 
	Hubbard-Stratonovich transformation. This is done in appendix (\ref{A.2}) and 
	 can be read out from (\ref{partitionatm})
	 	\begin{align} \label{gn3dparti}
		\log Z(m_{\rm th})&=-\frac{1}{6 \pi  \beta ^2}\Big[\beta ^3 m_{\rm th}^3+3 \beta  m_{\rm th} \left(\text{Li}_2(-e^{-m_{\rm th}\beta+\mu\beta})+\text{Li}_2(-e^{-m_{\rm th}\beta-\mu\beta})\right)\nonumber\\
		&\qquad\qquad\qquad+3 \text{Li}_3(-e^{-m_{\rm th}\beta+\mu\beta})+3 \text{Li}_3(-e^{-m_{\rm th}\beta-\mu\beta})\Big].
	\end{align}
	The saddle point  which dominates at large $N$ and strong coupling can be obtained by minimising this partition function 
	with respect to the thermal mass $m_{\rm th}$, which results in the gap equation
		\begin{align} \label{gap3dgn}
	&	\frac{\partial}{\partial{m_{\rm th}}}\log Z(m_{\rm th})=0,\nonumber\\
	&		m_{\rm th} \big[m_{\rm th}\beta+\log(1+e^{-m_{\rm th}\beta+\mu\beta})+\log(1+e^{-m_{\rm th}\beta-\mu\beta})\big]=0.
	\end{align}
	The only solution of this equation for real thermal mass
	 is at  $m_{\rm th}=0$, this is the trivial or the Gaussian fixed point of the theory. 
	However let us analytical continue the gap equation (\ref{gap3dgn})  in $m_{\rm th}$ and look for solutions 
	in the complex $m_{\rm th}$ plane. 
	Then we see that the equation admits the following solutions
	\begin{align}\label{thermal mass 3d fer}
		m_{\rm th}\beta= \log \frac{1-2 \cosh\mu\beta\pm i \sqrt{(4 -(1-2\cosh\mu\beta)^2 }}{2}.
	\end{align}
	Is is clear from this solution that $m_{\rm th}$ is purely imaginary as long as 
	\begin{equation}\label{range}
	1\leq \cosh\mu\beta < \frac{3}{2}.
	\end{equation}
	When the chemical potential vanishes we have
	\begin{equation} \label{petkoufix}
	m_{\rm th} \beta = \pm \frac{2\pi i }{3} .
	\end{equation}
	This solution was first discussed in \cite{Petkou:2000xx}. In this work, it was observed that though the thermal mass 
	is purely imaginary, the stress tensor is real and therefore physical.  
	 It was argued  this is a new  fixed point of the Gross-Neveu model   and is 
	 in the Lee-Yang class of CFT's and therefore non-unitary.  The reason essentially is attributed to the 
	 thermal mass  being  un-physical. 
	  Recently in \cite{Diatlyk:2023msc},  the fixed point   (\ref{petkoufix}) has been re-visited and $1/N$ corrections to 
	 both the Free energy and high spin one point functions in the class ${\cal O}_+[0, l ]$ has been evaluated. 
	 In this sub-section we would like to investigate the line of fixed points of the 3d Gross-Neveu 
	  parameterized by real chemical potential 
	   $\mu$ in the range (\ref{range}). 
	   
\begin{figure}\begin{center}
			\includegraphics[scale=.5]{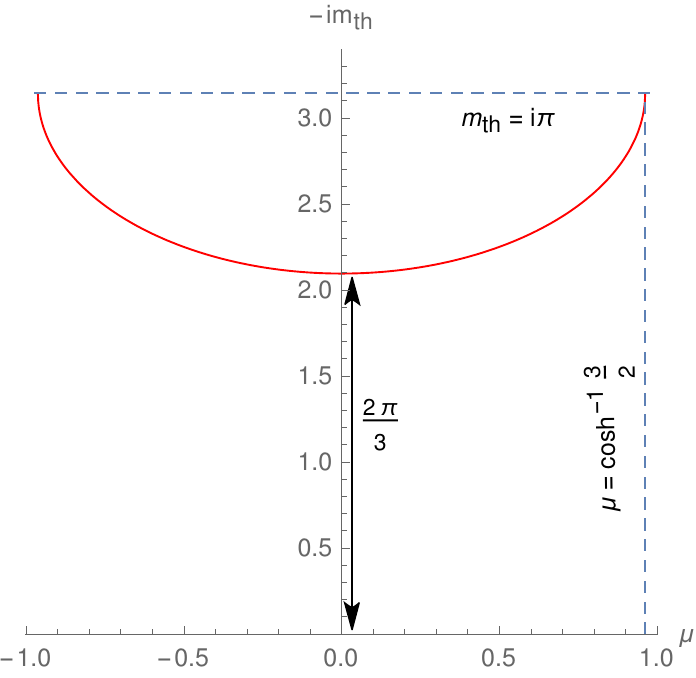}
			\caption{ The behaviour of the thermal mass $m_{\rm th}$ as a function of real chemical potential $\mu$ obtained from the solution of the gap equation \eqref{thermal mass 3d fer}.  $m_{\rm th}$ attains a finite value of $i\pi$ at the critical value of the chemical potential $\mu=\pm\cosh^{-1}\frac{3}{2}$, but the slope of the graph diverges at $\mu=\pm\cosh^{-1}\frac{3}{2}$.}\label{3dgnmth}
		\end{center}
	\end{figure}

	    As we have seen  the thermal mass  (\ref{thermal mass 3d fer}) is purely imaginary in this 
	  range of chemical potentials, let us restrict our attention to positive imaginary thermal mass 
	  \footnote{The theory at negative imaginary thermal mass is 
	   shown to be equivalent to the one at positive imaginary thermal mass appendix \ref{symmetry}.  }. 
	 The thermal mass at  small values of the chemical potential  $\mu$, is given by 
	  	\begin{align}
	\lim_{\mu \rightarrow 0}	m_{\rm th}\beta=\frac{2 i \pi }{3}+\frac{i (\mu\beta) ^2}{\sqrt{3}}+O((\mu\beta)^4)
	\end{align}
	We define the chemical potential at the end point of the range in (\ref{range}) by 
	\begin{equation}
	\mu_c \equiv \cosh^{-1} \Big( \frac{3}{2} \Big) , 
	\end{equation}
	where we take the positive root.  We will also choose  $\beta =1$ to de-clutter our expressions. 
	Then the expansion of the thermal mass  for $\mu>0$ close to $\mu_c$ is given by 
	\begin{align} \label{expmth3d1}
	\lim_{\mu\to\mu_c}	m_{\rm th} =i \pi-i5^{1/4} \sqrt{\mu_c-\mu}  +\frac{13 i }{24\times 5^{1/4}} \left(\mu_c-\mu\right)^{3/2}+O((\mu_c-\mu)^{5/2}).
	\end{align}
	Similarly the expansion for $\mu<0$ around  $-\mu_c$ is given by 
	\begin{align} \label{expmth3d2}
	\lim_{\mu\to\mu_c}	m_{\rm th} =i \pi-i5^{1/4} \sqrt{\mu_c+\mu}  +\frac{13 i }{24\times 5^{1/4}} \left(\mu_c+\mu\right)^{3/2}+O((\mu_c+\mu)^{5/2}).
	\end{align}
	The figure \ref{3dgnmth} shows the behaviour of the thermal mass as the chemical potential varies 
	 in the window (\ref{range}).  
	
	Form the expansion at $\mu_c$, we see that the thermal mass admits a branch cut in the $\mu$ plane.
	The point $\mu_c$ is therefore the branch point. It is also a special point from the following consideration. 
	The partition function in (\ref{gn3dparti})  is a function  $m_{\rm th}$ and $\mu$, the condition that $m_{\rm th}$ 
	is a extremum is given by
	\begin{equation}\label{1stder}
	\frac{\partial}{\partial{m_{\rm th}}}\log Z(m_{\rm th})=0.
	\end{equation}
	which results in the gap equation given (\ref{gap3dgn}). 
	If this point is also an inflexion point, then we must have 
	\begin{equation}\label{2ndder}
	\frac{\partial^2}{\partial{m_{\rm th}}^2}\log Z(m_{\rm th})=0.
	\end{equation}
	Conditions such as  (\ref{1stder}) and (\ref{2ndder}) together are usually used to identify second order phase transitions. 
	In \cite{Basar:2021hdf} these conditions  have been  used to identify Lee-Yang edge singularity in the Gross-Neveu model in (1+1)-dimensions.
	Let us examine  the condition (\ref{2ndder})
			\begin{align}
	&\frac{\partial^2}{\partial{m_{\rm th}}^2}\log Z(m_{\rm th})\nonumber\\
	&=	-\frac{  m_{\rm th} e^{  \mu } \left(e^{2   m_{\rm th}}-1\right)}{2 \pi  \left(e^{ \mu }+e^{ m_{\rm th}}\right) 
	\left(e^ {(\mu +m_{\rm th})}+1\right)}-\frac{\log \left(e^{  (\mu -m_{\rm th})}+1\right)+\log \left(e^{-  (\mu +m_{\rm th})}+1\right)+
	m_{\rm th}}{2 \pi }.
	\end{align}
	The second term in the above equation vanishes once the gap equation holds. 
	Therefore to satisfy (\ref{1stder}) and (\ref{2ndder}) simultaneously we must have 
	\begin{equation}
	m_{\rm th}  = i\pi, 
	\end{equation}
	in the principal branch. 
	From (\ref{2ndder}), we see that this occurs at $\mu=\mu_c$.  Since the thermal mass is imaginary  we are  therefore 
	in an un-physical  regime of parameter space, we are led to identify the point $(m_{\rm th} =i\pi, \mu=\mu_c)$ 
	as a Yang-Lee edge singularity.

	One test of a Yang-Lee edge singularity is to examine the behaviour of the pressure
	in the chemical potential plane. The pressure  usually 
	 around the Lee-Yang edge  singularity exhibits a branch cut, the critical exponent at this branch point is called $\phi$ and has been evaluated for various theories. 
	The general expansion of the pressure  in terms of the fugacity about a Yang-Lee edge singularity is of the form
	 \begin{eqnarray}
	 P (z)& =& p_0 + p_1 ( z-z_c) + p_2(z-z)^2 + \cdots , \\ \nonumber
&&  + \hat p( z -z_c)^{\phi} \Big[ 1+ p_\theta ( z- z_c)^\theta  + \cdots \Big] + \cdots .
	 \end{eqnarray}
	 See  \cite{10.1063/1.470178} for more details and a  list of references,  this expansion can be equivalently written in terms of the chemical potential by substituting 
	 $z= e^{\mu} $.  Here $z_c$ is the Yang-Lee edge singularity and $\phi$ is the relevant critical exponent. 
	 In general the exponents $\phi,  \theta$ need not be integers. 
	 Usually the fugacity is in some un-physical domain, say negative. In our situation this is not the case, but 
	 the theory is still in some un-physical domain due to the imaginary thermal mass. 
	We can perform such an  expansion for the 3d Gross-Neveu model and compare the exponent 
	with known results. 
	The pressure is proportional to the free energy  which in turn is proportional to the stress tensor
	\begin{equation}
	F= -\frac{\log Z}{ \beta N V} = -P = \frac{T_{00}}{2} .
	\end{equation}
	Due to these relations it is sufficient to study the stress tensor which can be read out from \eqref{T00 f}
		\begin{align} \label{3dstressgnanl}
		T_{00}=&\frac{1}{3\pi }\Big[-m_{\rm th}^2  \left(\log (1+e^{- m_{\rm th}- \mu})+\log (1+e^{-m_{\rm th}+\mu})\right)\\
		&+3   m_{\rm th} \left(\text{Li}_2(-e^{- m_{\rm th}- \mu})+\text{Li}_2(-e^{-  m_{\rm th}+ \mu})\right)+3 \text{Li}_3(-e^{-m_{\rm th}- \mu})+3 \text{Li}_3(-e^{-  m_{\rm th}+ \mu})\Big]. \nonumber
	\end{align}
	where we have set $\beta =1$ and $\hat \mu  = i \mu $ and analytically 
	 continued to real values of the chemical potential. 
	We first expand the stress tensor at small values of  values of the chemical potential 
		\begin{align}\label{3dstresszer}
			\lim_{\mu \rightarrow 0} T_{00}=-\Big(\frac{4}{3}{\rm Cl}_2(e^{\frac{i\pi}{3}})-\frac{2\zeta(3)}{3\pi}\Big)-\frac{\mu ^2}{\sqrt{3}}+O(\mu^4).
		\end{align}
		where, ${\rm Cl}_2(x)={\rm Im}[{\rm Li}_2(e^{i x})] $ for $x\in (0,2\pi).$ 
		The leading term precisely agrees with the expression for the stress tensor obtained 
		for the fixed point $( m_{\rm th} =  \frac{2\pi i }{3}, \mu =0) $
		 of the 3d  Gross-Neveu model first obtained in \cite{Petkou:2000xx}. 
		Now let us examine the expansion around $\mu_c$. 
	\begin{align} \label{3dstressmc}
		\lim_{\mu \rightarrow \mu_c}
		T_{00}=-\Big(\frac{4}{5} \pi  \cosh ^{-1}\frac{3}{2}-\frac{8 \zeta (3)}{5 \pi }\Big)+\frac{4\pi}{5} (\mu_c-\mu)
		-\frac{2\times5^{3/4}}{3} {(\mu_c-\mu)^{3/2}} + \cdots.
	\end{align}
	Observe that the exponent of the branch cut originating at $\mu_c$ is $\frac{3}{2}$. 
	At this point, we mention the equation (\ref{3dstressgnanl}) may appear to be analytic in the chemical potential, 
	however it is important to remember that the gap equation (\ref{thermal mass 3d fer}) determines the 
	thermal mass $m_{\rm th}$ at the critical point. 
	The gap equation depends non-analytically in the chemical potential, in fact 
	from (\ref{expmth3d1}), 
	 it is easy to see that there is a branch point at $\mu_c$.  This is the origin of the non-analytic behaviour of the 
	stress tensor.  As we will see subsequently, the same holds true for the higher spin currents.

\begin{figure}\begin{center}
		\includegraphics[scale=.54]{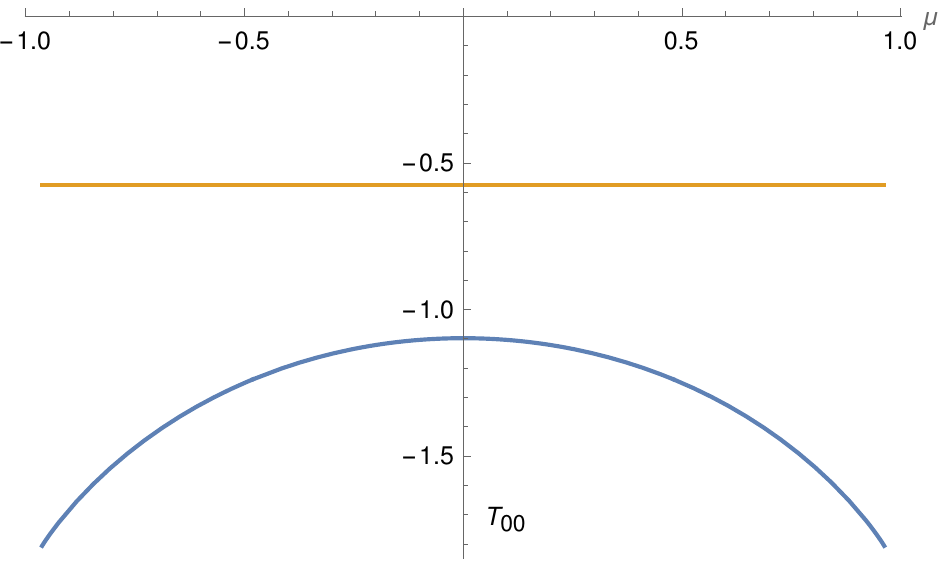}
		\caption{ The blue curve is the 
		stress tensor $T_{00}$(in units of $\beta$) for the critical theory plotted as a function of $\mu$ in the domain $|\mu|<\mu_c$.  The orange straight line refers to the value of the stress tensor for free massless fermions at zero chemical potential.  }\label{3dgnstressp}
	\end{center}
\end{figure}

	The exponent  $\phi = \frac{3}{2}$ 
	 characterises the Yang-Lee edge singularity. 
	It has been evaluated for various theories, the exponent $3/2$ coincides with the exponent
	 seen in mean field theory or for gases in  $d\geq 6$  with a repulsive core interaction \cite{10.1063/1.470178}
	It is also the exponent for the mean field description of the Ising model.  
	There is a similar expansion for $\mu<0$.  
	The behaviour of the stress tensor as the chemical potential is varied in the range (\ref{range}) is shown 
	in figure \ref{3dgnstressp}. The stress tensor monotonically decreases and reaches a finite value at $\mu_c$. 
	The figure also shows the value of the stress tensor for the free theory,  that is the fixed point
	with $m_{\rm th}=0$, which is always greater than the  non-trivial fixed point.  
	
	From the expansions in (\ref{3dstresszer}) and  (\ref{3dstressmc}) we see 
	 that the stress tensor is real in spite of the fact that $m_{\rm th}$ is purely imaginary. 
	This is because the stress tensor is an even function of $m_{\rm th}$, 
	This was noted in \cite{Petkou:2000xx}. In the appendix \ref{symmetry}
	 we have discussed this symmetry in detail and generalise this observation to the situation when the 
	 chemical potential is non-zero. 
	 We show that the expectation value of all operators in the class ${\cal O}_{+}[0, l]$ are real because of this symmetry 
	 of the one point functions in the complex $m_{\rm th}$ plane. 
	 This relies on non-trivial identities involving sums of Bernoulli polynomials  which is proved in the appendix \ref{symmetry}. 
	 We also show this symmetry persists  for Gross-Neveu models in arbitrary odd dimensions.

	The exponent $\phi =\frac{3}{2}$ in the expansion around $\mu_c$ is perhaps unique to the pressure or the stress tensor. 
	To see this we study the expansion of the currents ${\cal O}_{+}[0, l ]$ for spins $l = 1, 2,3, 4$ around $\mu_c$
	The spin-2 case of course is proportional to the stress tensor. 
	{\footnotesize
	\begin{flalign} \nonumber
		\lim_{\mu \rightarrow \mu_c} 
		a_{{\cal O}_+}[0,1]=\frac{1}{2}\Big[ \frac{2 \pi }{5}-\frac{1}{2} 5^{3/4} \sqrt{\mu_c-\mu}+O(\mu_c-\mu)\Big], &&
	\end{flalign}
		\begin{flalign} \nonumber
		\lim_{\mu\rightarrow \mu_c} 
		a_{{\cal O}_+}[0,2]=\frac{1}{3}\Big[-\Big(\frac{6\pi\cosh ^{-1}\left(\frac{3}{2}\right)}{5}   -\frac{12 \zeta (3)}{5 \pi }\Big)
		-\frac{6\pi}{5}(\mu-\mu_c)-5^{3/4} \left(\mu_c-\mu \right)^{3/2}+O\big((\mu_c-\mu)^2\big)\Big], &&
	\end{flalign}
	\begin{flalign}
	\lim_{\mu \rightarrow \mu_c} 
	a_{\mathcal{O}_+}[0,3]=	\frac{1}{10}\bigg[\frac{5 \left(48 \text{Li}_4\left(\frac{3}{2}-\frac{\sqrt{5}}{2}\right)+\log ^4\left(\frac{2}{\sqrt{5}+3}\right)\right)}{16 \pi }-\frac{\pi ^3}{15}+10\pi   \text{csch}^{-1}(2)^2 && \nonumber\\
\qquad\qquad	+\left(\frac{5^{3/4}}{2}-\frac{6}{5^{3/4}}\right) \pi ^2 \sqrt{\mu_c-\mu}+O(\mu_c-\mu)\bigg],&& \nonumber
	\end{flalign}
		\begin{flalign}
		\lim_{\mu\rightarrow \mu_c} 
		a_{\mathcal{O}_+}[0,4]=\frac{2}{105}\Bigg\{\Big[\frac{105 \text{Li}_5\left(\frac{1}{2} \left(3-\sqrt{5}\right)\right)}{\pi }-36 \pi  \zeta (3)-\frac{7 \log ^5\left(\frac{\sqrt{5}}{2}+\frac{3}{2}\right)}{16 \pi }+\frac{2\pi ^3\sinh ^{-1}\frac{11}{2}}{3}  && \nonumber\\
		-\frac{140}{3} \pi  \text{csch}^{-1}(2)^3\Big]-2\times 5^{1/4} \left(\pi ^2 \cosh ^{-1}(9)-6 \zeta (3)\right)\sqrt{\mu_c-\mu}+O(\mu_c-\mu)\Bigg\}.&& 
	\end{flalign}
	}
	It is interesting to note that  it is  only   the stress tensor whose branch cut has the exponent $3/2$. 
		All other one point functions till spin-4, 
	 have branch cut with $1/2$ as the exponent, it will be interesting to show that this is indeed the 
	case for all the other spins. 
	
	For completeness, we also provide the expansions of the  one point functions for small $\mu$. 
	{\footnotesize
	\begin{flalign} \nonumber
		\lim_{\mu\rightarrow 0} 
		a_{{\cal O}_+} [0,1]=\frac{\mu }{2 \sqrt{3}}+\frac{\left(3+2 \sqrt{3} \pi \right) \mu ^3}{36 \pi }+O\left(\mu ^5\right),  &&
	\end{flalign}
	\begin{flalign} \nonumber
		\lim_{\mu\rightarrow 0} 
		a_{{\cal O}_+} [0,2]=\Big(\frac{\zeta (3)-  2\pi \Im[\text{Li}_2(e^{\frac{i \pi }{3}})]}{3 \pi }\Big)-\frac{\mu ^2}{2 \sqrt{3}}-\frac{\left(3+2 \sqrt{3} \pi \right) \mu ^4}{72 \pi }+O\left(\mu ^5\right),  &&
	\end{flalign}
	\begin{flalign}\nonumber
		\lim_{\mu\rightarrow 0} 
		a_{{\cal O}_+} [0,3]=\mu  \Big(\Im\big[\text{Li}_2\big(e^{\frac{i \pi }{3}}\big)\big]-\frac{\zeta (3)}{2 \pi }-\frac{2 \pi ^2}{45 \sqrt{3}}\Big)+\Big(\frac{1}{405} \pi  (9-2 \sqrt{3} \pi )+\frac{1}{4 \sqrt{3}}\Big) \mu ^3+O\left(\mu ^5\right), &&
u	\end{flalign}
		\begin{flalign} \nonumber
			\lim_{\mu\rightarrow 0} 
		a_{{\cal O}_+} [0,4]=\frac{1}{567} \left(32 \pi ^2  \Im[\text{Li}_2(e^{\frac{i \pi }{3}})]-756 \Im[\text{Li}_4(e^{\frac{i \pi }{3}})]-72 \pi  \zeta (3)+\frac{525 \zeta (5)}{\pi }\right) &&\\ 
		+\mu^2\Big(\frac{(-24 \pi ^2 \sqrt{3}-378 \pi ) \Im\left(\text{Li}_2(e^{\frac{i \pi }{3}} ) \right) }{567 \pi }+\frac{\left(12 \pi  \sqrt{3}+189\right) \zeta (3)}{567 \pi }+\frac{8 \pi ^2}{189 \sqrt{3}}\Big). &&
	\end{flalign}
	}
	
	These expansions explicitly  show  that the  expectation values of the currents $ {\cal O}_+[0, l]$ for purely imaginary thermal mass $m_{\rm th}$. This is because these expectation values  are even functions of  $m_{th}$ as shown in the appendix \ref{symmetry}. 
	However, the expectation values of the currents 
	${\cal O}_0[0, l ]$ are purely imaginary. 
	From (\ref{relcurrent}) for $d=3, k=1$ we have the relation 
	\begin{eqnarray}
	a_{{\cal O}_0}[0, l] = \frac{1}{2l} m_{\rm th} a_{{\cal O}_+}[0, l].
	\end{eqnarray}
	This implies that since $ a_{{\cal O}_+}[0, l]$  is real and $m_{\rm th}$ is purely imaginary $a_{{\cal O}_0}[0, l] $ is imaginary. 
	The fact that one point functions of the operator 
	${\cal O}_0[0, l]$ are imaginary is not consistent with the hermiticity property of these operators. 
	Let us be more specific by examining the $l=1$ operator whose expectation value has been evaluated
	directly from in the partition function in (\ref{a_O0}). 
	In the Hamiltonian picture, this  the expectation value is given by 
	\begin{eqnarray}
	\tilde J =  \frac{i}{2} \Big\langle   \bar{\tilde \psi }\big(  \partial_\nu   - i \mu \delta_{\nu 0} \big)  \tilde  \psi  - 
  \big( \partial_\nu  + i \mu \delta_{\nu 0}  \big)  \bar{\tilde \psi}  \; \tilde \psi  \Big\rangle_{(\beta, \mu)},
	\end{eqnarray}
	here $\bar{\tilde \psi} \psi  = \tilde \psi^\dagger \gamma^0$ 
	with $\gamma^0$ Hermitian.   We have  used that it is only the time component that gets non-trivial expectation value. 
	Note that the operator is Hermitian, and therefore the expectation value should be real. 
	This is similar to the more familiar case of the $U(1)$ current whose expectation value in the Hamiltonian picture 
	is given by 
	\begin{eqnarray}
	J   = \langle \bar{\tilde \psi }\gamma^\mu  \tilde \psi\rangle_{\beta, \mu}.
	\end{eqnarray}
	This too is Hermitian and therefore the $U(1)$ charge is real. 
	That fact that both $J$ and $\tilde J$ are real is consistent with the relation  \ref{realrel}
	\footnote{This relation does not rely on the gap equation and also holds 
	for the free theory with massive fermions. This  is clear from its derivation in appendix \ref{A.2}. }
	\begin{eqnarray}
	\tilde J = -m_{\rm th} J,
	\end{eqnarray}
	for real masses as it should be, since there is nothing pathological for a theory of fermions with real mass. 
	However when $m_{\rm th}$ is purely imaginary, this relation contradicts the Hermiticity property of $\tilde J$ and therefore 
	we are in an un-physical domain of the theory. 
	The fact that all the currents $a_{{\cal O}_0}[0, l]$ are purely imaginary is therefore a reflection that theory is in an 
	un-physical domain and is consistent with the fact that the point $\mu=\mu_c$ is a Lee-Yang edge singularity.

	\subsection{The large spin limit} \label{gnlargel}
	
\begin{figure}[t]
		\begin{subfigure}{.475\linewidth}
			\includegraphics[width=.9\linewidth]{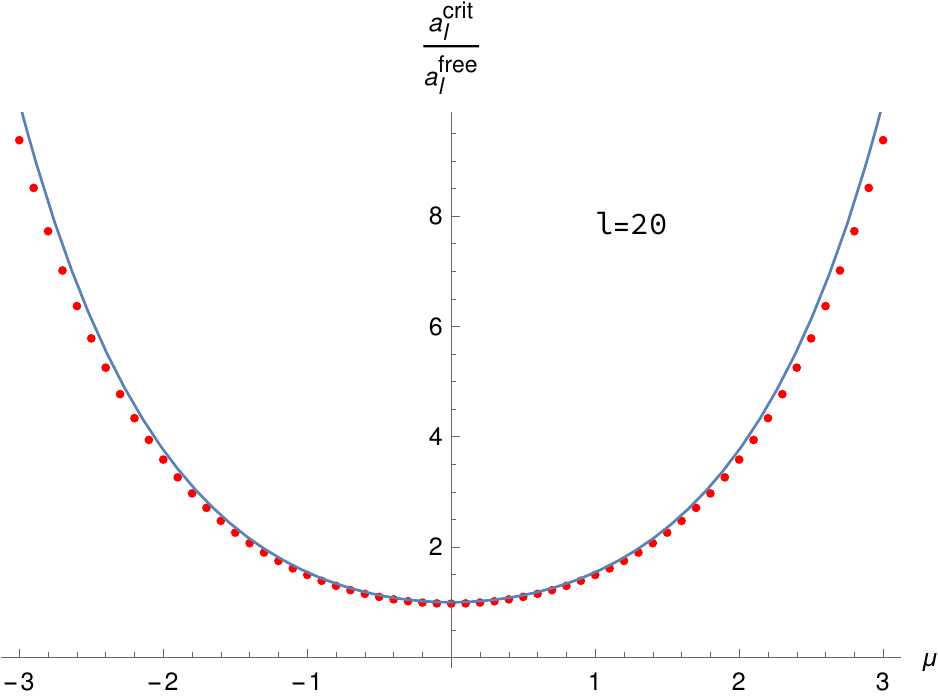}
			\caption{}
		\end{subfigure}\hfill 
		\begin{subfigure}{.475\linewidth}
			\includegraphics[width=.9\linewidth]{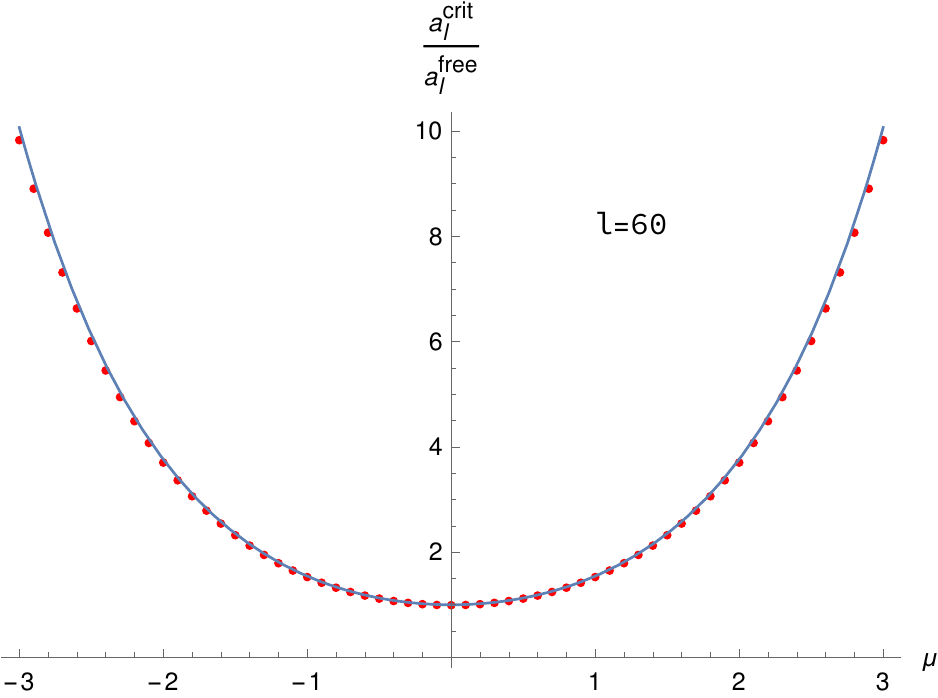}
			\caption{}
		\end{subfigure}
		\par\bigskip
		\par\bigskip
		\par\bigskip
		\begin{subfigure}{.475\linewidth}
			\includegraphics[width=.9\linewidth]{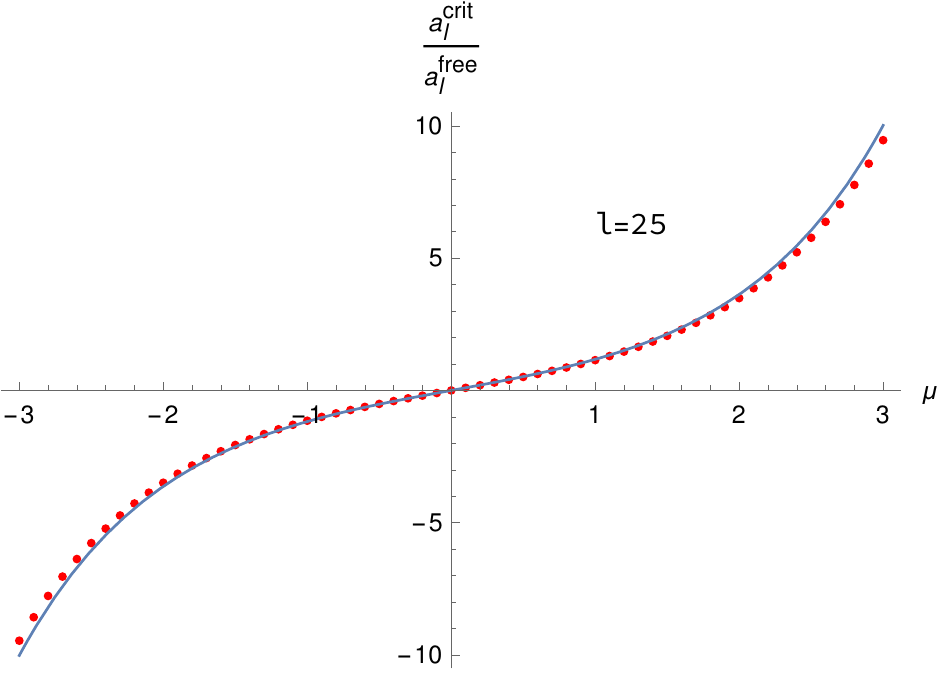}
			\caption{}
		\end{subfigure}\hfill 
		\begin{subfigure}{.475\linewidth}
			\includegraphics[width=.9\linewidth]{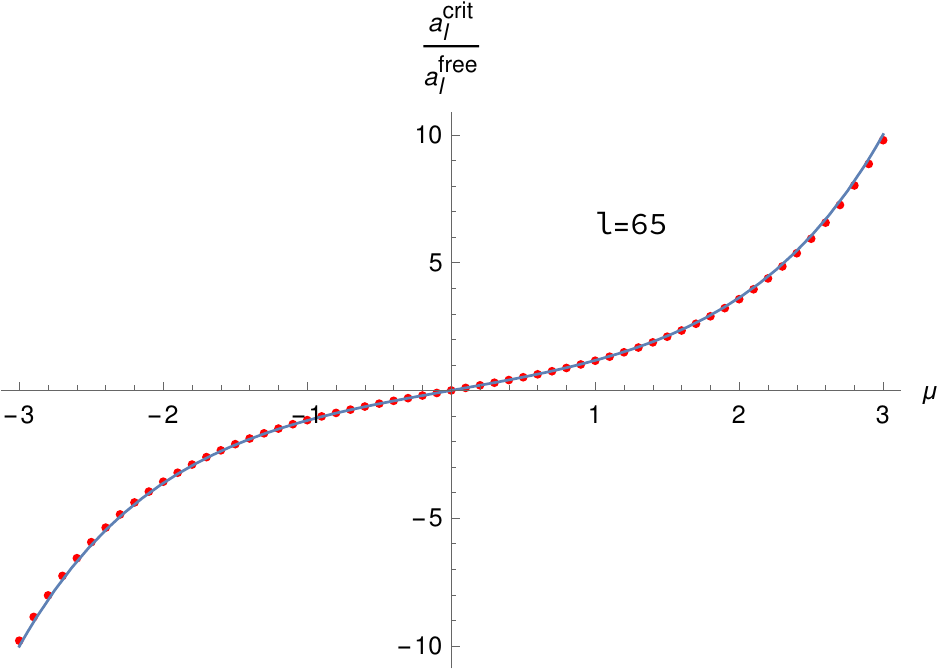}
			\caption{}
		\end{subfigure}
		
		\caption{This plot shows the agreement of the ratio 
		$\frac{a_{\mathcal{O}_+} [0, l] }{{a}_{\mathcal{O}_+}^{\rm free} [0, l ]}$ 
		by numerically evaluating the thermal mass $m_{\rm th}$ 
		  with the large $l$ asymptotic formula \eqref{cosh fer} for $k=2.$ The red dots refer to the numerically evaluated values and the solid blue curves are the plots of $\sinh\mu$ and $\cosh\mu$ for $l$ being odd and even respectively. $a_l^{\rm crit}$ and $a_l^{\rm free}$ denote the 1-point function for critical and free theory respectively in the figure.}
		\label{sinh}
	\end{figure}

	In this section we study the large spin limit of the one point functions for the critical 
	 Gross-Neveu model in $d=2k+1$ dimensions. 
	 Just as in the case of the bosonic model studied in section \ref{largel}, 
	 to obtain  the large spin limit 
	 it is easier to deal with the expressions of the 1-point functions in terms of the modified Bessel functions of 2nd kind.
	 This is given by 
	\begin{align}\label{1 pt bessel fer}
		a_{{\cal O}_+}[n=0, l] = \sum_{m=1}^\infty (e^{-im\mu}+(-1)^le^{im\hat \mu})
		\frac{  (-1)^{m}m^{\frac{1}{2} -k} 
			\pi^{\frac{1}{2}-k }K_l  m_{\rm th}^{l + k -\frac{1}{2}} }{ 2^{l + k -\frac{3}{2}} \Gamma(l)} K_{l + k-\frac{1}{2} } ( m m_{\rm th}).
	\end{align}
	Again we use the following asymptotic expression of Bessel functions of 2nd kind at large orders, but fixed argument 
	\begin{align}
		\lim_{l\rightarrow \infty}
		K_{k+l-\frac{1}{2}}(m m_{\rm th}) = \sqrt{\frac{\pi}{2(k+l-\frac{1}{2})}}\bigg(\frac{e m m_{\rm th}}{2(k+l-\frac{1}{2})}\bigg)^{-(k+l-\frac{1}{2})}.
	\end{align}
	Substituting the above expression 
	 into \eqref{1 pt bessel fer} and using the limit
	 \begin{eqnarray}
	  \lim_{l \rightarrow\infty}
	    {\rm Li}_{2k+l-\frac{1}{2}}(-e^{-i\hat \mu}) =  -e^{-i\hat \mu} .
	    \end{eqnarray}
	 we obtain
	\begin{align}\label{larglgn}
		\lim_{l \rightarrow \infty} a_{\mathcal{O}_+}[n=0,l]=-\frac{l\Gamma(k-\frac{1}{2})}{4\pi^{k+\frac{1}{2}}}(e^{-i\hat \mu}+(-1)^le^{i\hat \mu}).
	\end{align}
	Observe that the dependence on the thermal mass $m_{\rm th}$ drops out. 
	The pre-factor that occurs in the above expression can be identified with the one point function of the Gaussian theory 
	at large $l$ which is given by 
	\begin{equation}\label{free fer}
	a_{{\cal O}_+}^{\rm free}[0, l ]  =  l ( 2^{2-2k-l} -1)  \frac{\Gamma( k - \frac{1}{2})}{4 \pi^{ k +\frac{1}{2}  }}
	\zeta( l + 2k -1) ( 1+ (-1)^l ) .
	\end{equation}
	This result can be obtained by taking the $m_{\rm th}=0, \mu =0$ limit in the expression (\ref{aO+ fer}) for the one point functions. 
	Taking the large $l$ limit we obtain
	\begin{eqnarray} \label{gngaussian}
	\lim_{l\rightarrow \infty} a_{{\cal O}_+}^{\rm free}[0, l ]  &=& -\frac{l \Gamma( k - \frac{1}{2}) }{2\pi^{k  + \frac{1}{2} }}, 
	\qquad l \in 2 \, \mathbb{Z}, \\ \nonumber
	&=& a_{{\cal O}_+}^{\rm free}[0, \infty ].
	\end{eqnarray}
	Using this definition we can write the large spin limit of the one point functions  at the non-trivial fixed point 
	given in (\ref{larglgn}) as 
	\begin{eqnarray} \label{cosh fer}
		\lim_{l \rightarrow \infty} a_{ {\cal O}_+ } [n=0,l] &=&  a_{{\cal O}_+}^{\rm free}[0, \infty ] \cosh \mu ,
		 \qquad l
		 \in 2\mathbb{Z}, \\  \nonumber \quad  
		&=&  a_{{\cal O}_+}^{\rm free}[0, \infty ] \sinh \mu , \qquad l \in 2 \mathbb{Z}+1.
	\end{eqnarray}
where $\mu$ refers to the real chemical potential. 	
	This simplification obtained at large $l$ has been tested against the numerical values for the 
	one point functions  in figure \ref{sinh}.

	\subsection{The large  $d$ limit}  \label{gnlarged}
	
		To obtain the large $d$ limit for the one point functions of the Gross-Neveu model we follow 
		the same manipulations carried out  for the bosons in section \ref{largedun}. 
		The gap equation for the Gross-Neveu model admits a real solution for the thermal mass in $d=2k+1$ dimensions with 
		$k$ even.  In \cite{David:2023uya} these mass were evaluated till $k=40$.  Then the ratio of the one point functions of the 
		operator ${\cal O}_+[0, l]$  of the critical model to the Gaussian fixed  point  was evaluated. It was shown numerically 
		this ratio tends to zero on increasing the dimensions. 
		In this section we obtain the solution for the gap equation analytically for large $k$ and use it to show that 
		the ratio
		\begin{align}\label{r_f}
		r_f(l,k)=\frac{a_{\mathcal{O}_+}^{\rm crit}[0,l]}{a_{\mathcal{O}_+}^{\rm free}[0,l]}.
	\end{align}
		indeed vanishes even in the presence of chemical potential.  The one point function for the Gaussian model 
		is given in (\ref{free fer}). 
		
		The gap equation for the Gross-Neveu model is given by 
	\begin{eqnarray} \label{gngap1}
		 ( 2m_{\rm th})^k \sum_{n=0}^{k-1} \frac{ ( k-n)_{2n} }{ ( 2m_{\rm th})^n n! }  \Big[
		{\rm Li}_{k+n} ( - e^{-m_{\rm th}+ \mu })     + {\rm Li}_{k+n} ( - e^{-m_{\rm th}- \mu })   \Big]
		+ \frac{(m_{\rm th}) ^{2k}  \sqrt{\pi}}{\Gamma( \frac{1}{2} + k) \cos\pi k} =0. \nonumber \\
	\end{eqnarray}
	At large $k$ we use the limit $\lim_{k\rightarrow \infty} {\rm Li}_{k+n}(x)= x$, thus the above gap equation becomes,
	\begin{align}
		\frac{(m_{\rm th}) ^{k-\frac{1}{2}}  {\pi}}{\Gamma( \frac{1}{2} + k) \cos\pi k}-{2^{k+\frac{3}{2}} K_{k-\frac{1}{2}}(m_{\rm th})}\cosh (\mu )=0.
	\end{align}
	Here we have  used the representation of the Bessel function to perform the sum. 
	Once $k$ is set to even integers, this 
	  equation is identical to the equation for the bosonic model in (\ref{gapeqchempot}).  For the bosonic model $k$ was odd. 
	 Therefore the 
	large $k$ asymptotic solution to the  gap equation of the Gross-Neveu for is identical to the  bosonic model 
	but   with $k$   being an even integer 
	\begin{align}\label{mth k fer}
		\lim_{k\rightarrow \infty} 
		m_{\rm th}=0.69486k+0.313322 \log (\cosh (\mu ))+0.0560686, \qquad {\rm for}\ k\in2\mathbb{Z}.
	\end{align}
	
\begin{figure}\begin{center}
			\includegraphics[scale=.58]{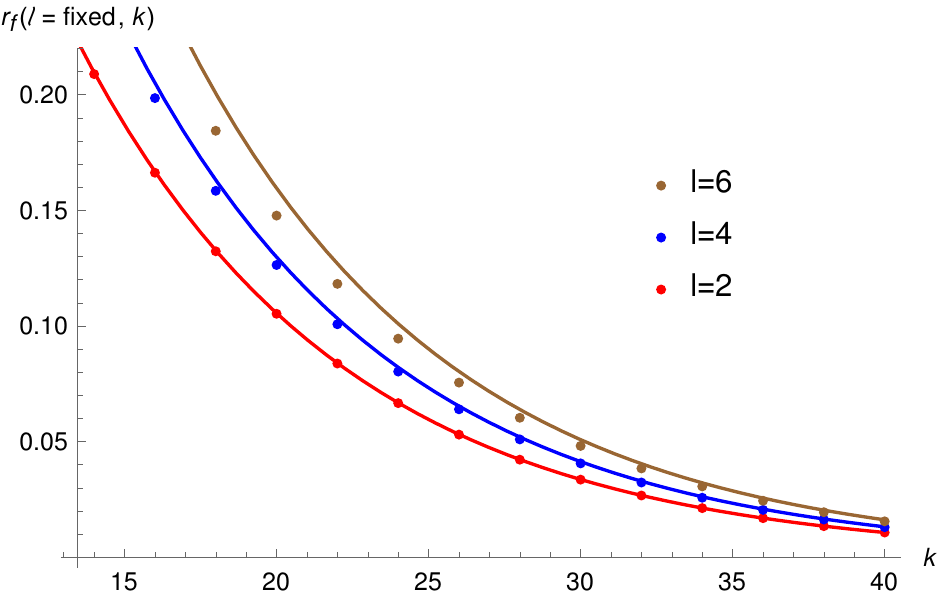}
			\caption{ For $\mu=0$, $r_f(l,k)$ is evaluated by solving the gap equation numerically and plotted against $k$ for various fixed values of $l$; the red, blue and brown dots represent the numerical values of $r_f$ for $ l=2,4,6 $ respectively.  The red, blue and brown solid curves in the figure refer to the plot of large $k$ the asymptotic formula of $r_f$ \eqref{gnrfk} for $ l=2,4 $ and $6$ respectively. The asymptotic formula works well for $l\ll k$.
}\label{large k fer}
		\end{center}
	\end{figure}

	Similarly we can approximate the 1-point function at large $k$ as,
	\begin{align}
		\lim_{k\rightarrow\infty}
		a_{\mathcal{O}_+}[0,l]=-\frac{ l   m_{\rm th}^{k+l-\frac{1}{2}} K_{k+l-\frac{1}{2}}(m_{\rm th})}{\pi^{k+\frac{1}{2}}2^{k+l+\frac{1}{2}}(k-\frac{1}{2})_l}(e^{\mu}+(-1)^l e^{- \mu }).
	\end{align}
	Again substituting the large $k$ asymptotic solution for the thermal mass we can evaluate the 
	ratio (\ref{r_f}), which results in 
	\begin{eqnarray} \nonumber \label{gnrfk} 
		\lim_{k\rightarrow\infty}r_f(l={\rm fixed},k)&=&1.53742 \exp\Big\{-0.114385 k+0.10333 l-0.0981709 \log (\cosh (\mu ))
		\\ && \qquad-0.597829+O(k^{-1})\Big\}
		\times \Big( \frac{e^{\mu}+(-1)^le^{-\mu} }{2} \Big) \Big(1+O(k^{-1}) \Big).  \nonumber \\ 
	\end{eqnarray}
	In figure \ref{large k fer}, we have compared the asymptotic expression for the ratio $r_f (l, k )$ against
	the value obtained by solving the gap equation numerically. 
	We see that the asymptotic expression indeed is a good approximation to the numerics, furthermore from (\ref{gnrfk}) we see that  the ratio tends to zero at large $k$ exponentially. 

\begin{figure}[t]
		
		\begin{subfigure}{.475\linewidth}
			\includegraphics[width=1\linewidth]{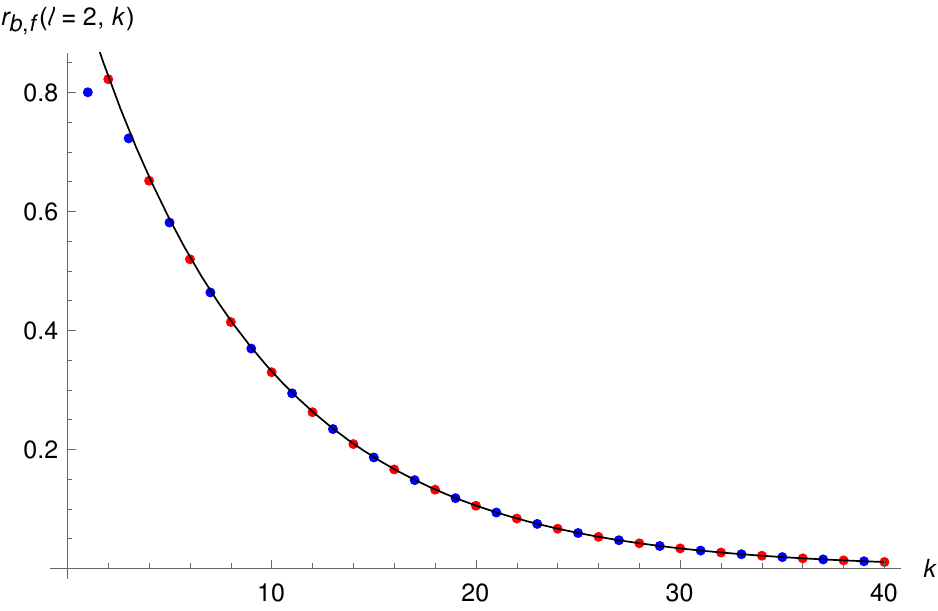}
			\caption{}
		\end{subfigure}\hfill 
		\begin{subfigure}{.475\linewidth}
			\includegraphics[width=1\linewidth]{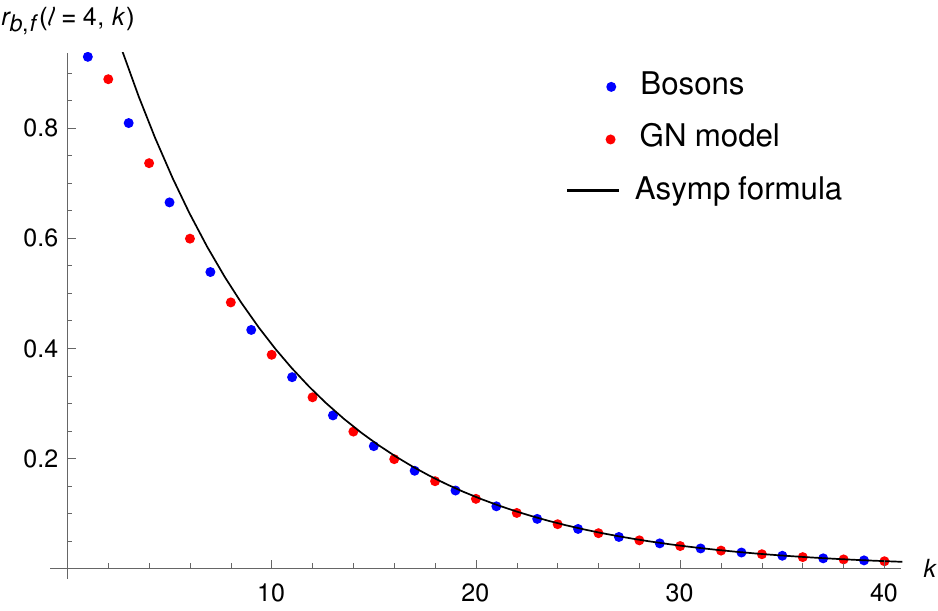}
			\caption{}
		\end{subfigure}
		\par\bigskip
		\begin{subfigure}{.475\linewidth}
			\includegraphics[width=1\linewidth]{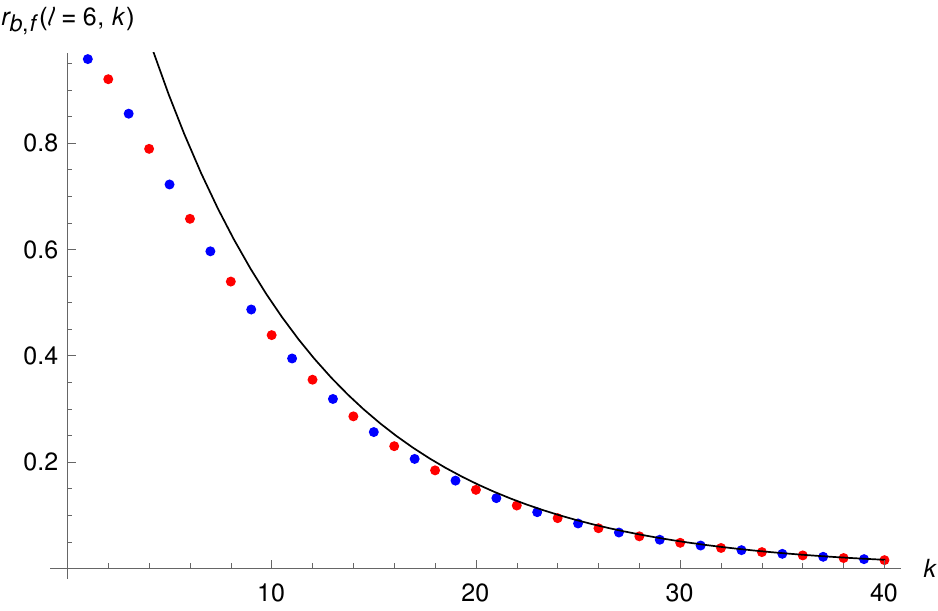}
			\caption{}
		\end{subfigure}\hfill 
		\begin{subfigure}{.475\linewidth}
			\includegraphics[width=1\linewidth]{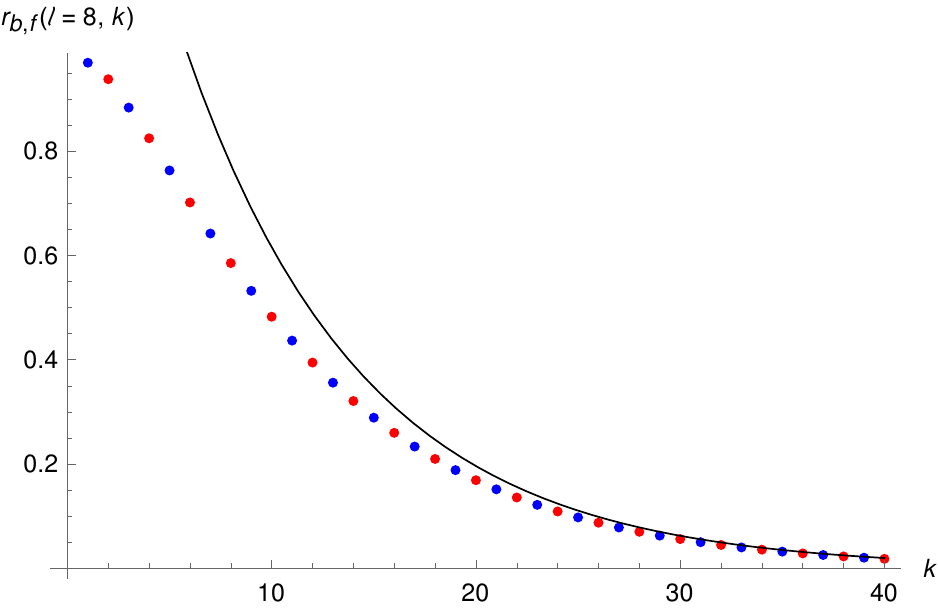}
			\caption{}
		\end{subfigure}
		
		\caption{We plot thermal one point functions at zero chemical potential for model of  bosons in the fundamental of $U(N)$
		and the  Gross-Neveu  model in the same graph as $k$ increases for different fixed values of $l.$ 
		The blue and red points refer to the bosonic  and fermionic model respectively.  The solid black curve represents the large $k$ asymptotic formula of the thermal one point function given in \eqref{exp} or \eqref{gnrfk}. 
			The asymptotic formula agrees well with the numerics when $l$ is comparably small with respect to $k$.
			The figures suggest that there is a unique analytical function in $k$  or dimensions for the one point function of large 
			$N$ vector models.  }
		\label{fig:sinh1}
	\end{figure}

In figure \ref{fig:sinh1} we have studied the numerical results for both the ratio $r_b(l, k ), r_f(l , k )$ against the large $k$ asymptotic 
formula (\ref{exp}), (\ref{gnrfk}), which is identical for both bosons and fermions. 
The figure suggests that  the one point functions of both the bosonic and fermionic models can be thought of a unique 
analytical function  in $k$. 
	
	\section{Conclusions} \label{conclude}
	
	In this paper we have generalised the use of the  Euclidean inversion formula to large $N$ vector models  at finite 
	values of chemical potentials. 
	We studied the resulting one point functions  of bi-linear in fields at the non-trivial fixed point of these models 
	at both large spins and large dimensions and 
showed that they simplify.  At large spins they are proportional to the one point functions of the free theory, while at large 
dimensions they vanish in comparison with the free theory. 
We studied both the bosonic and fermionic model  in detail for the special case of $d=3$ space-time dimensions. 
For the Gross-Neveu model in $d=3$ we studied a fixed point which is argued to exhibit  the Yang-Lee edge singularity. 
The exponent  of the branch cut  of the pressure at the singularity in the chemical potential plane is found to be 
$\frac{3}{2}$ which coincides with mean field theory description of  systems which have repulsive core interactions. 
As a byproduct of our investigations, we have found non-trivial identities satisfied by certain sums of Bernoulli polynomials.

There has been several interesting works which have developed the program of constraining both one 
point functions and two point functions in thermal CFT's using symmetries \cite{Iliesiu:2018zlz, Gobeil:2018fzy,Benjamin:2023qsc,Luo:2022tqy,Karlsson:2022osn,Marchetto:2023fcw,Marchetto:2023xap}.
It will be interesting to see how the results obtained for CFT's from vector models fit these general observations. 
One specific direction is to study the perturbative expansion of thermal one point functions for general 
CFT's at large spin obtained in \cite{Iliesiu:2018fao} to see if it can be generalised to situations when the chemical potential 
is turned on. 
A more ambitious program would be to see if a solution dual to a thermal state can be constructed in 
Vasiliev theories and these simplifications and large spin can be seen in the dual descriptions. 

Another direction is to develop the inversion formula for vector models on geometries $S^1\times S^{d-1}$ or $S^1\times AdS_{d-1}$
and connect with the general results of the ambient space formalism developed in \cite{Parisini:2022wkb,Parisini:2023nbd}. 
Finally, it is important to study how the results of this paper generalise to large $N$ Chern-Simmons theories with matter. 
The recent development in this subject  \cite{Minwalla:2023esg}
 of evaluating the partition function in temporal gauge should be useful 
in obtaining the Euclidean inversion formula in these theories. 

The results from the analysis at large dimensions suggest that there exists   a unique function of dimension  that describes 
one point functions of large $N$ vector models both bosonic and fermionic. 
It will be interesting to obtain this for arbitrary odd dimensions, rather than just 
in the asymptotic limit.

\acknowledgments

J.R.D thanks the organisers of ``Aspects of CFTs'' , January 8-11, 2024, IIT Kanpur, India, for  hospitality and  the opportunity 
to present  some of the preliminary results in this paper. S.K thanks the organisers of ``The 18th Kavli Asian Winter School on Strings, Particles and Cosmology'', December 5-14, 2023,
YITP, Kyoto University, Japan and ``Non-perturbative methods in Quantum Field Theory and String Theory'', Jan 29-Feb 2, 2024, HRI, Prayagraj, India for the warm hospitality and giving the opportunity for attending encouraging sessions on related topics.
	
	\appendix

	\section{
		Gap equation and stress tensor from the partition function}
	\label{gap part}
	In this appendix we discuss the derivation of the gap equation as a saddle point equation of the partition function at large $N$ for both the case of complex scalars and Gross-Neveu model at finite chemical potential. The partition function can be computed by linearising the theory at  the leading order in $N$. We also compute the stress tensor and the spin-1 conserved currents from the partition function in this appendix.
	
	\subsection{Complex Scalars at finite density} \label{unbfd}
	
	Consider the action for $ N$ massless complex scalar fields with imaginary chemical potential given by,
	\begin{align}
		S=\int d^{2 k+1} x \bigg[|(\partial_\tau-i\hat \mu){\tilde \phi}|^2+\partial_i \tilde\phi \partial_i \tilde\phi^*+\frac{\lambda}{N}(|{\tilde\phi}|)^4\bigg].
	\end{align}
	where ${\tilde\phi}$ is $N$ dimensional vector of complex scalars with $\tilde\phi^*\tilde\phi=\sum_{a=1}^N \tilde\phi^*_a\tilde\phi_a$ and $i=1,2,\cdots,2k$ denoting the spatial directions.
	At finite temperature the partition function can be obtained by the following path integral over $N$ scalar fields with the imaginary time direction being compactified on a circle of length $\beta.$
 	\begin{equation}
		\tilde Z= \int {\cal D}\tilde \phi^* {\cal D} \tilde\phi e^{ - S(\tilde\phi,\tilde\phi^*) }.
	\end{equation}
	Using the Hubbard-Stratonovich transformation, one obtains a quadratic action in $ \phi$ by introducing an auxiliary field $\zeta$,
	\begin{eqnarray}
		\tilde Z &=&   \int {\cal D}\tilde\phi^* {\cal D} \tilde\phi \exp \left[ { - \int_0^\beta \int d\tau d^{2k} x 
			\Big(|(\partial_\tau-i\hat \mu){\tilde \phi}|^2+|\partial_{\vec{x}}{\tilde\phi}|^2+\frac{\lambda}{N}(|{\tilde\phi}|)^4\Big) } \right], \\ \nonumber
		&=&  \int {\cal D}\tilde\phi^*{\cal D} \tilde\phi {\cal D} \zeta
		\left[  - \int_0^\beta \int  d\tau d^{2k}x  \Big(
		|(\partial_\tau-i\hat \mu){\tilde\phi}|^2+|\partial_{\vec{x}}{\tilde\phi}|^2  + \frac{\zeta^2 N}{4\lambda}  + i \zeta |{\tilde\phi}|^2
		\Big)  \right].
	\end{eqnarray}
	By isolating the zero mode $\zeta_0$ from the auxiliary field $\zeta$,
	\begin{eqnarray}
		\zeta = \tilde \zeta + \zeta_0 .
	\end{eqnarray}
	$\tilde{\zeta}$ is the non-zero mode, we substitute this into the partition function to obtain,
	\begin{eqnarray}
		&&\tilde Z = \int d\zeta_0  {\cal D} {\tilde\phi^*} {\cal D}\tilde\phi \left[ 
		\exp\Big(  -\frac{\zeta_0^2 N \beta V_{2k}}{4\lambda} 
		\Big)  \exp ( -S_0 - S_I) \right] ,\qquad{\rm where,} \\ \nonumber
		&&S_0 = \int_0^\beta \int  d\tau d^{2k}x  \Big(
		|(\partial_\tau-i\hat \mu){\tilde\phi}|^2+|\partial_{\vec{x}}{\tilde\phi}|^2  + \frac{\zeta^2 N}{\lambda}  + i \zeta_0 |{\tilde\phi}|^2
		\Big),  
		\quad S_I =  \int_0^\beta \int d\tau d^{2k} x \;  i \tilde\zeta  |{\tilde\phi}|^2 .
	\end{eqnarray}
	Now, at the leading order in large $N$ we can ignore $S_I$ as it is easy to show that the perturbative corrections due to this term appear in the order $\frac{1}{\sqrt{N}}$ when the non-zero mode of the auxiliary field is integrated out.
	Thus the non-zero mode of $\zeta$ is present as an Gausssian path integral in the leading order at large $N$ and this integral is easily performed resulting in an ordinary integral on the zero mode,
	\begin{eqnarray} \label{Z tilda}
		\tilde Z &=&  \int d\zeta_0  \exp\left[  -\beta V_{2k} N \Big( \frac{ \zeta_0^2}{4\lambda}  - \frac{1}{\beta} \log Z ( i\zeta_0)
		\Big)
		\right],
	\end{eqnarray}
	where,
	\begin{eqnarray}
		\log Z ( i\zeta_0) =-   \sum_{n =-\infty}^\infty \int \frac{d^{2k} p}{(2\pi)^{2k}}  \log
		\left[ \frac{  (2\pi n-\hat \mu\beta)^2 }{\beta^2}  + \vec p^2 + i\zeta_0 \right].
	\end{eqnarray}
	Now we can evaluate the partition function by looking for the saddle point of the above integral in $\zeta_0.$ If the saddle point value of $\zeta_0$ is $\zeta_0^*$, the thermal mass can be expressed as,
	\begin{align}
		m_{\rm th}^2=i\zeta_0^*.
	\end{align}
	The saddle point equation of the integral \eqref{Z tilda} is given by,
	\begin{equation}\label{saddle}
		m_{\rm th}^3  =- \lambda \frac{1}{\beta} \frac{\partial}{\partial m_{\rm th} } \log Z(m_{\rm th}) .
	\end{equation}
	After doing the matsubara sum, $Z(m_{\rm th}) $ is evaluated as follows ,
	\begin{align}\label{Z}
		\log {Z}(m_{\rm th})&=-\frac{1}{\beta^{d-1}}\int \frac{d^{d-1}k}{(2\pi)^{d-1}}\Big[\sqrt{\vec{k}^2+m_{\rm th}^2\beta^2}+2 {\rm Re}\Big(\log\big[1-e^{-\sqrt{\vec k^2+m_{\rm th}^2\beta^2}+i\beta\hat \mu}\big]\Big)\Big],\nonumber\\
		&=\beta   \frac{m_{\rm th} ^{2 k+1} \Gamma \left(-k-\frac{1}{2}\right)}{2^{2k+1}\pi^{k+\frac{1}{2}}}+\sum_{n=0}^k\frac{(k+1-n)_{2 n} (m_{\rm th}  \beta )^{k-n} {\rm Re}[\text{Li}_{n+k+1}\left(e^{-\beta  (m_{\rm th} +i \hat \mu )}\right)]}{n!\pi^k 2^{n+k-1}\beta^{2k}}.
	\end{align}
	The integral of the first term from the parenthesis is evaluated by analytically continuing the following standard integral,
	\begin{align} \label{analyticcont}
		\int_0^\infty  dx\; \frac{ x^{2k -1} }{ ( x^2+1)^a }  =  \frac{ \Gamma( a - k)  \Gamma( k ) }{ 2\Gamma (a) } .
	\end{align}
	Integral of the 2nd term is convergent and one of the standard approaches is to use the infinite series expansion of $ \log(1+x) $ for small $ x $ and each term is integrated in term of BesselK functions of half integer order and finally one can resum the infinite number of terms using the fact that BesselK of half integer order can be represented by finite order polynomial. And it can be expressed in terms of the Polylogarithm functions as given in \eqref{Z}.\\
	Using \eqref{Z} and \eqref{saddle}, the gap equation at the leading order in large $ \lambda $ is computed to be,
	\begin{align} \label{appengapeq}
		\frac{(m_{\rm th}\beta) ^{k} \Gamma \left(\frac{1}{2}-k\right)}{\sqrt{\pi }}+\sum_{n=0}^{k-1}\frac{ (k-n)_{2n}  {\rm Re}\left[\text{Li}_{n+k}\left(e^{-\beta  (m_{\rm th} +i \hat \mu )}\right)\right]}{(m_{\rm th}\beta)^{n}2^{n-k-1}\Gamma (n+1) }=0.
	\end{align}
	This model behaves as a CFT at the critical value of the thermal mass $m_{\rm th}$ which satisfies the above equation. The stress tensor at the critical point can be evaluated from the partition function by,
	\begin{equation}
		T_{00} =\frac{1}{NV_{2k}} \frac{ \partial}{\partial \beta } \log \tilde Z( m_{\rm th})\bigg|_{\hat \mu\beta={\rm fixed}} .
	\end{equation}
	Note that in presence of the chemical potential the stress tensor can be evaluated by differentiating the partition function keeping $\hat \mu\beta=$fixed, this definition is consistent with the more general definition of the stress tensor by metric variation of the action. Finally the stress tensor can be given by,
	\begin{align}
		T_{00}&= \frac{m_{\rm th} ^{2 k+1} \Gamma \left(-k-\frac{1}{2}\right)}{2^{2k+1}\pi^{k+\frac{1}{2}}}-
		\sum_{n=0}^{k+1}\frac{m_{\rm th}    \left((k+n)^2+k-n\right) (k-n+2)_{2n-2}  {\rm Re}\left[\text{Li}_{n+k}\left(e^{-m_{\rm th}  \beta +i \hat \mu \beta}\right)\right]}{n!\pi^k 2^{n+k-1}\beta^{2k}(m_{\rm th}\beta)^{n-k}}.
	\end{align}
	The stress tensor at the critical point or the conformal fixed point characterised by the thermal mass can be further simplified on the use of the gap equation \eqref{appengapeq} to have,
	\begin{align}\label{T00}
		T_{00}=&-\frac{1}{\beta^{2k+1}\pi^k}\Big(\frac{k}{2k+1}\Big)\sum_{n=0}^{k+1}\frac{ (k-n+2)_{2n}}{ n!2^{n+k-2} } (m_{\rm th}\beta)^{k+1-n}  {\rm Re}[\text{Li}_{n+k}(e^{-m_{\rm th}  \beta +i \hat \mu \beta})].
	\end{align}
	As $ (m_{\rm th}\beta) $ is a dimensionless number, the conformal stress tensor scales with the powers of temperature which is the only scale present in the critical theory.
	Also the use of gap equation in the partition function \eqref{Z} gives the free energy density for the critical theory to be,
	\begin{align}\label{F}
		F=-\frac{1}{\beta NV_{2k}}\log \tilde{Z}=-\frac{1}{\beta^{2k+1}\pi^k}\sum_{n=0}^{k+1}\frac{( m_{\rm th}\beta)^{k-n+1} (k-n+2)_{2 n}}{2^{n+k-1}(2 k+1) n!}{\rm Re}[\text{Li}_{n+k}(e^{-m_{\rm th}  \beta +i \hat \mu \beta})].
	\end{align}
	And from equations \eqref{T00} and \eqref{F}, it is easy to observe the signature relationship between the stress tensor and the free energy density of a conformal field theory at finite temperature is being satisfied,
	\begin{align} \label{stressfreerel}
		T_{00}=(d-1) F.
	\end{align}
		By putting $l=2$ in the equation \eqref{a l d}, we see that the thermal one point function of the conformal spin-2 current evaluated using the OPE inversion formula agrees with the stress tensor $T_{00}$ up to an overall constant factor given by,
	\begin{align}
		a_\mathcal{O}[0,2]=-\frac{1}{4k(2k-1)} T_{00}.
	\end{align}
	The spin-1 conserved current can also be obtained form the partition function by,
	\begin{align} \label{apenchargeb}
		J&=\frac{i}{\beta}\frac{\partial}{\partial{\hat\mu}} \log  Z(m_{\rm th})\\
		&=-\sum_{n=0}^k \frac{(k-n+1)_{2 n} (m_{\rm th}  \beta )^{k-n} \Big[\text{Li}_{k+n}\left(e^{-\beta  m_{\rm th} -i \beta\hat\mu }\right)-\text{Li}_{k+n}\left(e^{-\beta  m_{\rm th} +i \beta\hat\mu }\right)\Big]}{\pi^k\beta^{2k}2^{k+n}n!}.
	\end{align}
	Now, the thermal expectation of conformal spin-1 current is obtained by putting $l=1$ in the OPE inversion formula answer \eqref{a l d}, and this also agrees with $J$ up to a constant factor given as follows,
	\begin{align}
		a_\mathcal{O}[0,1]=-\frac{1}{2(2k-1)} J.
	\end{align}

	\subsubsection{Evaluation of the partition function using a cut off} \label{appendixa2}

The evaluation of the  partition in (\ref{Z}), involved using the analytical continuation of the integral in (\ref{analyticcont}). 
This approach was introduced in \cite{Giombi:2019upv}, but it is illustrative to relate it with the  approach of introducing 
a cutoff scale or using a  dimensional regulator which is found in the literature \cite{Hands:1992ck,Filothodoros:2016txa,Filothodoros:2018pdj,Romatschke:2019ybu}.  
For this purpose it is sufficient to 
restrict our discussions to $d=3, 4$ which are  well studied in the literature  and to bring out the differences 
between even and odd dimension. 

\subsubsection*{ $d=3$}

Consider the integral in (\ref{Z}), with $d=3, k =1$, it is the first term that needs a regulator. 
Let us perform the integral using the momentum cut off $\Lambda$, the first term in (\ref{Z}) which needs a regulator.
This  can be found by using the result
\begin{align} \label{3dintegralcut}
	\lim_{\Lambda\to\infty}\int_0^{\Lambda } k^{} \sqrt{k^2+m_{\rm th}^2} \, dk=\frac{\Lambda ^3}{3}+\frac{\Lambda  m_{\rm th}^2}{2}-\frac{m_{\rm th}^3}{3} + O\Big( \frac{1}{\Lambda} \Big).
\end{align}
Substituting this in (\ref{Z}), we obtain 
\begin{eqnarray}
	\lim_{\Lambda\rightarrow \infty}
	\log Z(m_{\rm th})&=&-\frac{1}{2\pi}\Bigg[ \beta  \left(\frac{\Lambda ^3}{3}+\frac{\Lambda  m_{\rm th}^2}{2}-\frac{m_{\rm th}^3}{3}\right) \\ \nonumber
	&& -2{\rm Re}\frac{ \left(\beta  m_{\rm th} \text{Li}_2\left(e^{-m_{\rm th} \beta+i\mu\beta  }\right)+\text{Li}_3\left(e^{-m_{\rm th} \beta +i\mu\beta}\right)\right)}{\beta ^2}\Bigg]  +  O\Big( \frac{1}{\Lambda} \Big). \\ \nonumber
\end{eqnarray}
Note that the second integral in  (\ref{Z}) is convergent as the cut off is taken to infinity, which results the expression involving 
poly-logarithims. 
Let us also obtain the derivative of this partition function, 
\begin{align}
	\partial_{m_{\rm th}} \log Z=-\frac{1}{2\pi}\Big(\beta  \left(\Lambda  m_{\rm th}-m_{\rm th}^2\right)-2m_{\rm th}{\rm Re}  \log (1-e^{-\beta  m_{\rm th}+i\mu\beta})\Big).
\end{align}
We can now substitute this result in the gap equation (\ref{saddle})  to obtain 
\begin{eqnarray}\label{gapphy}
\frac{m_{\rm th}^2}{\lambda} - \frac{\Lambda}{2\pi} = -\frac{m_{\rm th}}{2\pi}  -\frac{ 1}{\pi \beta} {\rm Re}  \log (1-e^{-\beta  m_{\rm th}+i\mu\beta}).
\end{eqnarray}
Let us now define  the renormalized coupling $\lambda_R$ in the following manner
\begin{align}\label{renor coup}
	 \frac{1}{\lambda}-\frac{\Lambda}{2\pi m_{\rm th}^2}=\frac{1}{\lambda_R}.
\end{align}
And the gap equation \eqref{gapphy} at zero temperature relates the renormalized coupling $\lambda_R$ to the mass of theory at zero temperature $m_{(0)}$ which represents a physical length scale in the theory, given as
\begin{align}\label{m_0}
	\lambda_R=-2\pi m_{(0)}.
\end{align}
Now using \eqref{renor coup} and \eqref{m_0} we can rewrite the gap equation \eqref{gapphy} for the thermal mass $m_{\rm th}$ in terms of the physical mass scale $m_{(0)}$ and the inverse temperature $\beta$ as
\begin{align}
	\frac{m_{\rm th}^2}{2\pi m_0}=\frac{m_{\rm th}}{2\pi}  +\frac{ 1}{\pi \beta} {\rm Re}  \log (1-e^{-\beta  m_{\rm th}+i\mu\beta}).
\end{align}
By solving this gap equation we see that the dimensionless combination $m_{\rm th}\beta$ depends on another scale
$m_{(0)}$ and therefore not a CFT.  
When the value of this scale is set to a critical value such that
$m_{(0)}\beta\to \infty$ the above equation  coincides with the gap equation in (\ref{gapeq3db}) obtained using 
the analytical continuation of the integral (\ref{analyticcont}) and setting the coupling to infinity. 
This results in the thermal mass given by (\ref{solgap3d}). 
Therefore we conclude that the approach developed by \cite{Giombi:2019upv} in evaluating the partition function in odd dimensions 
is designed to study the theory at the fixed point or the critical theory. 
It is interesting to observe that demanding the vanishing of the expectation value of $|\phi|^2$ obtained using the inversion 
formula also results in the gap equation for the critical theory. 
We have checked that this reasoning works for other odd dimensions and also for the Gross-Neveu model in odd dimensions.

\subsubsection*{ $d=4$}

It is interesting to perform the analysis  in even dimensions to demonstrate the
 qualitative difference between odd and even dimensions.  
 The relevant divergent integral in the partition function   
 given in (\ref{Z}). with $d=4, k = 3/2$ can be regulated using a momentum cut off which results in 
 \begin{align}
	\lim_{\Lambda\to\infty}\int_0^{\Lambda } k^{2} \sqrt{k^2+m_{\rm th}^2} \, dk=\frac{\Lambda ^4}{4}+\frac{\Lambda ^2 m_{\rm th}^2}{4}-\frac{1}{32} m_{\rm th}^4 \left(\log \frac{16 \Lambda ^4}{m_{\rm th}^4}-1\right) + O\Big( \frac{1}{\Lambda}  \Big) .
\end{align}
The crucial difference we note from (\ref{3dintegralcut}) is the presence of the logarithm which contains the cut off. 
This feature persists in all even dimensions. 
Substituting the above result in (\ref{Z}) and performing the second integral, we obtain
\begin{align}
\lim_{\Lambda\to \infty}	\log Z(m_{\rm th})=-\frac{1}{2\pi^2 }\Bigg(\beta  \left[\frac{\Lambda ^4}{4}+\frac{\Lambda ^2 m_{\rm th}^2}{4}-\frac{1}{32} m_{\rm th}^4 \left(\log \frac{16 \Lambda ^4}{m_{\rm th}^4}-1\right)\right]\nonumber\\
-2{\rm Re}\sum_{n=0}^\infty\frac{e^{i\mu\beta n} m_{\rm th}^2 K_2(m_{\rm th} n \beta )}{\beta  n^2}\Bigg) .
\end{align}
Proceeding to evaluate the derivative of the partition function we obtain 
\begin{align}
	\partial_{m_{\rm th}} \log Z=-\frac{1}{2\pi^2}\Bigg[\frac{1}{8} \beta  m_{\rm th} \left(4 \Lambda ^2-m_{\rm th}^2 \log \frac{16 \Lambda ^4}{m_{\rm th}^4}+2
	m_{\rm th}^2\right)+2{\rm Re}\sum_{n=1}^\infty\frac{e^{i\mu\beta n} m_{\rm th}^2 K_1(m_{\rm th} n \beta )}{n}\Bigg] .
\end{align}
Finally  substituting this result in the gap equation (\ref{saddle}),  we obtain
\begin{eqnarray}\label{gap4d}
m_{\rm th} \left( \frac{1}{\lambda} - \frac{\Lambda^2}{4\pi^2 m_{\rm th}^2} \right) 
= -\frac{m_{\rm th} }{8\pi^2} \log\Big( \frac{ 4\Lambda^2}{m_{\rm th}^2 e} \Big) +
\frac{ 1 }{\pi^2 \beta} {\rm Re}\sum_{n=1}^\infty\frac{e^{i\mu\beta n}  K_1(m_{\rm th} n \beta )}{n} .
\end{eqnarray}
We now define  renormalized coupling
\begin{eqnarray}
\frac{1}{\lambda_R(\Lambda) } =  \frac{1}{\lambda }  -   \frac{\Lambda^2}{4\pi^2 m_{\rm th}^2}  .
\end{eqnarray}
From the gap equation (\ref{gap4d}) at zero temperature, we get the relation 
\begin{eqnarray}\label{rgflow}
\frac{1}{\lambda_R(\Lambda) } =   - \frac{1 }{8\pi^2} \log\Big( \frac{ 4\Lambda^2}{m_{(0)}^2 e} \Big) ,
\end{eqnarray}
where $m_{(0)}$ is a physical length scale. 
Replacing the expression on the RHS of (\ref{gap4d}) by the renormalized coupling, we obtain 
\begin{eqnarray}
m_{\rm th} \left( \frac{8 \pi^2 }{\lambda_R(\Lambda) }  +   \log\Big( \frac{ 4\Lambda^2}{m_{\rm th}^2 e} \Big) \right) 
=  \frac{4  }{\beta } {\rm Re}\sum_{n=0}^\infty\frac{e^{i\mu\beta n}  K_1(m_{\rm th} n \beta )}{n}.
\end{eqnarray}
Observe that the cut off dependence on cancels using the relation (\ref{rgflow}) and we obtain the gap equation which depends on  the physical length $m_0$. 
\begin{eqnarray}
m_{\rm th}  \log\Big( \frac{ m_{(0)}}{m_{\rm th} } \Big) 
=  \frac{2  }{\beta } {\rm Re}\sum_{n=0}^\infty\frac{e^{i\mu\beta n}  K_1(m_{\rm th} n \beta )}{n} .
\end{eqnarray}
Similar equations have been obtained using  dimensional regularization in \cite{Romatschke:2019gck}. 
Due to the logarithmic dependence on the physical length scale $m_0$ the thermal mass $m_{\rm th}$ is always a function of this physical length scale. In fact as seen in \cite{Romatschke:2019gck}, 
the trace of the stress tensor for this theory does not 
vanish and therefore the theory is not a thermal CFT.
The prescription of the analytical continuation given in (\ref{analyticcont}) cannot be carried out to even dimensions directly since the 
Gamma function has a pole.

	\subsection{Gross-Neveu model with chemical potential}\label{A.2}
	We repeat the same large $N$ analysis for $U(N)$ Gross-Neveu model with four fermion interaction at finite temperature in presence of  an imaginary chemical potential characterised by $\hat\mu$. We start with the following Euclidean action with $N$ massless Dirac fermions $\psi_a$'s where $a=1,\cdots,N$. 
	\begin{equation}\label{action f}
		S= \int d^{2k +1} x \left[ 
		 {\tilde\psi}_a^\dagger (  \gamma^\mu \partial_\mu ) \tilde\psi_a -i\hat\mu {\tilde\psi}_a^\dagger \gamma^0 \tilde\psi_a- \frac{\lambda}{N} (  {\tilde\psi}_a^\dagger \tilde\psi_a)^2 \right].
	\end{equation}
	The interaction strength is controlled by the coupling constant $\lambda$.
	In Euclidean signature we have $\bar {\tilde\psi}=\tilde\psi^\dagger $ and the gamma matrices satisfy the following anti-commutation relation,
	\begin{equation}
		\{\gamma^\mu , \gamma^\nu \} = 2\delta^{\mu\nu}.
	\end{equation}
	Again the partition function can be expressed as a path integral of a quadratic action in $\psi$ by introducing an auxiliary field $\zeta$ applying the Hubbard-Stratonovich transformation,
	\begin{eqnarray}\label{path int}
		\tilde Z &=&   \int {\cal D}{\tilde\psi}^\dagger {\cal D} \tilde\psi \exp \left[ { - \int_0^\beta  d\tau d^{2k} x 
			\Big(
			 {\tilde\psi}^\dagger_a  \gamma^\mu \partial_\mu  \tilde\psi_a -i\hat\mu {\tilde\psi}^\dagger_a \gamma^0 \tilde\psi_a - \frac{\lambda}{N} ({\tilde \psi}^\dagger_a \tilde\psi_a)^2 \Big) } \right], \nonumber\\ 
		&=&  \int {\cal D}{\tilde\psi}^\dagger {\cal D} \tilde\psi {\cal D} \zeta
		\left[  - \int_0^\beta  d\tau d^{2k}x  \Big(
		 {\tilde\psi}^\dagger_a  \gamma^\mu \partial_\mu   \tilde\psi_a -i\hat\mu {\tilde\psi}^\dagger_a \gamma^0 \tilde\psi_a + \frac{\zeta^2 N}{4\lambda}  + \zeta ( {\tilde\psi}^\dagger_a \tilde\psi_a)
		\Big)  \right].\nonumber\\
	\end{eqnarray}
	Again path integral over all fermionic fields is denoted above.
	Normalisation constant of the Gaussian path integral is absorbed in the measure. Identical to the previous case we separate out the zero mode of the auxiliary field as $\zeta_0$ and call the rest to be $\tilde \zeta$,
	\begin{eqnarray}
		\zeta = \tilde \zeta + \zeta_0 .
	\end{eqnarray}
	Plugging this definition into \eqref{path int}, we get,
	\begin{eqnarray}
		&&\tilde Z = \int d\zeta_0  {\cal D} \bar\psi {\cal D}\psi \left[ 
		\exp\Big(  -\frac{\zeta_0^2 N \beta V_{2k}}{4\lambda} 
		\Big)  \exp ( -S_0 - S_I) \right] , \\ \nonumber
		&&S_0 = \int d \tau d^{2k}x\Big[ \tilde \psi^\dagger \gamma^\mu \partial_\mu \tilde\psi -i\hat\mu\tilde \psi^\dagger \gamma^0 \tilde\psi + \zeta_0 \tilde \psi^\dagger\tilde \psi  
		+ \frac{\tilde \zeta^2 N}{4\lambda} \Big], \qquad 
		\qquad S_I =   \int d\tau d^{2k} x \;   \tilde\zeta \tilde \psi^\dagger \tilde\psi .
	\end{eqnarray}
	Using the same argument  as before, the contribution due to $S_I$ in the partition function is suppressed by $\frac{1}{\sqrt{N}}$ and can be dropped off in our calculation restricted to the leading order in $N$. Thus ignoring $S_I$, we are left with the gaussian path integral in the non-zero mode of the auxiliary field which is evaluated trivially and the computation of the partition function is reduced to evaluating an ordinary integral over $\zeta_0$ as given below,
	\begin{eqnarray} \label{partwithz}
		\tilde Z &=&  \int d\zeta_0  \exp\left[  -\beta V_{2k} N \Big( \frac{ \zeta_0^2}{4\lambda}  - \frac{1}{\beta} \log Z ( \zeta_0)
		\Big)
		\right],
	\end{eqnarray}
	where 
	\begin{eqnarray}
		\log Z ( \zeta_0) =  2^{k-1} \sum_{n =-\infty}^\infty \int \frac{d^{2k} p}{(2\pi)^{2k}}  \log
		\left[ \frac{  \big((2n+1)\pi-\hat \mu\beta\big)^2 }{\beta^2}  + \vec p^2 + \zeta_0^2 \right].
	\end{eqnarray}
	Note that $2^{k-1}$ factor is due to the fact that the Dirac operator is a $2^k \times 2^k$ dimensional matrix. 
	After evaluating the Matsubara sum, we obtain 
	\begin{eqnarray} \label{aftermat}
		\log Z(\zeta_0) =  \frac{ 2^{k-1}}{\beta^{2k}}  \int \frac{d^{2k} p}{(2\pi)^{2k}} 
		\left[  \sqrt{ \vec p^2 + \zeta_0^2 \beta} +  2 {\rm Re}\log \Big(1+ e^{ - \sqrt{ \vec p^2 + \zeta_0^2 \beta^2 } +i\hat \mu\beta}  \big)
		\right].
	\end{eqnarray}
	And using similar steps as in the case of complex scalars this can be evaluated to give
	\begin{eqnarray} \label{partitionatm}
		\log Z(\zeta_0) =-\frac{\zeta_0 ^{2 k+1} \beta}{\pi^k 2^{k+2}} \left[\sum _{n=0}^k \frac{   (k-n+1)_{2 n} (\zeta_0  \beta )^{-k-n-1} 
			{\rm Re}[{\rm Li}_{k+n+1}\left(-e^{-\zeta_0  \beta -i\hat \mu\beta}\right)]}{2^{n-k-2}n!}
		+\frac{  \Gamma \left(-k-\frac{1}{2}\right)}{\sqrt{\pi }}\right]. \nonumber \\
	\end{eqnarray}
	Again we compute the partition function by approximating the integral \eqref{partwithz} by the value of the integrand at the saddle point $ \zeta_0=\zeta_0^* $, and it is related to the thermal by $ m_{\rm th}=\zeta_0^*. $ The saddle point equation is given by,
	\begin{equation}\label{fer gap}
		m_{\rm th}  = 2 \lambda \frac{1}{\beta} \frac{\partial}{\partial m_{\rm th} } Z(m_{\rm th}) .
	\end{equation}
	At the leading order in large $ \lambda $ it reduces to,
	\begin{equation}
		\frac{\partial }{\partial m_{\rm th}} Z(m_{\rm th})  =0.
	\end{equation}
	The above saddle point condition at large $ \lambda $ leads to the following gap equation in thermal mass $ m_{\rm th} $,
	\begin{align} \label{appengapeqf}
		2 (2m_{\rm th}\beta )^k \sum _{n=0}^{k-1} \frac{(k-n)_{2n} }{ (2 m_{\rm th}  \beta )^n n!}  {\rm Re}[\text{Li}_{k+n}(-e^{-m_{\rm th}  \beta -i\hat \mu\beta})] +\frac{(m_{\rm th} \beta)^{2k}}{\sqrt{\pi}} \Gamma\big(\frac{1}{2}-k\big) =0.
	\end{align}
	The theory at the critical value of the thermal $ m_{\rm th} $ satisfying the gap equation is a CFT.  The stress
	tensor  is given by 
	\begin{equation}
		T_{00} =\frac{1}{V_{2k}N} \frac{ \partial}{\partial \beta } \log \tilde Z( m_{\rm th})\bigg|_{\hat \mu\beta={\rm fixed}} .
	\end{equation}
	\begin{align}
		T_{00}=- \frac{m_{\rm th} ^{2 k+1} \Gamma \left(-k-\frac{1}{2}\right)}{2^{k+2}\pi^{k+\frac{1}{2}}} +\sum_{n=0}^{k+1} \frac{ \left((k+n)^2+k-n\right) (k-n+2)_{2 n-2}  {\rm  Re}[\text{Li}_{k+n}\left(-e^{-m_{\rm th}  \beta -i \mu\beta }\right)]}{(m_{\rm th}\beta)^{n-k-1}2^{n}\beta^{2 k+1}\pi^k n! }.
	\end{align}
	The above equation together with the gap  equation results in 
	\begin{align}\label{T00 f}
		T_{00}=\frac{1}{\beta^{2k+1}\pi^k}\Big(\frac{k}{2k+1}\Big)\sum_{n=0}^{k+1} \frac{  (k-n+2)_{2n} (\beta  m_{\rm th})^{k-n+1}}{ 2^{n-1}n! } {\rm Re} [\text{Li}_{k+n}\big(-e^{-m_{\rm th}  \beta -i\beta\hat\mu}\big)].
	\end{align}
	This expression is the stress tensor or the energy density per fermion at the conformal fixed point and again it scales with the powers of inverse temperature. 
	Using  the gap equation in the partition function it is straightforward to see that the stress tensor is proportional free energy density 
	\begin{align}
		T_{00}=(d-1)F, \qquad {\rm where}\ \ F=-\frac{1}{\beta V_{2k}N}\log \tilde Z.
	\end{align}
	
In the Hamiltonian picture the expectation value of the spin-1 current is given by 
\begin{equation}
J = \langle\bar{ \tilde \psi } \gamma^\mu \tilde \psi \rangle_{(\beta, \mu )},
\end{equation}
Here $\tilde \psi $ is an operator, $\gamma^0$ is Hermitian.  Therefore as an operator, the current is Hermitian. 
We have also used the fact that it is only the time component that 
gets non-trivial expectation value.  From the path integral, this expectation value is obtained by the following
	\begin{align}\label{J}
	 J &=-\frac{i}{\beta }\frac{\partial}{\partial\hat\mu}\log  Z(m_{\rm th})\nonumber\\
		&=\sum_{n=0}^k \frac{ (k-n+1)_{2 n} (m_{\rm th}  \beta )^{k-n} \Big[\text{Li}_{k+n}\left(-e^{-\beta  m_{\rm th} -i \hat\mu\beta }\right)-\text{Li}_{k+n}\left(-e^{-\beta m_{\rm th} +i \hat\mu\beta }\right)\Big]}{\beta^{2k}\pi^k2^{n+1}n!}.
	\end{align}
	Observe that this charge is purely imaginary when the chemical potential is purely imaginary and real when the 
	chemical potential is real as it should be. 

We can also consider the expectation of the spin-1 current in the class ${\cal O}_0[0, 1]$. 
In the Hamiltonian picture it is given by 
\begin{eqnarray}
\tilde J =  \frac{i}{2} \Big\langle   \bar{\tilde \psi }\big(  \partial_\nu   - i \mu \delta_{\nu 0} \big)  \tilde  \psi  - 
  \big( \partial_\nu  + i \mu \delta_{\nu 0}  \big)  \bar{\tilde \psi}  \; \tilde \psi  \Big\rangle_{(\beta, \mu)},
\end{eqnarray}
Note that the current  is a  Hermitian operator.  
Furthermore,  the expectation value of the sum of the terms  occurring in the above equation  vanishes
\begin{eqnarray}
  \frac{i}{2} \Big\langle   \bar{\tilde \psi }\big(  \partial_\nu   - i \mu \delta_{\nu 0} \big)\tilde\psi  +  
   \big( \partial_\nu  + i \mu \delta_{\nu 0}  \big)  \bar{\tilde \psi}  \;  \tilde \psi)  \Big\rangle_{(\beta, \mu)} =0
\end{eqnarray}
This is because the operator involved is a descendant of a primary, 
note here the chemical potential term cancels and the operator is just a total derivative 
of the primary $\bar{\tilde \psi} \psi $.
Therefore we can write $\tilde J$ as 
\begin{eqnarray}
\tilde J =  i  \Big\langle   \bar{\tilde \psi }\big(  \partial_\nu   - i \mu \delta_{\nu 0} \big) \tilde \psi   \Big\rangle_{(\beta, \mu)}.
\end{eqnarray}
Now to obtain this expectation value from the path integral we deform the Euclidean action
	\eqref{action f} in the following manner,
	\begin{equation}
		S[\alpha]= \int d^{2k +1} x \left[ 
		\tilde \psi_a^\dagger (  \gamma^\mu \partial_\mu )\tilde \psi_a -i\hat\mu \tilde\psi_a^\dagger \gamma^0 \tilde\psi_a- \frac{\lambda}{N} ( \tilde \psi_a^\dagger \psi_a)^2-\alpha \tilde \psi^\dagger_a(\partial_\tau -i\hat\mu)\tilde \psi_a \right].
	\end{equation}
	The thermal one point function $J$ 
	 at large $\lambda$ can be evaluated by taking  the derivative of the free energy of the above action with respect to 
	  $\alpha$ evaluated at $\alpha=0$,
	\begin{align}\label{a_O0}
		\tilde J=\langle\tilde \psi^\dagger(\partial_\tau-i\hat\mu)\tilde\psi\rangle=\frac{1}{\beta V_{2k}} \partial_\alpha\log Z[m_{\rm th},\alpha]\Big|_{\alpha=0}.
	\end{align}
The partition function is evaluated by repeating the similar steps of Hubbard-Stratanovich transformation shown earlier in this appendix,
\begin{align}
	\log Z ( \alpha,\zeta_0) =  2^{k-1} \sum_{n =-\infty}^\infty \int \frac{d^{2k} p}{(2\pi)^{2k}}  \log
	\Big[ \frac{1-\alpha^2}{\beta^2}\Big((2n+1)\pi-\frac{i\zeta_0\alpha}{1-\alpha^2}-\hat\mu\Big)^2+\vec p^2+\frac{\zeta_0^2}{1-\alpha^2} \Big],
\end{align}
where $\zeta_0$ is the zero mode of the auxiliary field.
Now, we evaluate $\log Z(\alpha,\zeta_0)$ following similar steps as shown before, and finally from \eqref{a_O0} we obtain,
\begin{align}\label{a_0}
\tilde J 
=\frac{1}{\beta^{2k+1}}\sum _{n=0}^k \frac{  (k-n+1)_{2 n} \big[\text{Li}_{k+n}\left(-e^{ -m_{\rm th} \beta+i \beta \hat \mu }\right)-\text{Li}_{k+n}\left(-e^{-\beta  m_{\rm th}-i \hat\mu\beta }\right)\big]}{(m_{\rm th }\beta)^{n-k-1}\pi^k2^{n+1}n!}.
\end{align}
And it is easy to see from \eqref{a_0} and \eqref{J},
\begin{align} \label{realrel}
		\tilde J =-m_{\rm th} J.
\end{align}

\subsubsection{Evaluation of the partition function using a cut off}
\label{A.2.1}
The divergent integral in  \eqref{aftermat} can also be regularised using a cutoff in the similar approach used for $U(N)$ complex scalars in \ref{appendixa2}. We restrict the discussion here for $d=3$ but it can be straightforwardedly generalised for any higher dimension for both odd and even $d.$ For $d=3$ the 1st term in  \eqref{aftermat} involves the same diverging integral as  it was for the $U(N)$ complex scalars given in \eqref{3dintegralcut} with a momentum cut off $\Lambda$,
\begin{align} \label{3dferintegralcut}
	\lim_{\Lambda\to\infty}\int_0^{\Lambda } k^{} \sqrt{k^2+m_{\rm th}^2} \, dk=\frac{\Lambda ^3}{3}+\frac{\Lambda  m_{\rm th}^2}{2}-\frac{m_{\rm th}^3}{3} + O\Big( \frac{1}{\Lambda} \Big).
\end{align}
Now the integral in the 2nd term of \eqref{aftermat}
is convergent and adding the 1st term regularised using the cutoff  with it and finally differentiating the resulting expression  w.r.t $m_{\rm th} $ we obtain
\begin{align}
	\partial_{m_{\rm th}} \log Z(m_{\rm th})=\frac{\beta  \left(\Lambda  m_{\rm th}-m_{\rm th}^2\right)-2 m_{\rm th}{\rm Re}  \log \left(1+e^{-\beta  m_{\rm th}+i\mu}\right)}{2 \pi }.
\end{align}
Now plugging this in the gap equation \eqref{fer gap} we have
\begin{align}\label{gapf}
	\frac{1}{2\lambda}-\frac{\Lambda}{2\pi}=-\frac{m_{\rm th}}{2\pi}-\frac{1}{\beta\pi}{\rm Re}\log \left(1+e^{-\beta  m_{\rm th}+i\mu}\right).
\end{align}
Now we define the renormalised coupling $\lambda_R$ as
\begin{align}\label{l r fer}
	\frac{1}{2\lambda}-\frac{\Lambda}{2\pi}=\frac{1}{\lambda_R}.
\end{align}
Using this in \eqref{gapf} at zero temperature we have,
\begin{align}\label{fer m 0}
	\frac{1}{\lambda_R}=-\frac{m_{(0)}}{2\pi},
\end{align}
where $m_{(0)}$ is the mass of the theory at zero temperature which serves as a physical scale in the theory. And finally combining \eqref{gapf},\eqref{l r fer} and \eqref{fer m 0} we can express the gap equation for the thermal mass $m_{\rm th} $ in terms of the physical scale $m_{(0)}$ and the inverse temperature $\beta$ as given below.
\begin{align}
	\frac{m_{(0)}}{2\pi}-\frac{m_{\rm th}}{2\pi}=\frac{1}{\beta\pi}{\rm Re}\log (1+e^{-\beta  m_{\rm th}+i\mu}).
\end{align}
The gap equation in this form was also obtained in \cite{Hands:1992ck}, using the similar approach used here. Now setting the physical length scale at its critical value $m_{(0)}=0 $, the dimensionless combination $m_{\rm th}\beta$ turns out to be completely independent of any length scale and thus the theory behaves as a thermal CFT.
	\section{$m_{\rm th}\to-m_{\rm th}$ symmetry in the Gross-Neveu model} \label{symmetry}

		In this appendix we show that  the one point functions of 
		higher spin currents in the class $a_{\mathcal{O}_+}[0,l]$ for Gross-Neveu model in any odd space-time dimension are even functions of the thermal mass $m_{\rm th}$ in the complex $m_{\rm th}$ plane with $-\pi<{\rm Im}(m_{\rm th})<\pi\ {\rm for}\ \mu\in R.$ 
	The proof of this statement boils down to proving a pair of non-trivial mathematical identities of Bernoulli polynomials. 
	Thermal one point functions of the spin-$l$ operators of the class $a_{\mathcal{O}_+}[0,l]$ was given in \eqref{aO+ fer}, replacing $m_{\rm th }$ by $-m_{\rm th}$ we have,
	
		\begin{align}\label{mth to -mth}
		a_{{\cal O}_+}[n=0, l]\Big|_{m_{\rm th}\to -m_{\rm th}} = &\frac{l  }{ 4 \pi ^k (  k - \frac{1}{2} )_l} \left(- \frac{m_{\rm th}}{2} \right)^{l +k -1}\times\nonumber\\ 
		&\sum_{n=0}^{l +k -1} \frac{(-1)^n (l +k -n)_{2n}}{ (2 m_{\rm th}) ^n n!} [{\rm Li}_{k+n} ( - e^{m_{\rm th}+\mu} ) 
		+(-1)^l{\rm Li}_{k+n} ( - e^{m_{\rm th}-\mu} ) ].
	\end{align}
	Now we have to show that the expression above is equal to the original expression of the one point function in equation \eqref{aO+ fer}.
	Now to do so we will use the following standard polylogarithm identity,
	\begin{align}\label{identity}
		i^{-s} {\rm Li}_{s}(e^{2\pi iz})+i^s{\rm Li}_s(e^{-2 \pi i z})=\frac{(2\pi)^s\zeta(1-s,z)}{\Gamma(s)},
	\end{align}
	for $0\le {\rm Re} (z)<1$ if ${\rm Im}(z)\ge 0$ and $0<{\rm Re}(z)\le 1$ if ${\rm Im}(z)<0$; and $ \zeta(1-s,z)$ refers to the Hurwitz zeta function. 
	Using this identity we can recast \eqref{mth to -mth} into the following form,
	\begin{align}\label{after Hurwitz}
		a_{\mathcal{O}_+}&[n=0,l]\Big|_{m_{\rm th}\to-m_{\rm th}}\nonumber\\&=\frac{l m_{\rm th}^{l+k-1} }{ 2^{l+k+1} \pi ^k (  k - \frac{1}{2} )_l}\nonumber
		\sum_{n=0}^{l +k -1} \frac{ (l +k -n)_{2n}}{ (2 m_{\rm th}) ^n n!} [{\rm Li}_{k+n} ( - e^{-m_{\rm th}+\mu} ) 
		+(-1)^l{\rm Li}_{k+n} ( - e^{-m_{\rm th}-\mu} ) ]\nonumber\\
		&\qquad\qquad\qquad\qquad-\frac{(-i)^{k}l {m_{\rm th}}^{k+l-1} }{2^{l+1}  \left(k-\frac{1}{2}\right)_l}[(-1)^lg_{k,l}(m_{\rm th},\mu)+g_{k,l}(m_{\rm th},-\mu)],
	\end{align}
	Where,
	\begin{align}
		g_{k,l}(m_{\rm th},\mu)=\sum _{n=0}^{k+l-1} \frac{\pi ^{n} (k+l-n)_{2 n} \zeta \left(1-k-n,\frac{1}{2}+\frac{m_{\rm th}}{2 \pi  i}+\frac{\mu }{2 \pi  i}\right)}{i^{n} n!  m_{\rm th}^n \Gamma (k+n)},
	\end{align}
	with $-\pi< {\rm Im}(m_{\rm th})<\pi$ and $\mu\in R.$
	Note that the term in the first line of \eqref{after Hurwitz} is exactly the original expression for the thermal one point function $a_{\mathcal{O}_+}[0,l]$ given in \eqref{aO+ fer}, thus for $a_{\mathcal{O}_+}[0,l]$ to be an even function of $m_{\rm th}$ the rest of the terms in \eqref{after Hurwitz} should vanish.

	We will see that indeed $[(-1)^lg_{k,l}(m_{\rm th},\mu)+g_{k,l}(m_{\rm th},-\mu)]$ is zero for all values of $m_{\rm th}$ and $\mu$,
	based on the existence of the non-trivial mathematical identities as given below. Using the relation between Hurwitz zeta functions and Bernoulli polynomials $\zeta(1-n,x)=-\frac{B_n(x)}{n}$, the $[(-1)^lg_{k,l}(m_{\rm th},\mu)+g_{k,l}(m_{\rm th},-\mu)]$ 
	vanishes if the following identity is true
	\begin{eqnarray}\label{y ne 0}
 \sum _{n=0}^{k+l-1} \frac{(-1)^n (k+l-n)_{2 n} }{ n! (2 x)^n \Gamma (k+n+1)} \Big[B_{k+n}\Big(x+y+\frac{1}{2}\Big)+(-1)^lB_{k+n}\Big(x-y+\frac{1}{2}\Big) \Big]=0,
	\end{eqnarray}
	for $k,l=1,2,3,\ldots$.  In the absence of chemical potential, which relates to $y=0$ we must have 
	the following identity for $l$ even 
	\begin{align}\label{y=0}
		\sum _{n=0}^{k+l-1} \frac{(-1)^n (k+l-n)_{2 n} }{ n! (2 x)^n \Gamma (k+n+1)} B_{k+n}\Big(x+\frac{1}{2}\Big)=0.
	\end{align}
	These identities imply that 
	for $\mu=0$ using \eqref{y=0} we obtain $g_{k,l}(m_{\rm th},\mu)=0$ and for $\mu>0$ using \eqref{y ne 0} one gets that $[(-1)^lg_{k,l}(m_{\rm th},\mu)+g_{k,l}(m_{\rm th},-\mu)]=0$. This implies  the $a_{\mathcal{O}_+}[0,l] $ is shown to be an even function of $m_{\rm th} $ in the complex $m_{\rm th } $ plane
 with $-\pi<{\rm Im}(m_{\rm th})<\pi\ {\rm for}\ \mu\in R$ given that the identities \eqref{y ne 0} and \eqref{y=0} hold true.
 We have first verified in Mathematica that these identities are true by substituting $k=1, 2, 3, 4$ and various values of $l$.

We will now prove the identity \eqref{y=0},  we have also proved the  identity \eqref{y ne 0},  but
the proof is more tedious and can be reproduced using the same approach.

	\subsection*{Proof of identity \eqref{y=0}}

	$B_{k+n}(x+\frac{1}{2})$ can be expanded as a polynomial in powers of $x$ using the following finite expansion of the 
	Bernoulli polynomial, 
	\begin{align} \label{bfinitesum}
	B_{n}\left(x+h\right)=\sum_{k=0}^{n}{n\choose k}B_{k}\left(x\right)h^{n-k}.
	\end{align}
	Substituting the above identity in the LHS of  \eqref{y=0} gives, 
\begin{align}\label{lhs}
	\sum _{n=0}^{k+l-1} \frac{(-1)^n (k+l-n)_{2 n} }{ n! (2 )^n \Gamma (k+n+1)}\sum_{m=0}^{n+k} {n+k \choose m} B_m\Big(\frac{1}{2}\Big) x^{k-m}. 
\end{align}
	We organise the expansion in powers  of $x$ by interchanging the order of $n$ and $m$ sums. Thus the expression \eqref{lhs} becomes,
	\begin{align}\label{interchange sum}
		\sum_{m=0}^k x^{k-m} B_m\Big(\frac{1}{2}\Big) \sum_{n=0}^{k+l-1} \frac{(-1)^n(k+l-n)_{2 n}}{2^n n! \Gamma(k+n+1)} {n+k \choose m}\nonumber\\
		+\sum_{m=k+1}^{2k+l-1} x^{k-m} B_m\Big(\frac{1}{2}\Big)\sum_{n=m-k}^{k+l-1} \frac{(-1)^n (k+l-n)_{2 n}}{2^n n! \Gamma(k+n+1)} {n+k\choose m}.
	\end{align}

	If the equation \eqref{y=0} holds, the coefficients at each order in $x$ in \eqref{interchange sum} must vanish, we will show that this indeed is true. 	
	For $ m$ being any odd integer $B_m(\frac{1}{2})$ is zero, thus from now on we will restrict our attention for even integer values of $m$. The coefficients of $x^{k-m}$ in the first $k+1$ terms given in the first line of \eqref{interchange sum} can be identified as
	a   Gauss Hypergeometric function at special values of both  its argument and the order. 
	This function further reduces to the  associated Legendre polynomial at zero argument as shown below,
	\begin{align} 
		\sum_{n=0}^{k+l-1} \frac{(-1)^n(k+l-n)_{2 n}}{2^n n! \Gamma(k+n+1)} {n+k \choose m}=\frac{P_{k+l-1}^{m-k}(0)}{m!}.
	\end{align}
The associated Legendre polynomials   obey the following parity relation
\begin{align}
	P_\alpha^\beta(x)=(-1)^{\alpha-\beta} P_\alpha^\beta(x).
\end{align}
Since both $l$ and $m$ are even integers it is easy to see that,
	\begin{align}
		P_{k+l-1}^{m-k}(0)=0. 
	\end{align}
	This implies the  coefficients in the first $k+1$ terms  in (\ref{interchange sum}) vanish. 
	Now the coefficients of $x^{k-m}$ in the rest of terms given in the second line of \eqref{interchange sum} can also 
	be evaluated to be a Gauss Hypergeometric function  with specific arguments, as given below,
	\begin{align}
		\sum_{n=m-k}^{k+l-1} \frac{(-1)^n (k+l-n)_{2 n}}{2^n n! \Gamma(k+n+1)} {n+k\choose m}=\frac{2^{k+l-1} (-1)^{m-k} \Gamma \left(\frac{l+m}{2}\right)}{m! \Gamma (2 k+l-m) \Gamma \left(\frac{1}{2} (-2 k-l+m+2)\right)}.
	\end{align}
	Now as both $m$ and $l$   are even integers and $m\le 2k+l-2$, due to the last gamma function in the denominator of the above expression it vanishes.
	
	Thus we have shown that the coefficient in each power of $x$ in the expression \eqref{interchange sum} vanishes. 
	this concludes the proof of \eqref{y=0}.

	The same procedure can be repeated for the proof of the identity in (\ref{y ne 0}), we need to use  a similar expansion in $y$  as
	(\ref{bfinitesum}) 
	to write the Bernoulli polynomial in terms of a finite sum, and then perform the  expansion in $x$. 
	The steps are tedious, therefore we refrain from presenting the proof here.

	\bibliographystyle{JHEP}
	\bibliography{references} 	
\end{document}